\documentclass[12pt]{article}
\usepackage[dvips]{color}
\usepackage{epsfig}
\usepackage{amsmath}
\usepackage{graphicx}

\begin{document}
\title{\bf STU/QCD Correspondence}
\author{{B. Pourhassan$^{a}$\thanks{Email: b.pourhassan@du.ac.ir} and J. Sadeghi$^{b}$\thanks{Email: pouriya@ipm.ir}\hspace{1mm}}\\
$^{a}${\small {\em  School of Physics, Damghan University, Damghan, Iran}}\\
$^{b}${\small {\em  Department of Physics, University of Mazandaran,
Babolsar, Iran}}} \maketitle
\begin{abstract}
\noindent In this review article we consider a special case of
$D=5$, $\mathcal{N}=2$ supergravity called the STU model. We apply
the gauge/gravity correspondence to the STU model to gain insight
into properties of the quark-gluon plasma. Given that the
quark-gluon plasma is in reality described by QCD, therefore we call
our study STU/QCD correspondence. First, we investigate the
thermodynamics and hydrodynamics of the STU background. Then we use
dual picture of the theory, which is type IIB string theory, to
obtain the drag force and jet-quenching
parameter of an external probe quark.\\\\
{\bf Keywords:} Gauge/Gravity duality; STU model; String theory;
QGP; QCD.\\
{\bf Pacs:} 04.65.+e, 11.25.Mj, 12.38.Mh.
\end{abstract}
\newpage
\tableofcontents
\newpage
\section{Introduction}
The relation between gauge theories and string theory has been the
subject of many important studies in the last three decades. First,
Maldacena [1] proposed the AdS/CFT correspondence, therefore the
AdS/CFT correspondence sometimes called Maldacena duality. According
to this conjecture there is a relation between a conformal field
theory (CFT) in $d$-dimensional space and a supergravity theory in
$(d+1)$-dimensional anti-de Sitter (AdS) space. Maldacena suggests
that a quantum string in $(d+1)$-dimensional AdS space,
mathematically is equivalent to the ordinary quantum field theory
with conformal invariance in $d$-dimensional space-time which lives
on the boundary of $AdS_{d+1}$ space. The preliminary formulation of
Maldacena are developed and completed by independent works of Witten
[2] and Gubser, et al. [3]. The famous example of AdS/CFT
correspondence is the relation between type IIB string theory in
$AdS_{5}\times S^{5}$ space and $\mathcal{N}=4$ super Yang-Mills
gauge theory on the 4-dimensional boundary of $AdS_{5}$ space. For
more studying about the AdS/CFT correspondence and its applications
see Refs. [4-9]. One of the most interesting application of the
AdS/CFT correspondence is to study of quark-gluon plasma (QGP). A
QGP or quark soup is a phase of quantum chromodynamics (QCD) which
exists at extremely high temperature or density. This phase consists
of free quarks and gluons, which are several of the basic building
blocks of matter. The QGP created at CERN's super proton synchrotron
(SPS) firstly. Current experiments at Brookhaven national
laboratory's relativistic heavy ion collider (RHIC) are continuing
this effort. Nowadays scientists at Brookhaven RHIC have tentatively
claimed to have created a QGP with an approximate temperature of 4
trillion degrees Celsius. The study of the QGP is a testing ground
for finite temperature field theory. Such studies are important to
understand the early evolution of our universe. Already, there are
many attempt to study QCD by using gauge/gravity duality which
usually called AdS/QCD correspondence where the $\mathcal{N}=4$
super Yang-Mills (SYM) plasma considered. The most important
quantities of QGP are the shear viscosity, drag force and
jet-quenching parameter. The shear viscosity is one of the important
hydrodynamical quantities of QGP which relates to the important
thermodynamical quantity so-called entropy, specially it is found
that the ratio of shear viscosity $\eta$ to the entropy density $s$
had a universal value: $\eta/s=1/4\pi$ [10-27]. However, for the
several cases, this value may be enhanced or reduced [28-39]. For
example, $\alpha^{\prime}$ corrections in string theory enhance the
value of $\eta/s$, but higher derivative corrections may be reduced
it. In this paper we use diffusion constant [10, 11] to obtain the
ratio of shear viscosity to entropy density for the three-charged
black hole in the STU model. Also we include higher derivative
correction. The STU model admits a chemical potential for the
$U(1)^{3}$ symmetry and this makes it more interesting. For
instance, presence of a baryon number chemical potential for heavy
quark in the context of AdS/CFT correspondence yields to introducing
a macroscopic density of heavy quark baryons. Already the shear
viscosity in the STU background computed [16, 17] and higher
derivative effects of the five-dimensional gauged supergravity [40]
applied on the ratio of shear viscosity to entropy [41, 42, 43]. We
should note that our paper is extension of the Refs. [17] and [41,
42, 43] because we are going to consider the STU model with three
different charges [44], which corresponds to three different
chemical potential, and arbitrary space curvature. The STU model is
an example of $D=5$, $\mathcal{N}=2$ gauged supergravity theory
which is dual to the $\mathcal{N}=4$ SYM theory with finite chemical
potential. The solutions of ${\mathcal{N}}=2$ supergravity may be
solutions of supergravity theory with more supersymmetry. Already
the duality between gravity and ${\mathcal{N}}=2$ gauged theory
investigated and found that $\mathcal{N}=2$ supergravity is an ideal
laboratory [45-51]. Therefore, it may be to consider the STU model
as a gravity dual of a strongly coupled plasma. In order to avoid
naked singularity of the BPS black holes [50], non-extremal black
holes of five dimensional $\mathcal{N}=2$ AdS supergravity analyzed
in the Ref. [51] and found a lower bound on the non-extremality
parameter where the corresponding non-extremal black hole has
regular horizon. On the other hand the ${\mathcal{N}}=2$
supergravity theory in five dimensions can be obtained by compaction
of the eleven dimensional supergravity in a three-fold Calabi-Yau
[52]. The advantage of Kaluza-Klein reductional dimension and
reduction of supersymmetry to obtain five-dimensional
$\mathcal{N}=2$ gauged supergravity is better understanding the
nature, also some calculation such as quantum correction is very
difficult in the theory with more supersymmetry. Moreover, the
$D=5$, $\mathcal{N}=2$ gauged supergravity theory is a natural way
to explore gauge/gravity duality, and three-charge non-extremal
black holes are important thermal background for this
correspondence. Now, we called this duality as STU/QCD
correspondence. The STU model describes a five-dimensional
space-time which its four-dimensional boundary includes QCD. For
these reasons we focused on the STU background and studied the
problem of the drag force and jet-quenching parameter [53-56]. The
calculation of energy loss of moving heavy charged particle through
a thermal medium known as the drag force. One can consider a moving
heavy quark (such as charm and bottom quarks) through the thermal
plasma with the momentum $P$, mass $m$ and constant velocity $v$,
which is influenced by an external force $F$. So, one can write the
equation of motion as $\dot{P}=F-\zeta P$, where in the
non-relativistic motion $P=mv$, and in the relativistic motion
$P=mv/\sqrt{1-v^{2}}$, also $\zeta$ is called friction coefficient.
In order to obtain drag force, one can consider two special cases.
The first case is the constant momentum ($\dot{P}=0$). So, for the
non-relativistic motion, one can obtain $F=(\zeta m)v$. In this case
the drag force coefficient $(\zeta m)$ will be obtained. In the
second case, external force is zero, so one can find
$P(t)=P(0)exp(-\zeta t)$. In another word, by measuring the ratio
$\dot{P}/P$ or $\dot{v}/v$ one can determine friction coefficient
$\zeta$ without any dependence on mass $m$. These methods lead us to
obtain the drag force for a moving heavy quark in the thermal
plasma. The moving heavy quark in context of QCD has dual picture in
the string theory where an open string attached to the D-brane and
stretched to the horizon of the black hole. The existence of the
black hole is necessary for considering the finite temperature field
theory. Also the existence of the D-brane is necessary for
considering the quark flavor. Moreover the existence of the rotating
black holes in the five dimensional space is necessary for
considering the finite chemical potential field theory. Already the
issue of the drag force considered in the ${\mathcal{N}}=4$ super
Yang-Mills thermal plasma with several interesting backgrounds
[57-64]. In the Ref. [58] the problem of the drag force for the
arbitrary metric studied, and the R-charged black D3-brane
background as an example considered. This is just STU model with
three different charges after the special re-scaling which explain
later in this paper. Therefore our work differs from the Ref. [58],
so we don't like to use any re-scaling on the original metric.
Another important property of the QGP is called the jet-quenching
parameter ($\hat{q}$). The knowledge about this parameter increases
our understanding about the QGP. In that case the jet-quenching
parameter obtained by calculating the expectation value of a closed
light-like Wilson loop and using the dipole approximation [64]. In
order to calculate this parameter in QCD one needs to use
perturbation theory. But, by using AdS/CFT correspondence the
jet-quenching parameter calculated in non-perturbative quantum field
theory. This calculations were already performed in the
$\mathcal{N}=4$ SYM thermal plasma with several interesting
backgrounds [65-72]. Also the effect of higher derivative
corrections such as Gauss-Bonnet on the drag force and the
jet-quenching parameter has been studied [72, 73]. Hence, in the
Ref. [68] we calculated the jet-quenching parameter in STU model
include higher derivative correction and external electric field. We
represent also our results in this paper. In the Ref. [53] we
considered the moving quark at $\mathcal{N}=2$ supergravity and
obtained the drag force for the first time. In that paper we
considered the non-extremal black hole with three equal charge and
have shown that our results at near-extremal limit agree with the
case of $\mathcal{N}=4$ SYM theory. Then in the Ref. [54] we
considered the non-extremal black hole with one charge and
calculated the drag force for the three different spaces: three
dimensional sphere, a pseudo-sphere and a flat space. These cases
are just special case of STU model. So, in the Ref. [55] we extended
our previous works to the general case of STU background, where the
non-extremal black hole has three different charges. Also we studied
the quark-anti quark ($q\bar{q}$) configuration and introduced
rotating $q\bar{q}$ pair in the STU background. Finally in the Ref.
[56] we compute the jet quenching parameter for the case of the
non-extremal black hole with three different charge. We generalize
that work to the case of arbitrary curvature and obtain general
expression of the jet-quenching parameter in this review article.
There are also interesting hydrodynamical quantity such as thermal
and electrical conductivity which can be calculated from
gauge/gravity duality. In the recent work [74] the thermal and
electrical conductivity calculated in presence of non-zero chemical
potential and found that conductivities for gauge theories dual to
R-charged black hole in $d=4$ behaves in a universal manner. In the
Ref. [74] R-charged black holes in arbitrary dimension considered
and electrical conductivity computed. We use results of the Refs.
[62, 63] to write an expression for electrical conductivity as a
hydrodynamical property of the QGP. In the Ref. [75] the STU model
used to describe a relativistic fluid with multiple charges, and
some transport coefficients relevant to the physics of the QGP
calculated. Also in the Ref. [75] a time-dependent version of the
STU model dual to a boost-invariant expanding plasma presented which
may be useful for future studies based on this paper. In this paper
we shall investigate some important properties of the QGP in the STU
model with non-extremal black hole and three different charges.
Indeed, we review some of the previous results and also add some new
things, and collect all of them in this review article. Therefore,
in section 2 we review basic properties of the STU model and obtain
corresponding general relativity equations. In section 3 we extract
thermodynamical quantities of the STU model, and in section 4 we
compute the ratio of shear viscosity to entropy density. Then, in
section 5 we consider the problem of the drag force for the several
configurations. In section 6 we generalized computation of the
jet-quenching parameter to the case of STU black hole with arbitrary
curvature space. Finally in section 7 we summarized our results and
give conclusion.
\section{STU Model}
\subsection{Metric}
The STU model is the special form of the ${\mathcal{N}}=2$
supergravity in several dimensions. This model has generally
8-charged (4 electric and 4 magnetic) non-extremal black hole.
However, there are many situations with less than charges such as
four-charged and three-charged black holes. In that case there is
great difference between the three-charged and four-charged black
holes. For example if there are only 3 charges, then the entropy
vanishes (except in the non-BPS case). So, one really needs four
charges to get a regular black hole. In 5 dimensions the situation
is different and actually much simpler, there is no distinction
between BPS and non-BPS branch. So, in 5 dimensions the
three-charged configurations are the most interesting ones [76].
Therefore, we begin with the three-charged non-extremal black hole
solution in ${\mathcal{N}}=2$ gauged supergravity which is called
STU model and described by the following solution [77],
\begin{equation}\label{s1}
ds^{2}=-\frac{f_{k}}{{\mathcal{H}}^{\frac{2}{3}}}dt^{2}
+{\mathcal{H}}^{\frac{1}{3}}(\frac{dr^{2}}{f_{k}}+\frac{r^{2}}{R^{2}}d\Omega_{3,k}^{2}),
\end{equation}
where,
\begin{eqnarray}\label{s2}
f_{k}&=&k-\frac{\mu}{r^{2}}+\frac{r^{2}}{R^{2}}{\mathcal{H}},\nonumber\\
{\mathcal{H}}&=&\prod_{i=1}^{3} H_{i},\nonumber\\
H_{i}&=&1+\frac{q_{i}}{r^{2}}, \hspace{10mm} i=1, 2, 3,\nonumber\\
A_{t}^{i}&=&\sqrt{\frac{kq_{i}+\mu}{q_{i}}}(1-H_{i}^{-1}),
\end{eqnarray}
where $R$ is the constant AdS radius and relates to the coupling
constant via $R=1/g$ (also, coupling constant relates to the
cosmological constant via $\Lambda=-6g^2$), and $r$ is the radial
coordinate along the black hole, so the boundary of AdS space
located at $r\rightarrow\infty$ (or $r=r_{m}$ on the D-brane). The
black hole horizon specified by $r=r_{h}$ which is obtained from
$f_{k}=0$. In the STU model there are three real scalar fields,
which is also solution of the metric (1), as
$X^{i}={\mathcal{H}}^{\frac{1}{3}}/H_{i}$, which satisfy the
following condition, $\prod_{i=1}^{3}X^{i}=1$. In another word, if
we set $X^{1}=S$, $X^{2}=T$, and $X^{3}=U$, then there is the
$STU=1$ condition. For the three R-charges $q_{i}$, in the equation
(2), there is an overall factor such as
$q_{i}=\mu\sinh^{2}\beta_{i}$, where $\mu$ is called non-extremality
parameter and $\beta_{i}$ are related to the three independent
electrical charges of the black hole. Finally, the factor of $k$
indicates the space curvature, so the metric (1) includes a $S^{3}$
(three dimensional sphere) for $k=1$, a pseudo-sphere for $k=-1$ and
a flat space for $k=0$. So, for $k=1$, $k=0$ and $k=-1$ one can
write, respectively,
\begin{eqnarray}\label{s3}
d\Omega_{3,k}^{2}\equiv\big\{\begin{array}{ccc}
R^{2}(d\rho^{2}+\sin^{2}\rho d\theta^{2}+\sin^{2}\rho\sin^{2}\theta d\phi^{2}) \\
dx^{2}+dy^{2}+dz^{2} \\
R^{2}(d\rho^{2}+\sinh^{2}\rho d\theta^{2}+\sinh^{2}\rho\sin^{2}\theta d\phi^{2}) \\
\end{array}
\end{eqnarray}

\subsection{Equations}
By introducing the new variable,
\begin{equation}\label{s4}
u=\frac{1}{6}\ln(H_{1}H_{2}H_{3}),
\end{equation}
one can obtain the following independent Christophel symbols,
\begin{eqnarray}\label{s5}
\Gamma_{rt}^{t}&=&\frac{1}{2}(-4u^{\prime}+\frac{f_{k}^{\prime}}{f_{k}}),\nonumber\\
\Gamma_{ri}^{i}&=&u^{\prime}+\frac{1}{r},\nonumber\\
\Gamma_{rr}^{r}&=&u^{\prime}-\frac{f_{k}^{\prime}}{2f_{k}},\nonumber\\
\Gamma_{tt}^{r}&=&\frac{1}{2}e^{-6u}f_{k}(f_{k}^{\prime}-4f_{k}u^{\prime}),\nonumber\\
\Gamma_{ii}^{r}&=&-f_{k}r(1+ru^{\prime}),
\end{eqnarray}
where the index $i$ refers to the angular components. This yields us to the following non-zero components of Riemann tensor,
\begin{eqnarray}\label{s6}
R_{iti}^{t}&=&\frac{r}{2}(4f_{k}u^{\prime}+4f_{k}ru^{\prime2}-f_{k}^{\prime}-rf_{k}^{\prime}u^{\prime}),\nonumber\\
R_{rtr}^{t}&=&\frac{1}{2}(4u^{\prime\prime}-\frac{f_{k}^{\prime\prime}}{f_{k}}-12u^{\prime2}+7\frac{f_{k}^{\prime}u^{\prime}}{f_{k}}),\nonumber\\
R_{tit}^{i}&=&\frac{1}{2}f_{k}e^{-6u}(f_{k}^{\prime}u^{\prime}-4f_{k}u^{\prime2}+\frac{f_{k}^{\prime}}{r}-4\frac{f_{k}u^{\prime}}{r}),\nonumber\\
R_{rir}^{i}&=&-u^{\prime\prime}-\frac{u^{\prime}}{r}-\frac{f_{k}^{\prime}u^{\prime}}{2f_{k}}-\frac{f_{k}^{\prime}}{2f_{k}r},\nonumber\\
R_{trt}^{r}&=&\frac{1}{2}f_{k}e^{-6u}(12f_{k}u^{\prime2}+f_{k}^{\prime\prime}-7f_{k}^{\prime}u^{\prime}-4f_{k}u^{\prime\prime}),\nonumber\\
R_{iri}^{r}&=&-\frac{r}{2}(2f_{k}u^{\prime}+f_{k}^{\prime}+rf_{k}^{\prime}u^{\prime}+2rf_{k}u^{\prime\prime}).
\end{eqnarray}
Hence, we can extract the following components of the Ricci tensor,
\begin{eqnarray}\label{s7}
R_{i}^{i}&=&e^{-2u}(\frac{f_{k}u^{\prime}-f_{k}^{\prime}}{r}+2f_{k}u^{\prime2}-f_{k}^{\prime}u^{\prime}-f_{k}u^{\prime\prime}),\nonumber\\
R_{t}^{t}&=&e^{-2u}(\frac{4f_{k}u^{\prime}-f_{k}^{\prime}}{2r}+3f_{k}^{\prime}u^{\prime}-4f_{k}u^{\prime2}-\frac{f_{k}^{\prime\prime}}{2}+2f_{k}u^{\prime\prime}),\nonumber\\
R_{r}^{r}&=&e^{-2u}(f_{k}u^{\prime\prime}-\frac{f_{k}^{\prime\prime}}{2}-6f_{k}u^{\prime2}-\frac{f_{k}u^{\prime}}{r}+3f_{k}^{\prime}u^{\prime}-\frac{f_{k}^{\prime}}{2r}).
\end{eqnarray}
Finally one can find the Ricci scalar as the following,
\begin{equation}\label{s8}
\mathcal{R}=e^{-2u}\left[2\frac{f_{k}u^{\prime}-f_{k}^{\prime}}{r}-8f_{k}u^{\prime2}+5f_{k}^{\prime}u^{\prime}-2f_{k}u^{\prime\prime}-f_{k}^{\prime\prime}\right].
\end{equation}
In order to study complete field equations which product metric (1)
see Ref. [38].
\subsection{Horizon structure}
Now, we would like to discuss horizon structure of the metric (1).
In the Ref. [51] the appropriate conditions for the existence of
horizon in the STU model with $k=1$ extracted. Here, we give similar
discussion for arbitrary $k$ and obtain exact relation for the black
hole horizon. The $f_{k}=0$ reduced to the following equation,
\begin{equation}\label{s9}
r^{6}+\mathcal{A}r^{4}-\mathcal{B}r^{2}+q_{1}q_{2}q_{3}=0,
\end{equation}
where $\mathcal{A}\equiv q_{1}+q_{2}+q_{3}+kR^{2}$, and
$\mathcal{B}\equiv \mu R^{2}-q_{1}q_{2}-q_{2}q_{3}-q_{1}q_{3}$. A
possible solutions of the equation (9) is given by,
\begin{equation}\label{s10}
r_{\pm}=\pm\left(\frac{W^{2}-2\mathcal{A}W+4(3\mathcal{B}+\mathcal{A}^{2})}{6W}\right)^\frac{1}{2},
\end{equation}
where we defined,
\begin{eqnarray}\label{s11}
W^{3}&=&-36\mathcal{A}\mathcal{B}-108\prod_{i=1}^{3}q_{i}-8\mathcal{A}^{3}\nonumber\\
&+&12\sqrt{-12\mathcal{B}^{3}-3\mathcal{A}^{2}\mathcal{B}^{2}+54\mathcal{A}\mathcal{B}\prod_{i=1}^{3}q_{i}+81(\prod_{i=1}^{3}q_{i})^{2}+12\mathcal{A}^{3}\prod_{i=1}^{3} q_{i}}.
\end{eqnarray}
The $r_{+}$ denotes outer horizon, while the $r_{-}$ denotes inner
horizon. The equation (9) has generally six solutions. Other
solutions of the equation (9) are imaginary therefore we neglect
them. In order to see behavior of the black hole horizon we give
plots of $f_{k}$ in terms of radius for possible values of $k$ in
the Fig. 1 and Fig. 2.

\begin{figure}[th]
\begin{center}
\includegraphics[scale=.29]{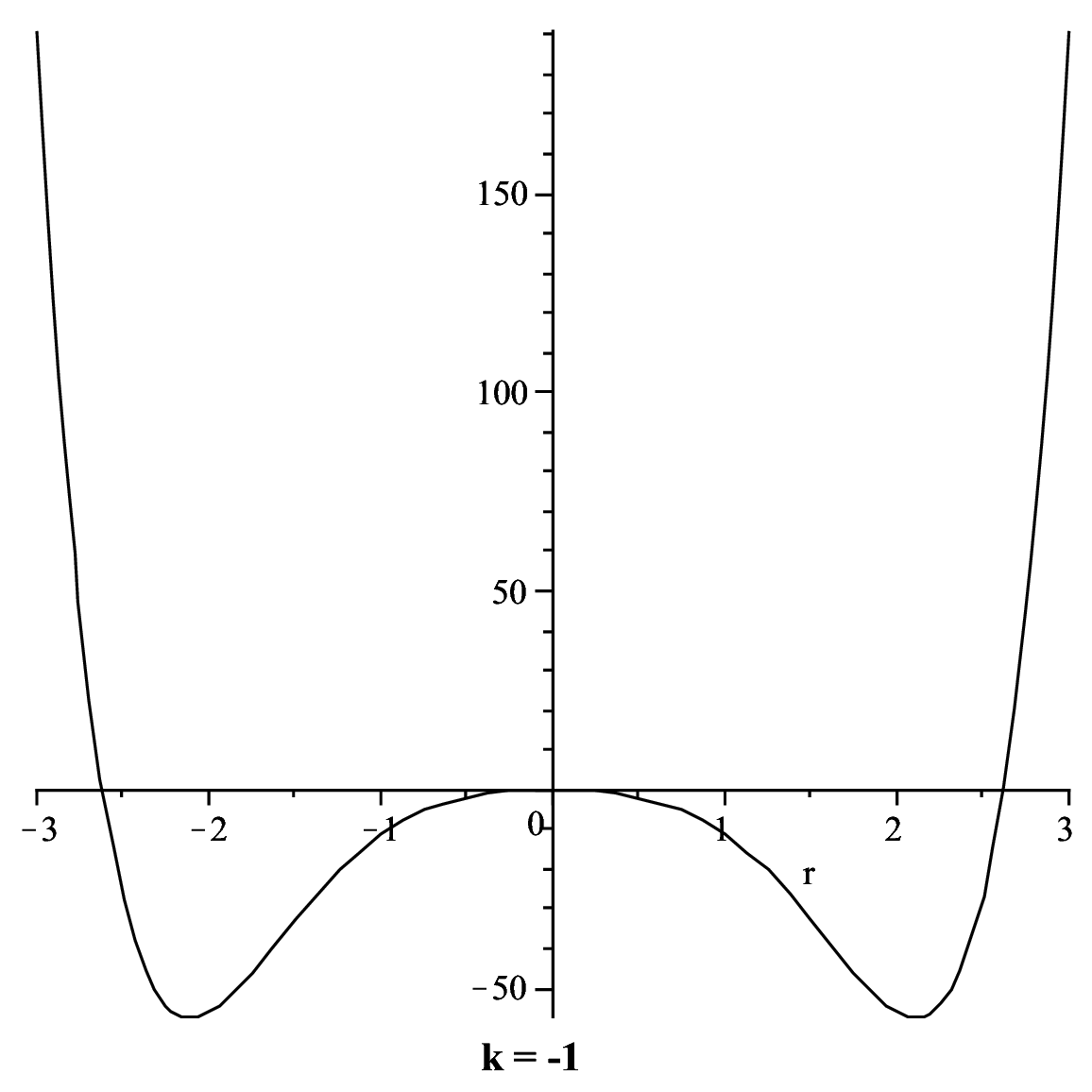}
\caption{Typical horizon situation of STU black hole with $k=-1$ for
small black hole charges.}
\end{center}
\end{figure}

\begin{figure}[th]
\begin{center}
\includegraphics[scale=.29]{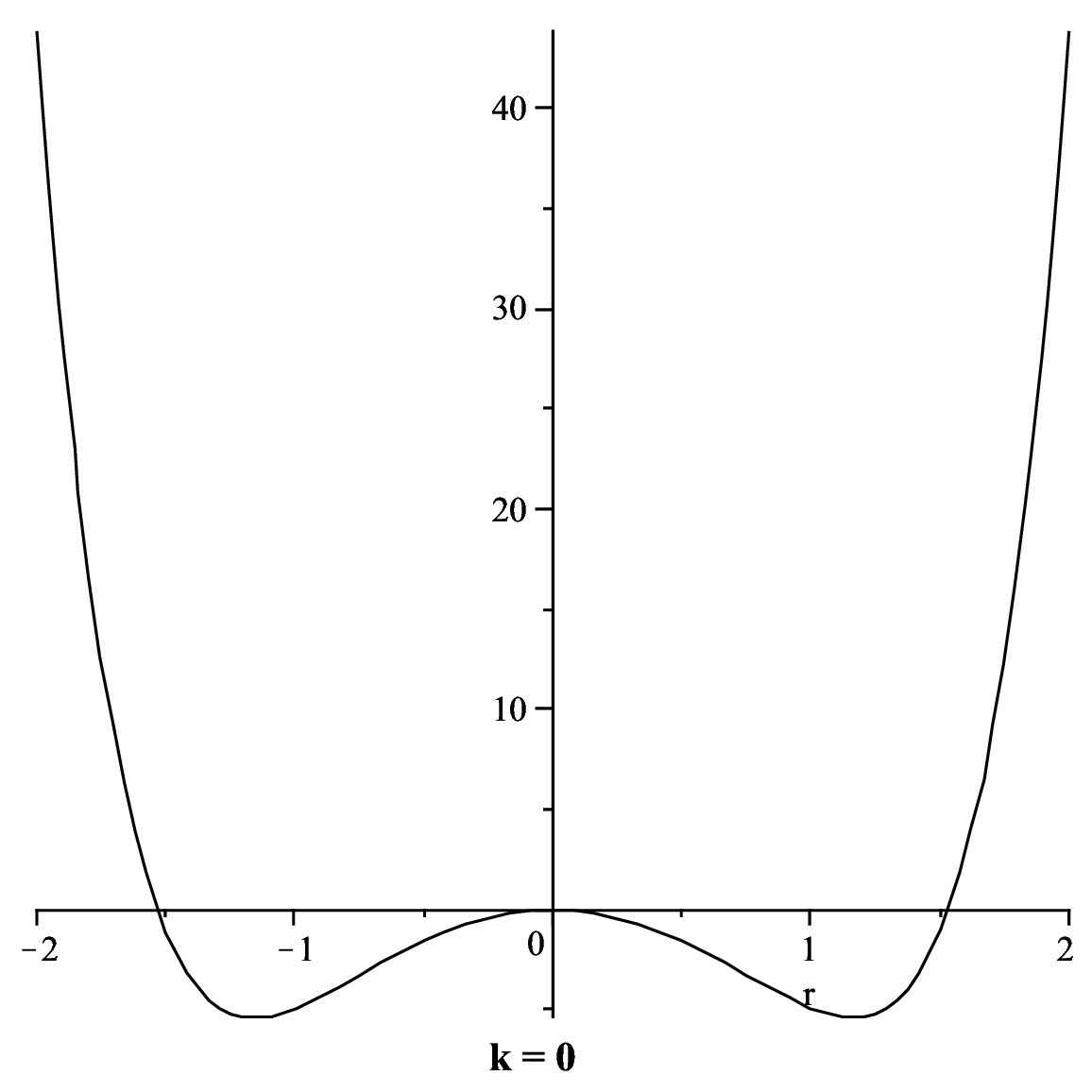}\hspace{1cm}\includegraphics[scale=.29]{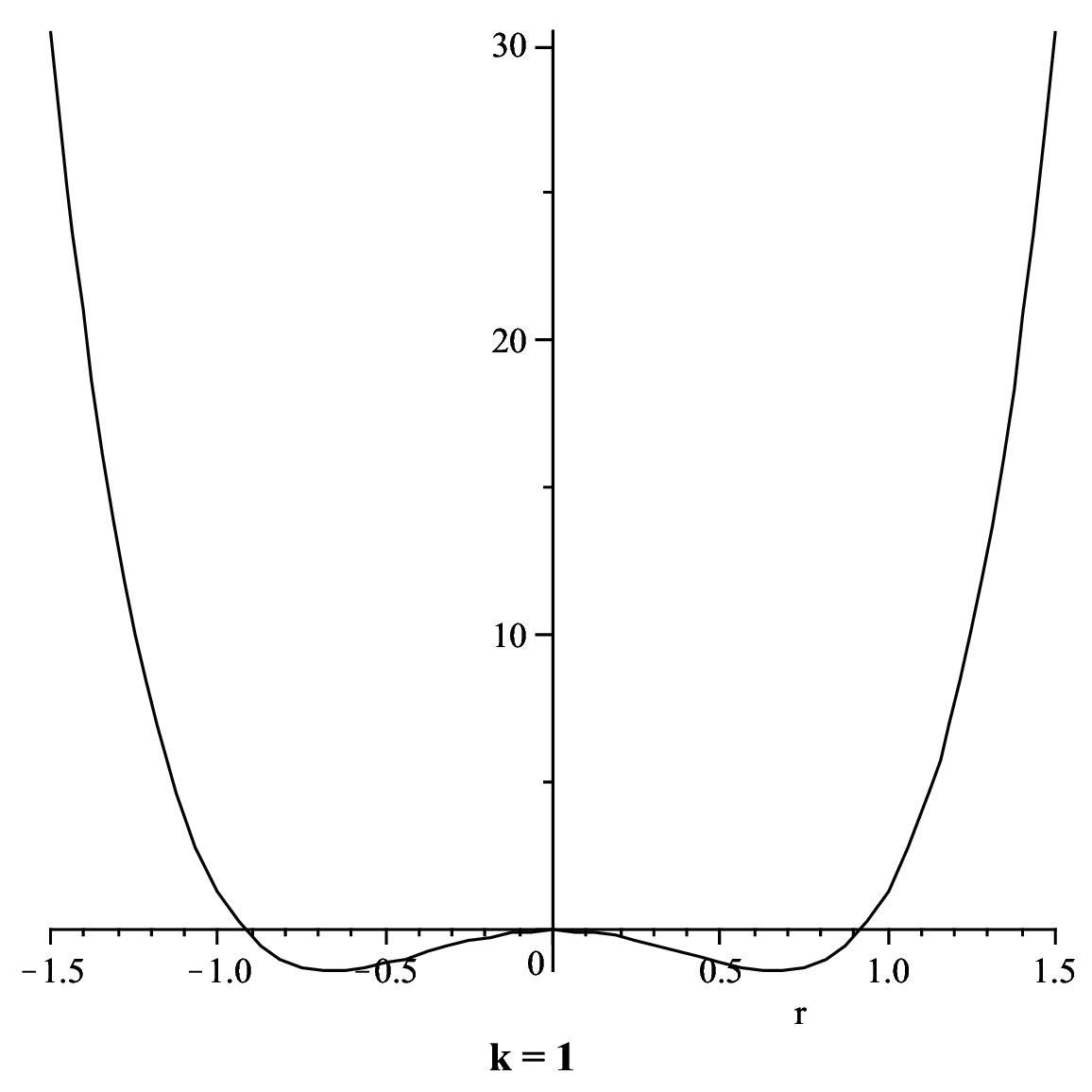}
\caption{Typical horizon situations of STU black hole with $k=0$ and
$k=1$ for small black hole charges.}
\end{center}
\end{figure}

However, with $r^{2}\equiv x$ the function
$x^{2}R^{2}f(x)=x^{3}+\mathcal{A}x^{2}-\mathcal{B}x+\prod_{i=1}^{3}q_{i}$,
has two extremum at $x_{\pm}=\frac{\mathcal{A}}{3}(-1\pm y)$ where
$y=1+z\equiv\sqrt{1+\frac{3\mathcal{B}}{\mathcal{A}^{2}}}>1$, so
$x_{-}<0$ is not acceptable region. Therefore, in order to have at
least one horizon in the positive region it should be to have
$x_{+}^{2}f(x_{+})\leq0$ which implies that $-2z^{3}-3z^{2}+c\leq0$,
where $c\equiv(\frac{3}{\mathcal{A}})^{3}\prod_{i=1}^{3}q_{i}\leq1$.
In order to find $z$, we restrict ourself to the following cases:\\
(I) $\frac{q_{i}}{R^{2}}\ll1$ which implies $c\ll1$,
$\mathcal{A}\simeq kR^{2}$ and $\mathcal{B}\simeq\mu\mathcal{A}$.\\
(II) $\frac{q_{i}}{R^{2}}\gg1$ and $q_{i}\sim q$ which implies
$c-1\ll1$, $\mathcal{A}\simeq 3q$ and $\mathcal{B}\simeq\mu
R^{2}-3q^{2}$.\\
In the first case one can obtain
$z=\sqrt\frac{c}{3}$ which yields to the following critical value
for the non-extremality parameter,
\begin{equation}\label{s12}
\mu_{c}=2\sqrt{\frac{k}{R^{2}}q_{1}q_{2}q_{3}}+\frac{q_{1}q_{2}+q_{2}q_{3}+q_{1}q_{3}}{R^{2}}.
\end{equation}
It tells us that the first approximation is only valid for the cases of $k=0$ and $k=1$. We can see that the space-time including pseudo sphere ($k=-1$)
yields to imaginary non-extremality parameter at critical point. In the second case one can obtain $z=1/2$ which yields to the following critical value for
the non-extremality parameter,
\begin{equation}\label{s13}
\mu_{c}=\frac{27}{4}\frac{q^{2}}{R^{2}}+\frac{5}{2}kq+\frac{5}{12}k^{2}R^{2}.
\end{equation}
Therefore, we success to obtain exact expression for horizon radius
and calculate approximate values for the non-extremality
parameter.\\
In the next step we add higher derivative terms and give horizon
radius, and try to obtain critical value of the non-extremality
parameter.
\subsection{Higher derivatives}
The higher derivative corrections to R-charged $AdS_{5}$ black holes
studied originally for the black hole with three equal charges [78]
where four-derivative corrections to the bosonic sector of
five-dimensional $\mathcal{N}=2$ gauged supergravity considered.
Then, the same problem in the STU model to linear order of the four
derivative terms constructed [40]. Now, we would like to extend this
work to the case of three different charges [55]. These solutions
generalize the Gauss-Bonnet black holes to Einstein-Maxwell theory.
In that case the metric (1) reminds unchange but,
\begin{eqnarray}\label{s14}
f_{k}&=&k-\frac{\mu}{r^2}+\frac{r^{2}}{R^{2}}\prod_{i}(1+\frac{q_i}{r^2})+c_{1}
\left(\frac{\mu^{2}}{96r^{6}\prod_{i}(1+\frac{q_i}{r^2})}-\frac{\prod_{i}q_{i}(q_{i}+\mu)}{9R^{2}r^{4}}\right),\nonumber\\
{\mathcal{H}}&=&\prod_{i=1}^{3} H_{i},\nonumber\\
H_{i}&=&1+\frac{q_i}{r^2}-\frac{c_{1}q_{i}(q_{i}+\mu)}{72r^{2}(r^{2}+q_{i})^{2}},
\hspace{10mm} i=1, 2, 3,\nonumber\\
A_{t}^{i}&=&\sqrt{\frac{kq_{i}+\mu}{q_{i}}}(1-\frac{1+c_{1}a_{1}}{H_{i}}),
\end{eqnarray}
where $c_{1}$ is the small constant parameter corresponding to the
higher derivative terms and $a_{1}$ is $q_{i}$-dependent quantity
which parameterize the corrections to the background geometry [40].
In that case the modified horizon radius for the case of $k=1$ is
given by the following expression,
\begin{eqnarray}\label{s15}
r_{h}&=&r_{0h}\nonumber\\
&+&\frac{c_{1}\prod_{i}(1+\frac{q_i}{r_{0h}^2})\left(\sum q_{i}^{2}-\frac{26r_{0h}^{2}}{3}\sum q_{i}+3r_{0h}^{4}\right)}{576R^{2}
\left[(\prod_{i}(1+\frac{q_i}{r_{0h}^2}))^{\frac{2}{3}}(\frac{1}{3}\sum
q_{i}-2r_{0h}^{2})-R^{2}\right]}\nonumber\\
&+&c_{1}\frac{2(\prod_{i}(1+\frac{q_{i}}{r_{0h}^2}))^{\frac{1}{3}}(\frac{13}{3}\sum
q_{i}-3r_{0h}^{2})+3R^{2}}{576
\left[(\prod_{i}(1+\frac{q_{i}}{r_{0h}^2}))^{\frac{2}{3}}(\frac{1}{3}\sum
q_{i}-2r_{0h}^{2})-R^{2}\right]}
\end{eqnarray}
where $r_{0h}$ is the horizon radius without higher derivative
corrections which is given by the equation (10). We should note
that, in order to obtain the expression (15) we removed $\mu$ by
using $f_{k}=0$. The $k=1$ solutions is more appropriate to studies
of the thermodynamic and hydrodynamic regimes of the theory, and
also have interesting application in the horizon structure of the
small black holes.\\
Just as previous subsection it is interesting
to find a critical value $\mu_{c}$. By using
$\frac{q_{i}}{R^{2}}\ll1$ approximation we find that all previous
relations are valid just we find a difference in the $c$, so one can
obtain,
\begin{equation}\label{s16}
c\equiv(\frac{3}{\mathcal{A}})^{3}[\prod_{i=1}^{3}q_{i}-\frac{c_{1}}{9}\prod_{i=1}^{3}q_{i}(q_{i}+\mu)].
\end{equation}
In that case the critical value $\mu_{c}$ is root of the following equation,
\begin{equation}\label{s17}
4c_{2}\mu^{3}+R^{2}\mu^{2}+(4c_{2}-2)(q_{1}q_{2}+q_{2}q_{3}+q_{1}q_{3})\mu+4(c_{2}-1)q_{1}q_{2}q_{3}=0,
\end{equation}
where we defined $c_{2}\equiv\frac{c_{1}}{9}q_{1}q_{2}q_{3}$. It is clear that $c_{1}=0$ yields to the relation (12).
\section{Thermodynamics}
Here, we study thermodynamics of STU black hole and extract some
important thermodynamical quantities such as temperature and entropy
and extend them to the higher derivative theory. Also we discuss
dual picture of the STU model which is $\mathcal{N}=4$ SYM with
finite chemical potential.
\subsection{Quantities}
In this section we are going to compute some thermodynamical
quantities in the STU model with three different black hole charges
for the arbitrary spaces. Some of these quantities such as
temperature and entropy will be useful to study QGP in the next
sections. The thermodynamics of the STU model has been studied for
the special cases [17, 51, 79, 80]. So, the main goal of this
section is to generalize previous studies and review the
thermodynamics of the STU black hole solution generally.
Also, we recall special re-scaling where the metric (1) changes to the dual picture namely $\mathcal{N}=4$ SYM with finite chemical potential.\\
According to the previous works, the Hawking temperature of the
black hole solution (1) will be as [77],
\begin{equation}\label{s18}
T=\frac{r_{h}}{2\pi
R^{2}}\frac{2+\frac{1}{r_{h}^{2}}\sum_{i=1}^{3}{q_{i}}-\frac{1}{r_{h}^{6}}\prod_{i=1}^{3}{q_{i}}}{{\sqrt{\prod_{i=1}^{3}(1+\frac{q_{i}}{r_{h}^{2}})}}}.
\end{equation}
There is also a chemical potential which is given by the following relation,
\begin{equation}\label{s19}
\phi_{i}^{2}=q_{i}(r_{h}^{2}+q_{i})\left(\frac{1}{R^{2}r_{h}^{2}}\prod_{j\neq
i}(r_{h}^{2}+q_{j})+k\right).
\end{equation}
Also, the entropy density in $d=4$ dimension is given by the
following expression, which is valid for $k=\pm1$ and $k=0$,
\begin{equation}\label{s20}
s=\frac{1}{4GR^{3}}\left(r_{h}^{3}\sqrt{\mathcal{H}(r_{h})}\right),
\end{equation}
where $G$ is Newton's constant and relates to the AdS curvature as
$G=\frac{\pi R^{3}}{2N^{2}}$, where $N$ is the number of colors. By
combining relations (18) and (20) and relation
$C_{v}=T\frac{\partial s}{\partial T}$ one can obtain the specific
heat of the theory, which is also valid for $k\pm1$ and $k=0$,
\begin{equation}\label{s21}
C_{v}=\frac{r_{h}^{2}\sqrt{\prod_{i=1}^{3}(r_{h}^{2}+q_{i})}}{4GR^{3}}\frac{\bar{M}}{\bar{N}},
\end{equation}
where we defined,
\begin{eqnarray}\label{s22}
\bar{M}&=&6r_{h}^{10}
+7\sum_{i}{q_{i}}r_{h}^{8}+2((\sum_{i}{q_{i}})^{2}+\sum_{i\neq j}{q_{i}q_{j}})r_{h}^{6}+(\sum_{i}{q_{i}}\sum_{i\neq j}{q_{i}q_{j}}-3\prod_{i}q_{i})r_{h}^{4}\nonumber\\
&-&2\sum_{i}{q_{i}}\prod_{i}q_{i}r_{h}^{2}-\sum_{i\neq j}{q_{i}q_{j}}\prod_{i}q_{i},\nonumber\\
\bar{N}&=&2r_{h}^{12} +3\sum_{i}{q_{i}}r_{h}^{10}+6\sum_{i\neq j}{q_{i}q_{j}}r_{h}^{8}
+(16\sum_{i}{q_{i}}+\sum_{j}\sum_{i\neq j}{q_{j}q_{i}^{2}})r_{h}^{6}\nonumber\\
&+&6\sum_{i}{q_{i}}\prod_{i}q_{i}r_{h}^{4}
+3\prod_{i}q_{i}\sum_{i\neq j}{q_{i}q_{j}}r_{h}^{2}+2\prod_{i}q_{i}^{2}.
\end{eqnarray}
For the case of $q=0$ one can obtain
$C_{v}=\frac{3\pi^{2}N^{2}}{2}T^{3}$. Above relations show that the
three cases of $k=-1,0,1$ yield to the same thermodynamical
quantities.\\
Another important thermodynamical quantity is the free energy
($F=-\int{sdT}$) of the theory,
\begin{eqnarray}\label{s23}
F=&-&\frac{4r_{h}^{6}\sum_{i\neq j}(q_{i}q_{j}^{2}-q_{j}q_{i}^{2})
+3r_{h}^{4}\sum_{i\neq j}(q_{i}q_{j}^{3}-q_{j}q_{i}^{3})}{12R^5\prod_{i\neq j}(q_{i}-q_{j})}\nonumber\\
&-&\frac{6r_{h}^{2}(\sum_{i\neq j}(q_{i}q_{j}^{4}-q_{j}q_{i}^{4})
+3\sum_{i\neq
j}(q_{i}^{2}q_{j}^{3}-q_{j}^{2}q_{i}^{3}))}{12R^5\prod_{i\neq
j}(q_{i}-q_{j})},
\end{eqnarray}
up to $\mathcal{O}(\ln{r_{h}})$. Importance of the free energy is its relation with the total energy and partition function, so by using the free energy
(23) one can obtain,
\begin{eqnarray}\label{s24}
E\sim\frac{4r_{h}^{8}
+3\sum_{i}q_{i}-6(\sum_{i}q_{i}^{2}+2\sum_{i}q_{i}-2\sum_{i\neq
j}q_{i}q_{j})r_{h}^{4}+2\sum_{i}q_{i}-12\prod_{i}q_{i}}{12R^{5}r_{h}^{2}},
\end{eqnarray}
where we used $E=F+sT$. Then the partition function specifies by using the relation $F=-T\ln Z$.\\
Now, we can discuss the above thermodynamical quantities for three
different cases of one, two, and three-charged black holes.
\subsubsection{one-charged black hole} In the case (i) we set
$q_{1}=q, q_{2}=q_{3}=0$, so the specific heat (21) reduced to the
following expression,
\begin{equation}\label{s25}
C_{v}=\frac{N^{2} r_{h}^{4}\sqrt{(r_{h}^{2}+q)}}{2\pi R^{6}}\frac{6r_{h}^{4}+7qr_{h}^{2}+2q^{2}}{2r_{h}^{6}+3qr_{h}^{4}+16q},
\end{equation}
where the horizon radius in terms of the temperature obtained from the relation (18) as the following,
\begin{equation}\label{s26}
r_{h}^{2}=\frac{1}{4}(-2q+2\pi^{2}R^{4}T^{2}+2\sqrt{2q\pi^{2}R^{4}T^{2}+\pi^{4}R^{8}T^{4}}).
\end{equation}
The free energy in this case vanishes, hence partition function has
unit value ($Z=1$), and the total energy becomes
$E=Ts=\frac{r_{h}^{2}}{R^{5}}(2r_{h}^{2}+q)$. In the Fig. 3 we plot
the specific heat in terms of the temperature (solid line of the
Fig. 3).
\subsubsection{two-charged black hole} In the case (ii) we
set $q_{1}=q_{2}=q, q_{3}=0$, so the specific heat (21) reduced to
the following expression,
\begin{equation}\label{s27}
C_{v}=\frac{N^{2}(r_{h}^{2}+q)}{2\pi R^{6}}\frac{2r_{h}^{7}+7qr_{h}^{5}+5q^{2}r_{h}^{3}+q^{3}r_{h}}{r_{h}^{6}+3qr_{h}^{4}+3q^{2}r_{h}^{2}+16q+q^{3}},
\end{equation}
where $r_{h}=\pi R^{2} T$.\\
The free energy, and hence partition function and total energy will
be infinite in this case. It is important to note that the
temperature of this situation is similar to the temperature of the
zero-charge limit ($q=0$), which is
corresponding to the $\mathcal{N}=4$ SYM plasma.\\
In the Fig. 3 we plot the specific heat in terms of the temperature
(dotted line of the Fig. 3). It is clear that the specific heat
increases by the black hole temperature.\\
We find that also the
specific heat for the case of two-charge black hole (the case of ii)
is larger than the case of one-charge black hole (the case of i).
\subsubsection{three-charged
black hole} In the case (iii) we set $q_{1}=q_{2}=q_{3}=q$, so the
specific heat (21) reduced to the following expression,
\begin{equation}\label{s28}
C_{v}=\frac{N^{2} r_{h}^{2}(r_{h}^{2}+q)^{\frac{3}{2}}}{2\pi R^{6}}\frac{6r_{h}^{10}
+21qr_{h}^{8}+24q^{2}r_{h}^{6}+6q^{3}r_{h}^{4}-6q^{4}r_{h}^{2}-3q^{5}}{2r_{h}^{12} +9qr_{h}^{10}+18q^{2}r_{h}^{8} +(48q+3q^{3})r_{h}^{6}+18q^{4}r_{h}^{4}
+9q^{5}r_{h}^{2}+2q^{6}},
\end{equation}
where,
\begin{equation}\label{s29}
r_{h}^{2}=\frac{1}{6}\left[\Psi+\frac{9q^{2}+4\pi^{4}R^{8}T^{4}}{\Psi}+2\pi^{2}R^{4}T^{2}\right],
\end{equation}
with,
\begin{equation}\label{s30}
\Psi^{3}=8\pi^{6}R^{12}T^{6}+27q^{2}\pi^{2}R^{4}T^{2}-27q^{3} +3q\pi
R^{2}T\sqrt{162q^{3}+27q^{2}\pi^{2}R^{4}T^{2}+48q\pi^{4}R^{8}T^{4}}.
\end{equation}
In the Fig. 3 we give plot the specific heat in terms of the black
hole temperature (dashed line of the Fig. 3).\\
We find that the specific heat for the case of three-charge black
hole (the case of iii) is larger than the case of one-charge black
hole (the case of i) and two-charge black
hole (the case ii). It tells that the number of the black hole charge increases the specific heat.\\
Before end of this subsection it is interesting to recall that the
domain of thermodynamical stability is given by the inequality
$\frac{q_{1}+q_{2}+q_{3}}{r_{h}^{2}}-\frac{q_{1}q_{2}q_{3}}{r_{h}^{6}}<2$
[69]. It is clear that the one-charged black hole has condition
$q/r_{h}^{2}<2$ and two-charged black hole has condition
$q/r_{h}^{2}<1$. We will use these conditions later to fix the black
hole charges.\\
In the next step we introduce interesting transformation which
changes the STU background to the $\mathcal{N}=4$ SYM with finite
chemical potential.

\begin{figure}[th]
\begin{center}
\includegraphics[scale=.3]{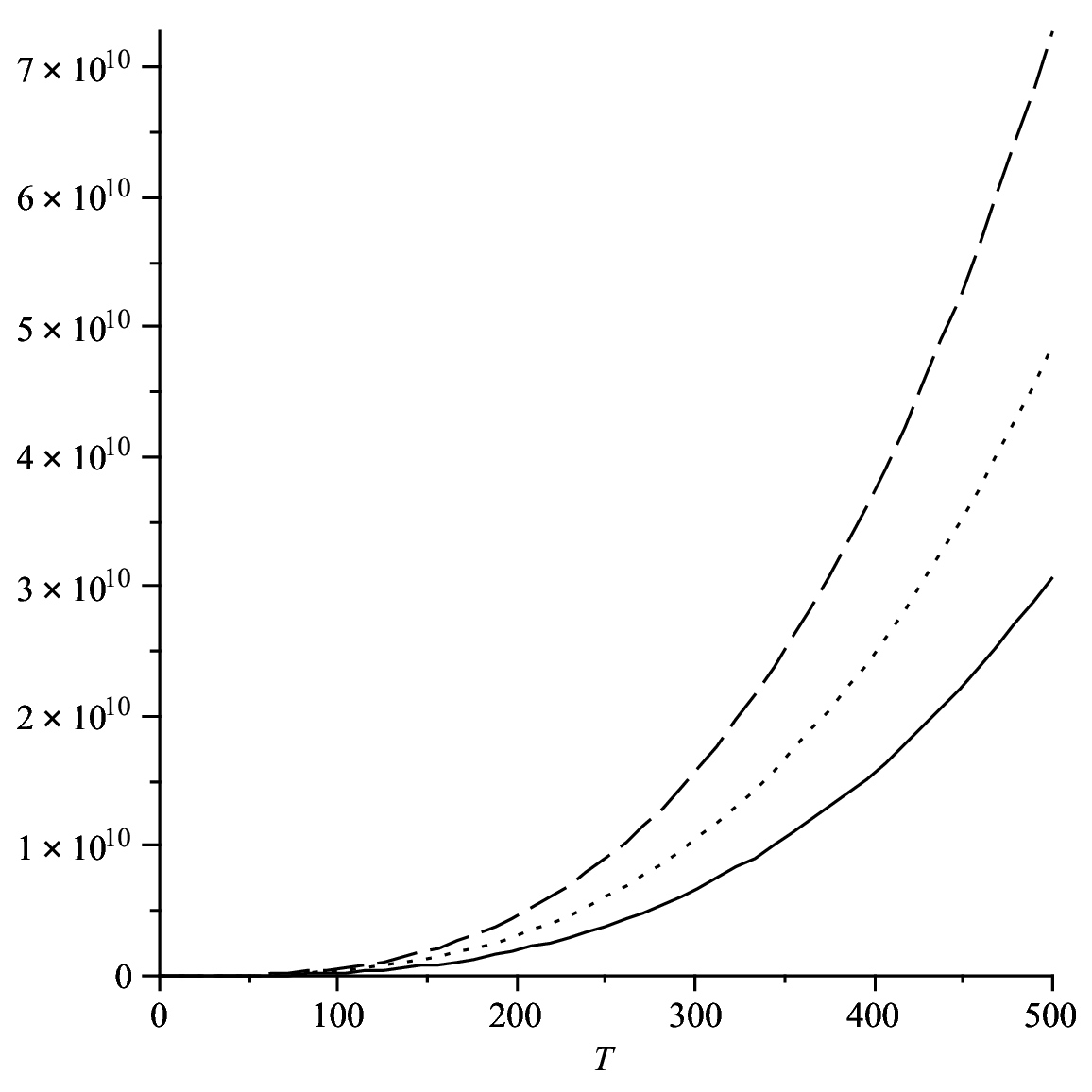}
\caption{Specific heat in terms of the temperature for $q=1$. The solid line represents the case (i). The dotted line represents the case (ii). The dashed
line represents the case (iii). We find that the value of the specific heat increases by number of the black hole charge.}
\end{center}
\end{figure}

\subsection{Dual picture}
As we mentioned already, the $\mathcal{N}=2$ $AdS_{5}$ supergravity
solution (1) is dual to the $\mathcal{N}=4$ SYM with finite chemical
potential in Minkowski space. It can be shown by the following
re-scaling [77],
\begin{equation}\label{s31}
r\rightarrow\lambda^{\frac{1}{4}}r, \hspace{5mm} t\rightarrow\frac{t}{\lambda^{\frac{1}{4}}}, \hspace{5mm} \mu\rightarrow\lambda\mu, \hspace{5mm}
q_{i}\rightarrow\lambda^{\frac{1}{2}}q_{i},
\end{equation}
and taking $\lambda\rightarrow\infty$ limit while,
\begin{equation}\label{s32}
d\Omega_{3,k}^{2}\rightarrow\frac{1}{R^{2}\lambda^{\frac{1}{2}}}(dx^{2}+dy^{2}+dz^{2}),
\end{equation}
and also we set $r_{0}^{4}\equiv\mu R^{2}$. Then, the solution (1) reduces to the following,
\begin{eqnarray}\label{s33}
ds^{2}&=&e^{2A(r)}\left[-\frac{f}{\mathcal{H}^{\frac{2}{3}}}dt^{2}+\mathcal{H}^{\frac{1}{3}} d{\vec{X}}^{2}+\frac{\mathcal{H}^{\frac{1}{3}}}{f}dr^{2}\right],\nonumber\\
f&=&\mathcal{H}-\frac{r_{0}^{4}}{r^{4}} ,\nonumber\\
\mathcal{H}&=&\prod_{i}(1+\frac{q_{i}}{r^{2}}),
\end{eqnarray}
where the geometric function $A(r)$ defined as $A(r)\equiv\ln{\frac{r}{L}}$, and $r_{0}$ is the horizon radius in the $\mathcal{N}=4$ SYM theory. In that case the chemical potential conjugate to the physical charge for the $U(1)$ R-charges is given by,
\begin{equation}\label{s34}
\phi_{i}=\frac{r_{h}^{2}}{R^{2}}\frac{2q_{i}}{r_{h}^{2}+q_{i}}\sqrt{\prod_{j}(1+\frac{q_{j}}{r_{h}^{2}})}.
\end{equation}
This is dual expression of the chemical potential which is given by the relation (19). For the special case of $q_{1}=q_{2}=q_{3}=q$ the Hawking
temperature reads as,
\begin{equation}\label{s35}
T_{H}=\frac{q+2r_{h}^{2}}{2\pi R^{2}\sqrt{q+r_{h}^{2}}},
\end{equation}
where the radius of the horizon (root of $f=0$) is given by,
\begin{equation}\label{s36}
r_{h}^{2}=\frac{1}{2}\left(\sqrt{4r_{0}^{4}+q^{2}}-q\right).
\end{equation}
In that case one can rewrite the chemical potential (34) in terms of the black hole charge and horizon radius,
\begin{equation}\label{s37}
\phi=\frac{r_{h}}{R^{2}}\sqrt{\frac{2q}{q+r_{h}^{2}}}.
\end{equation}
Therefore the $q=0$ limit is equal to the zero chemical potential
limit. In that case the specific heat of the $AdS_{5}$ black hole is
important parameter to find the phase transition which obtained as the following relation,
\begin{equation}\label{s38}
C_{v} \propto\frac{{\tilde{T}}^{2}}{\sqrt{{\tilde{T}}^{2}-q}}\left[(6\bar{c}-9q+\frac{2\bar{c}{\tilde{T}}^{2}}{{\tilde{T}}^{2}-q}+12{\tilde{T}}^{2})
e^{-\frac{\bar{c}}{{\tilde{T}}^{2}-q}}-6{\tilde{T}}^{2}\right],
\end{equation}
where $\bar{c}$ plays role of a mass scale (relates to the dilaton
field), and we defined $\tilde{T}\equiv\pi R^{2} T_{H}$. It is clear
that the case of $q=0$ recovers results of the Ref. [81]. In the
$\bar{c}\rightarrow0$ limit the sign of the specific heat is
positive for $q=0$, but in our case with $q\neq0$ the sign of the
specific heat is depends to the black hole charge. So, if
${\tilde{T}}^{2}>1.5 q$ then the charged black hole is in stable
phase. In the Ref. [66] it is found that the specific heat changes
the sign at ${\tilde{T}}^{2}\simeq0.75$ (for $\bar{c}\neq0$ and
$q=0$). In presence of the dilaton field ($\bar{c}\neq0$), and in
unit of $\bar{c}$, one can find that the phase transition
temperature from unstable to stable black hole increases for the
case of charged black hole. For example, in the case of $q=1$ we
find unstable/stable phase transition happen at
${\tilde{T}}^{2}\simeq2.4$, so the charged black hole is in stable
phase for ${\tilde{T}}^{2}>2.4$.
\subsection{Higher derivative correction}
The effect of the higher derivative corrections on the
thermodynamical quantities such as Hawking temperature and entropy
for the case of $k=0$ and a black hole with three equal charges
studied in the Ref. [78]. Here, we give extension to the case of
arbitrary space curvature and three different charges. In that case
the Hawking temperature obtained as,
\begin{eqnarray}\label{s39}
T&=&\frac{r_{h}}{2\pi R^{2}}\frac{\left[\frac{\mu
R^{2}}{r_{h}^{4}}-\frac{1}{r_{h}^{2}}\sum_{i}(q_{i}\prod_{j\neq
i}(1+\frac{q_{j}}{r_{h}^{2}}))+\prod_{i}(1+\frac{q_{i}}{r_{h}^{2}})\right]}{\sqrt{\prod_{i}(1+\frac{q_{i}}{r_{h}^{2}}
-\frac{c_{1}q_{i}(q_{i}+\mu)}{72r_{h}^{2}(q_{i}+r_{h}^{2})^{2}})}}\nonumber\\
&+&\frac{c_{1}}{4\pi r_{h}^{5}}\frac{\left[\frac{\mu^{2}}{48r_{h}^{4}}(\sum_{i}\frac{q_{i}}{1+\frac{q_{i}}{r_{h}^{2}}})
(\prod_{i}(1+\frac{q_{i}}{r_{h}^{2}})^{-1})-\frac{\mu^{2}}{16r_{h}^{2}\prod_{i}(1+\frac{q_{i}}{r_{h}^{2}})}+
\frac{4}{9R^{2}}\prod_{i}q_{i}(q_{i}+\mu)\right]}{\sqrt{\prod_{i}(1+\frac{q_{i}}{r_{h}^{2}})
-\frac{c_{1}q_{i}(q_{i}+\mu)}{72r_{h}^{2}(q_{i}+r_{h}^{2})^{2}}}},
\end{eqnarray}
where $r_{h}$ is given by the relation (15). Inserting $\mu$ from
$f_{k}=0$ into the relation (39) yields to the Hawking temperature
in terms of horizon radius and charges of the black hole. In that
case if we set $k=0$ and $q_{i}=q$ then solution (39) agree with the
result of the Ref. [78], ie,
\begin{equation}\label{s40}
T_{q_{i}=q, k=0}=\frac{(q_{i}+r_{0h}^{2})^{2}}{2\pi
L^{2}}\left[\frac{2r_{0h}^{2}-q}{r_{0h}^{2}}
+\frac{c_{1}(3q^{3}+4q^{2}r_{0h}^{2}+59qr_{0h}^{4}-10r_{0h}^{6})}{192R^{2}r_{0h}^{4}(2r_{0h}^{2}-q)}\right].
\end{equation}
Also $c_{1}=0$ limit of the relation (39) reduced to the relation (18). The modified entropy is obtained by using relations (14) and (20), so one can
obtain,
\begin{equation}\label{s41}
s=\frac{\sqrt{2}N^{2}}{1728\pi
R^{6}}\sqrt{\prod_{i}\frac{72r_{h}^{6}+216q_{i}r_{h}^{4}
+216q_{i}^{2}r_{h}^{2}+72q_{i}^{3}-c_{1}q_{i}^{2}-c_{1}q_{i}\mu}{(q_{i}+r_{h}^{2})^{2}}},
\end{equation}
where $r_{h}$ is given by the relation (15). Then, the specific heat
can be obtained by using $C_{v}=T\frac{\partial s}{\partial
r_{h}}(\frac{\partial T}{\partial r_{h}})^{-1}$. Numerically, we
find that the specific heat enhanced due to the higher
derivative terms.\\
In the next section we study some hydrodynamics aspects of the STU model and extract several interesting transport coefficients.
\section{Hydrodynamics}
\subsection{Ratio of shear viscosity to entropy}
In this subsection we are going to study universality of the shear
viscosity to entropy density ratio, $\eta/s$. As we know the shear
viscosity ($\eta$) is one of the important hydrodynamical quantities
of QGP which relates to the thermodynamical quantity, so-called
entropy. In the previous section we obtained the entropy of the
theory. Let us now review some important studies about the shear
viscosity.\\
The ratio of shear viscosity to entropy density of the strongly
coupled $\mathcal{N}=4$ SYM thermal plasma investigated [10, 11],
and found that $\eta=\frac{\pi}{8}N^{2}T^{3}$, where $N$ is the
number of coincident branes (number of colors). Also
$s=\frac{\pi^{2}}{2}N^{2}T^{3}$, therefore $\eta/s=1/4\pi$ verified.
Then, in the Refs. [12, 13] argued that this value of $\eta/s$
always saturated for gauge theories at large 't Hooft coupling.\\
The Ref. [14] showed that this value is a lower bound for a wide
class of systems, so $\eta/s\geq1/4\pi$. In that case in the Ref.
[82] the leading correction to the shear viscosity in the inverse
powers of 't Hooft coupling using the $\alpha^{\prime}$-corrected
low-energy effective action of type IIB string theory computed.\\
In the Ref. [17] by using the Kubo formula [81, 82] the shear
viscosity in the SYM theory dual to the STU model computed for the
case of the flat space ($k=0$). In the Refs. [16, 18] the viscosity
of gauge theory plasma with a chemical potential obtained. They used
the five-dimensional Reissner-Nordstorm AdS black hole, where the
chemical potential has unit value for the R-charges $U(1)^{3}$.\\
In the Refs. [19, 22] the effect of curvature squared corrections,
such as Gauss-Bonnet, on the $\eta/s$ bound computed and found that
the conjectured lower bound of $1/4\pi$ is violated for finite $N$.
These works generalized to the case of Gauss-Bonnet in arbitrary
higher dimensions [33, 34, 35], and showing that $\eta/s$ reduced in
these theories, but there is still a lower bound due to causality
which may arise for the large Gauss-Bonnet coupling limit. Finite 't
Hooft coupling corrections to the shear viscosity computed in the
Ref. [20] and found that it disagrees with the equilibrium
correlation function computations. This disagreement resolved in the Ref. [23].\\
There are several ways to compute the shear viscosity such as the
Kubo formula, which relates the shear viscosity to the correlation
function of the stress-energy tensor at zero spatial momentum. In
this paper we would like to use diffusion constant to extract the
ratio of shear viscosity to entropy density [10, 11, 15].\\
In that case, for a general situation with the given metric,
$$ds^2=g_{tt}dt^2+g_{rr}dr^2+g_{xx}d\vec{x}^2$$
the diffusion constant becomes,
\begin{equation}\label{s42}
D=\frac{\sqrt{-g(r_{h})}}{\sqrt{-g_{tt}(r_{h})g_{rr}(r_h)}}\, \int^{\infty}_{r_h}dr\frac{-g_{tt}\,g_{rr}}{g_{xx}\sqrt{-g}}.
\end{equation}
Then, we can use the following relation to investigate the
universality of the ratio of shear viscosity to entropy density,
\begin{equation}\label{s43}
\frac{\eta}{s}=TD,
\end{equation}
or a more general relation,
\begin{equation}\label{s44}
\frac{\eta}{D}=sT+\mu\rho,
\end{equation}
where density of the physical charge, $\rho$, is given by the
following equation,
\begin{equation}\label{s45}
\rho=\frac{\sqrt{2\sum_{i}q_{i}}N^{2}r_{h}^{2}}{8\pi^{2}R^{6}}\sqrt{\prod_{i}(1+\frac{q_{i}}{r_{h}^{2}})},
\end{equation}
It is fact that both relations (43) and (44) yield to the same
result. We should note that the relation (42) works in the flat
space only, therefore we should set $k=0$ in our calculations. In
that case one can obtain,
\begin{equation}\label{s46}
\eta=\frac{1}{16\pi GR^{3}}r_{h}^{3}\sqrt{\mathcal{H}(r_{h})},
\end{equation}
which shows that $\eta/s=1/4\pi$ is valid only for the case of flat
space. In order to discuss the shear viscosity we consider three
different cases of one, two and three-charged black holes.\\
In the first case we assume $q_{1}=q, q_{2}=q_{3}=0$. In that case
we have,
\begin{equation}\label{s47}
\eta=\frac{1}{16\pi GR^{3}}r_{h}^{3}\sqrt{1+\frac{q}{r_{h}}},
\end{equation}
where $r_{h}=\frac{1}{2}\,\sqrt {-2\,q+2\,\sqrt {{q}^{2}+8\ \mu}}$.
In the Fig. 4 we draw shear viscosity in terms of the black hole
charge. The Fig. 4 shows that the shear viscosity of the
$\mathcal{N}=2$ plasma, dual of one-charged black hole, decreased by
increasing the black hole charge. In that case thermodynamical
stability let us to choose $q<1.6$. So, the shear viscosity never
vanishes.

\begin{figure}[th]
\begin{center}
\includegraphics[scale=.4]{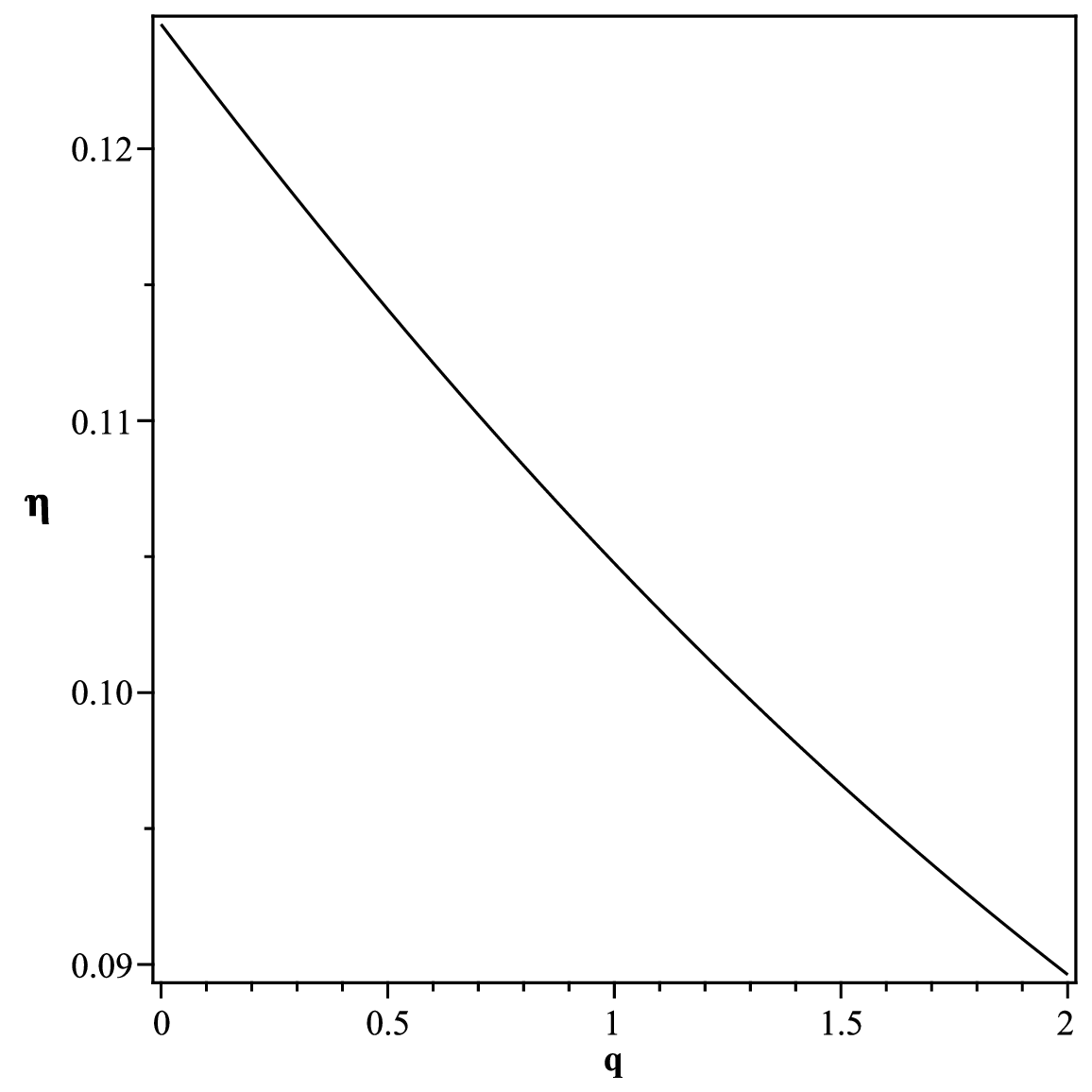}
\caption{The graph of $\eta$ for the case of one-charged black
hole}\label{ieta}
\end{center}
\end{figure}

In the second case we assume $q_{1}=q_{2}=q, q_{3}=0$. In that case
we have,
\begin{equation}\label{s48}
\eta=\frac{1}{16\pi GR^{3}}r_{h}^{3}(1+\frac{q}{r_{h}}),
\end{equation}
where $r_{h}=\sqrt{-q+\sqrt{2\mu}}$. We give plot of the shear
viscosity as a function of black hole charge in the Fig. 5.\\
According to the Fig. 5 the shear viscosity for two-charged black
hole increased by charge at the interval $0<q<0.8$, and decreased by
charge at the interval $0.8<q<1.4$. But, thermodynamical stability
tell us that allowed value of the black hole charge, in this case,
is $q<0.4$. Therefore the shear viscosity is completely increasing
by $q$ which is totally different with the previous case.\\
It is interesting result that the value of a parameter is depend on
the number of black hole charge. A black hole with odd number of
black hole charge yields to decreasing function of $q$ for shear
viscosity, on the other hand, a black hole with even number of black
hole charge yields to increasing function of $q$ for shear
viscosity. This assertion illustrated by studying three-charged
black hole.
We expect that three-charged black hole yields to decreasing function of $q$ for shear viscosity.\\
In the last case we assume that black hole has three equal charges
($q_{1}=q_{2}=q_{3}=q$). In that case we have,
\begin{eqnarray}\label{s49}
\eta=\frac{1}{16\pi GR^{3}}r_{h}^{3}\sqrt{(1+\frac{q}{r_{h}})^{3}},
\end{eqnarray}

\begin{figure}[th]
\begin{center}
\includegraphics[scale=.4]{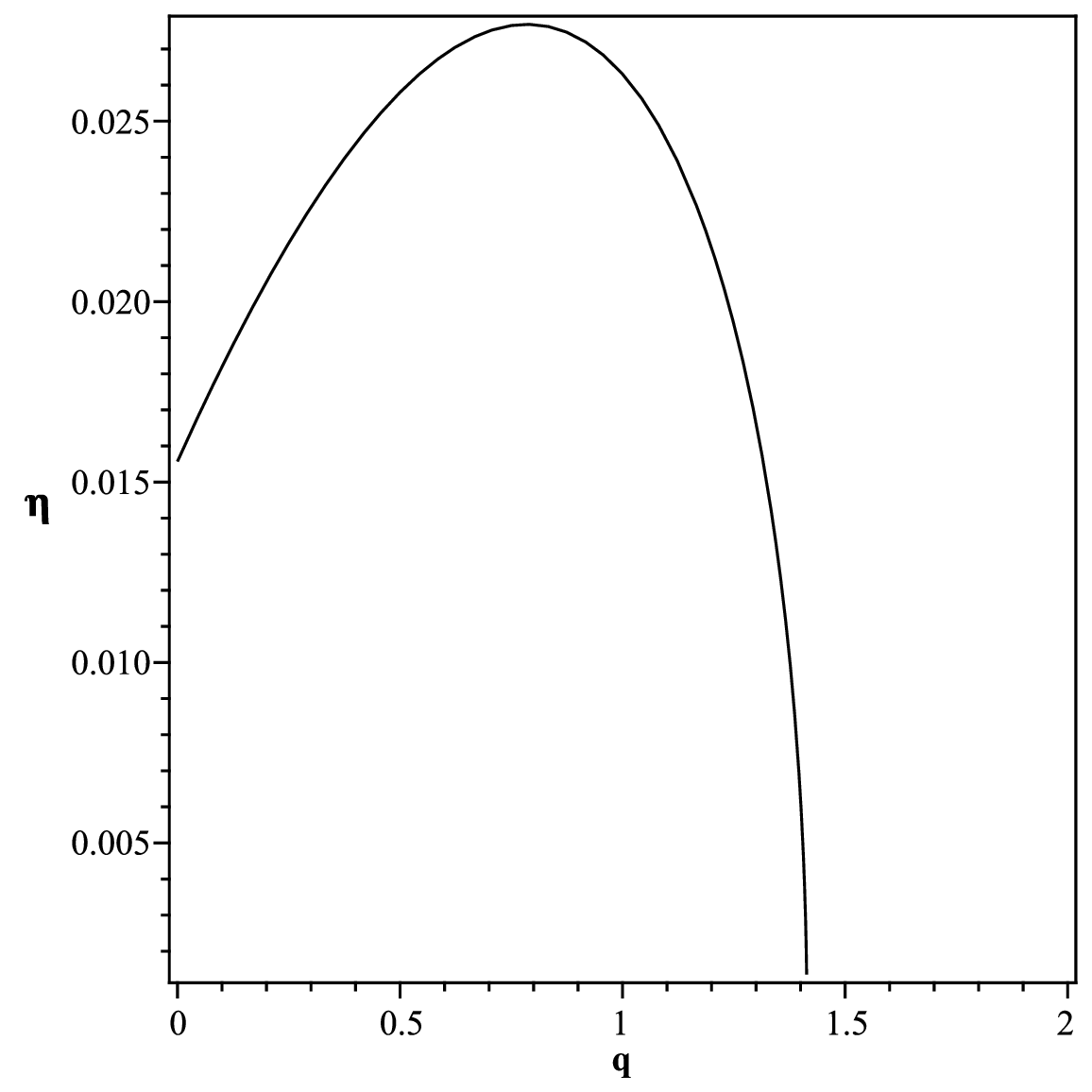}
\caption{The graph of $\eta$ for the case of two-charged black
hole.}\label{iieta}
\end{center}
\end{figure}

where,
\begin{equation}\label{s50}
r_{h}^2=\frac{1}{6}\,\mathcal{M}-\frac{2\,b-\frac{2}{3}a^2}{\mathcal{M}}-\frac{1}{3}
a,
\end{equation}
and we defined,
\begin{eqnarray}\label{s51}
a&\equiv&3q, \nonumber \\
b&\equiv&3q^{2}-2\mu, \nonumber \\
\mathcal{M}^3&\equiv&36a b-108{q}^{3}-8{a}^{3} \nonumber \\
&+&12\,\sqrt {12{b}^{3}-3{b}^{2}{a}^{2}-54ba{q}^{3}+81{q}^{6}+12{q}^{3}{a}^{3}},
\end{eqnarray}
We give plot of the shear viscosity (49) as a function of black hole
charge in the Fig. 6. The Fig. 6 shows that the shear viscosity for
the case of three-charged black hole decreased by the black hole
charge $q$.\\
We see in the Fig. 6 that the shear viscosity goes to
infinity for $q\sim0.8$, but thermodynamical stability
tell us that $q$ has lower value than 0.8, so the shear viscosity has finite value.\\
We conclude that the shear viscosity is strongly depend on the black
hole charges. The shear viscosity decreased by the black hole charge
in the case of one-charged and three-charged black hole, but
increased in the cases of two-charged black hole. However, the ratio
of the shear viscosity to entropy density has universal value.\\
In the next step we study thermal and electrical conductivities, and
the effect of higher derivative correction in STU model.

\begin{figure}[th]
\begin{center}
\includegraphics[scale=.4]{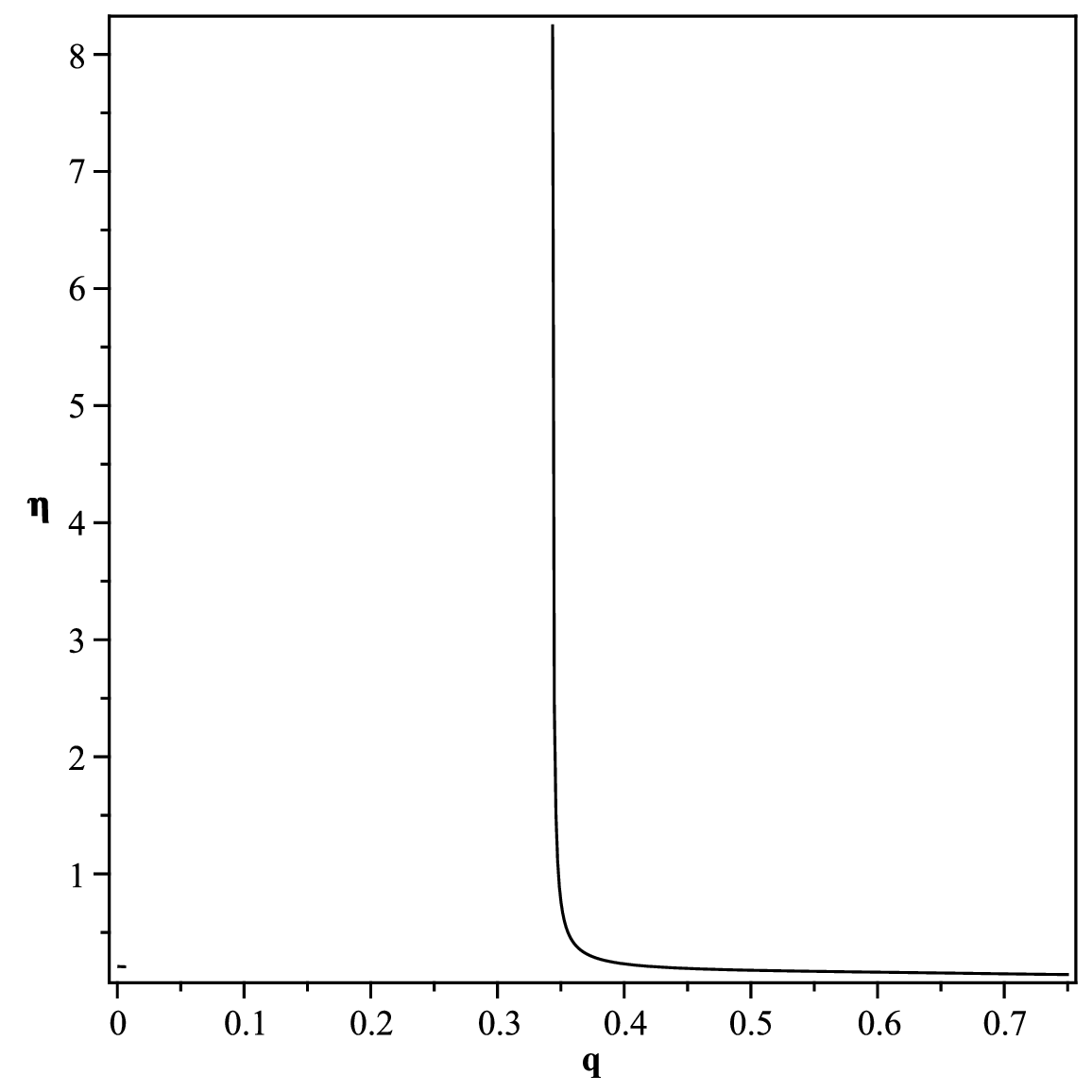}
\caption{The graph of $\eta$ for the case of three-charged black hole.}\label{iiieta}
\end{center}
\end{figure}

\subsection{Conductivity}
Now, we would like to use results of the Ref. [74] to obtain the
thermal and electrical conductivity. In the Ref. [74] it is found
that the conductivities for gauge theories dual to R-charge black
hole in 4, 5 and 7 dimensions behaves in a universal manner.
According to [74] one can obtain,
\begin{equation}\label{s52}
\sigma_{H}=r_{h}R^{2}(\prod_{i}(1+\frac{q_{i}}{r_{h}^{2}}))^{\frac{3}{2}}.
\end{equation}
And thermal conductivity is obtained as the following expression,
\begin{equation}\label{s53}
\kappa_{T}=(\frac{\epsilon+p}{\rho})^{2}\frac{\sigma_{H}}{T},
\end{equation}
where the energy density $\epsilon$, and pressure $P$ are defined as
[83, 84],
\begin{equation}\label{s54}
\epsilon=\frac{3N^{2}r_{h}^{4}}{8\pi^{2}R^{8}}\prod_{i}(1+\frac{q_{i}}{r_{h}^{2}}),
\end{equation}
\begin{equation}\label{s55}
P=\frac{N^{2}r_{h}^{4}}{8\pi^{2}R^{8}}\prod_{i}(1+\frac{q_{i}}{r_{h}^{2}}),
\end{equation}
where $N^{2}=8\pi^{2}R^{3}$ and we used $8\pi G=1$. In the Fig. 7 we draw graph of $\kappa_{T}$ in terms of the temperature for the simplest case of
$q_{1}=q, q_{2}=q_{3}=0$. It shows that the thermal conductivity vanishes at $T\approx28$ MeV for the large black hole charge and $T\approx0.9$ MeV for the
small black hole charge. It means that the thermal conductivity decrease with the black hole charge.

\begin{figure}[th]
\begin{center}
\includegraphics[scale=.29]{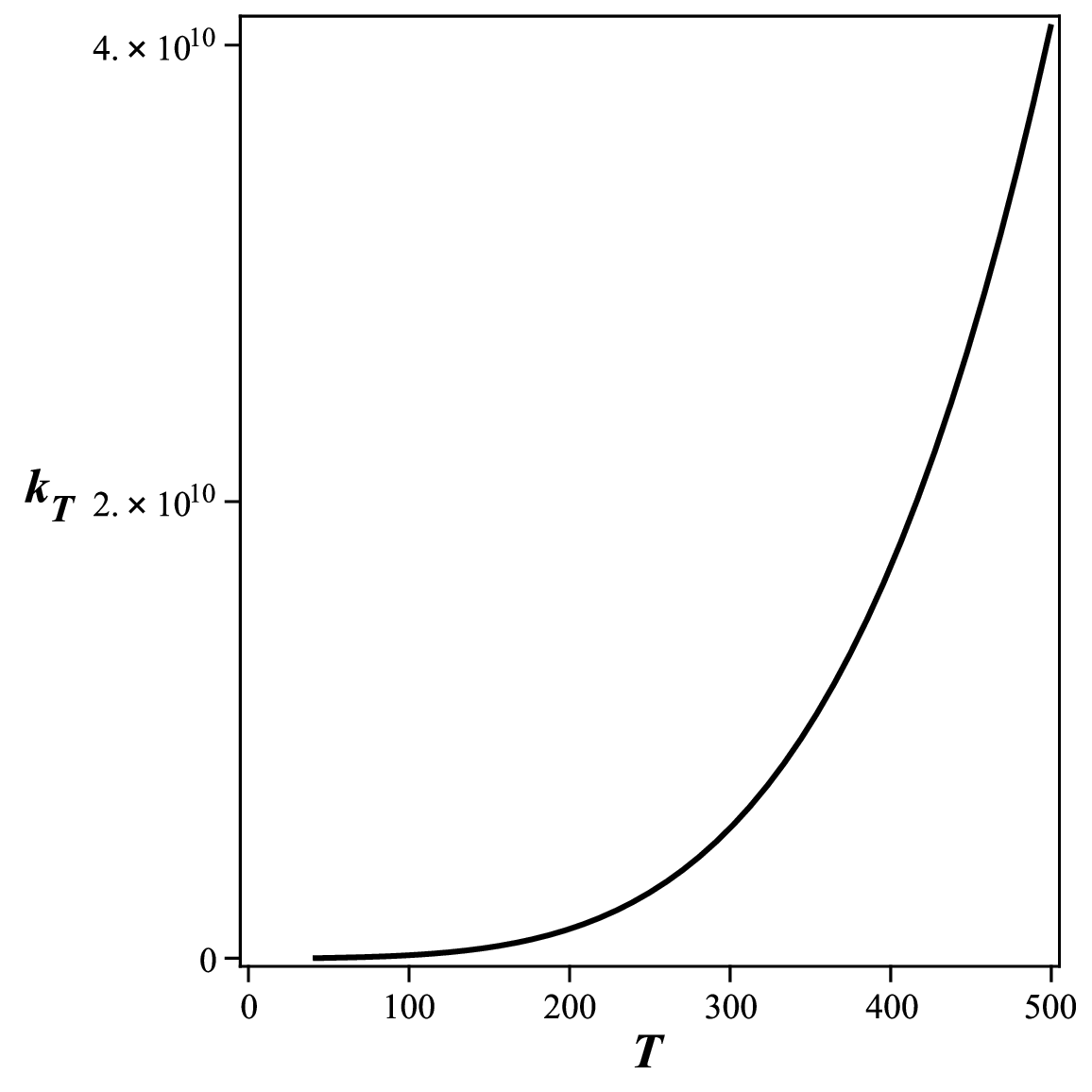}
\caption{The graph of $\kappa_{T}$ for the one charged black hole
with $q=10^6$ and $\mu=0.5$.}
\end{center}
\end{figure}

\subsection{Higher derivative correction}
If one include the higher derivative terms in STU model, then the
value of $\eta/s$ increased which is agree with the results of Refs.
[41, 42, 43]. In order to obtain effect of higher derivative terms
exactly, we focus on the special case of one-charged black hole. To
the first order of higher derivative correction the shear viscosity
to entropy ratio takes the following form [41],
\begin{equation}\label{s56}
\frac{\eta}{s}=\frac{1}{4\pi}\left(1+4c_{1}(\frac{q}{r_{h}^{6}}-2)\right),
\end{equation}
where $r_{h}$ obtained by the relation (15) for the special case of
$q_{1}=q, q_{2}=q_{3}=0$, with
\begin{equation}\label{s57}
r_{0h}^{2}=\frac{q}{2}(-1+\sqrt{1+\frac{4\mu R^{2}}{q^{2}}}),
\end{equation}
which is obtained by the equation (10). In the Fig. 8 we give plots
of $\eta /s$ for special case of one-charged black holes. It shows
that the higher derivative terms increases the value of $\eta /s$,
so there is no condition for choosing small black hole charge. As
expected the $c_{1}=0$ limit of the $\eta /s$ coincides with the
results of subsection 4.1. The left side of Fig. 8 shows that the
first order of the higher derivative terms increases the value of
$\eta /s$ for $q^{2}/r_{h}^{6}>2$. On the other hand the right side
of Fig. 8 tells that, for the fixed $c_{1}$, the black hole charge
increased the value of
the $\eta /s$.\\
The extension of the relations (52) and (53) to include the higher
derivative terms is more complicated. Therefore, just we draw graph
of the thermal conductivity in terms of the higher derivative
parameter $c_{1}$ (the left plot of the Fig. 9), and in terms of the
temperature (the right plot of the Fig. 9).\\
These figures tell us that the large value of the higher derivative
parameter yields to the negative thermal conductivity, which is not
acceptable. For example in the case of $T=250$ MeV one can obtain
$c_{1}\leq0.6$. In this situation by choosing $c_{1}=0.3$ the
thermal conductivity becomes negative for $T>450$ MeV.

\begin{figure}[th]
\begin{center}
\includegraphics[scale=.25]{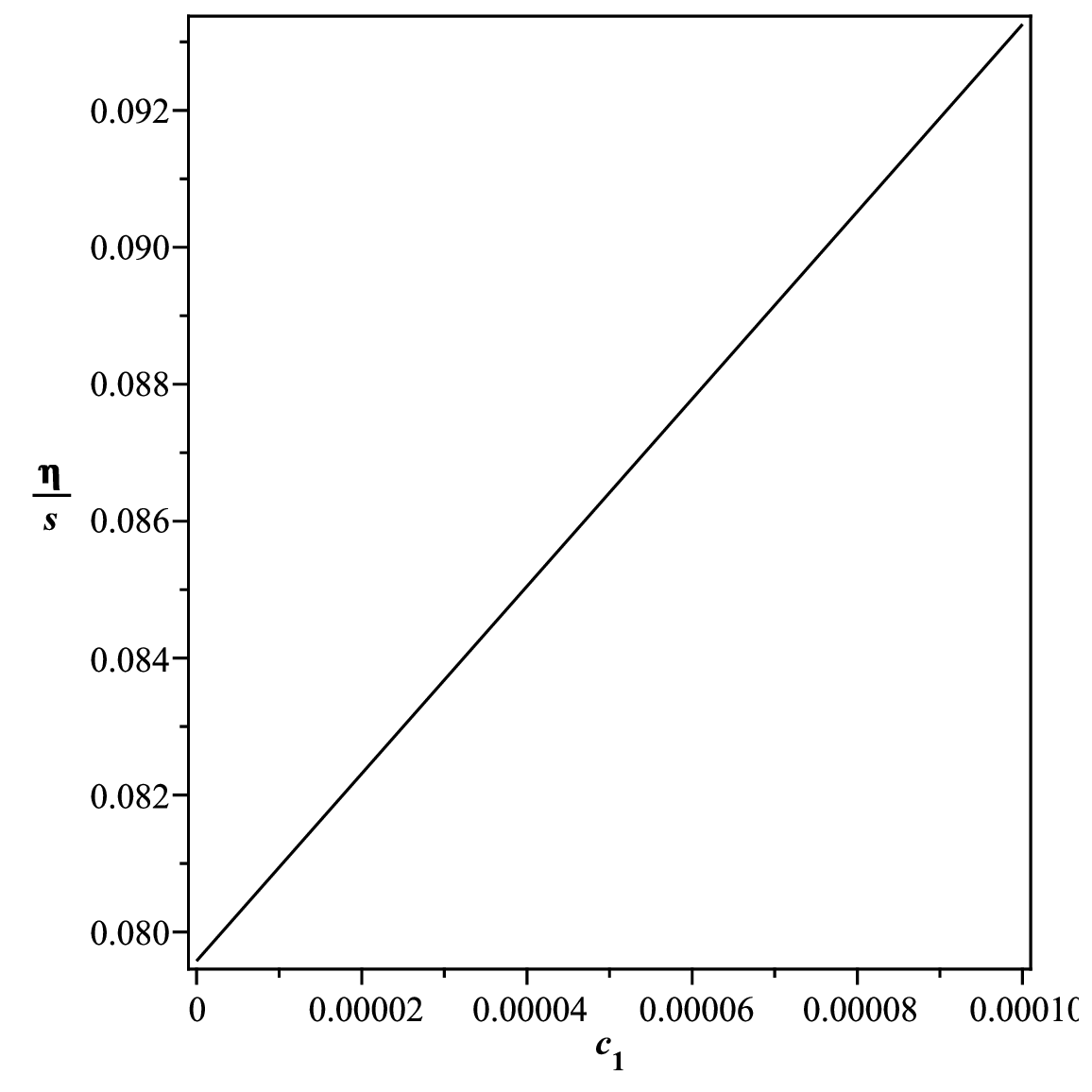}\includegraphics[scale=.25]{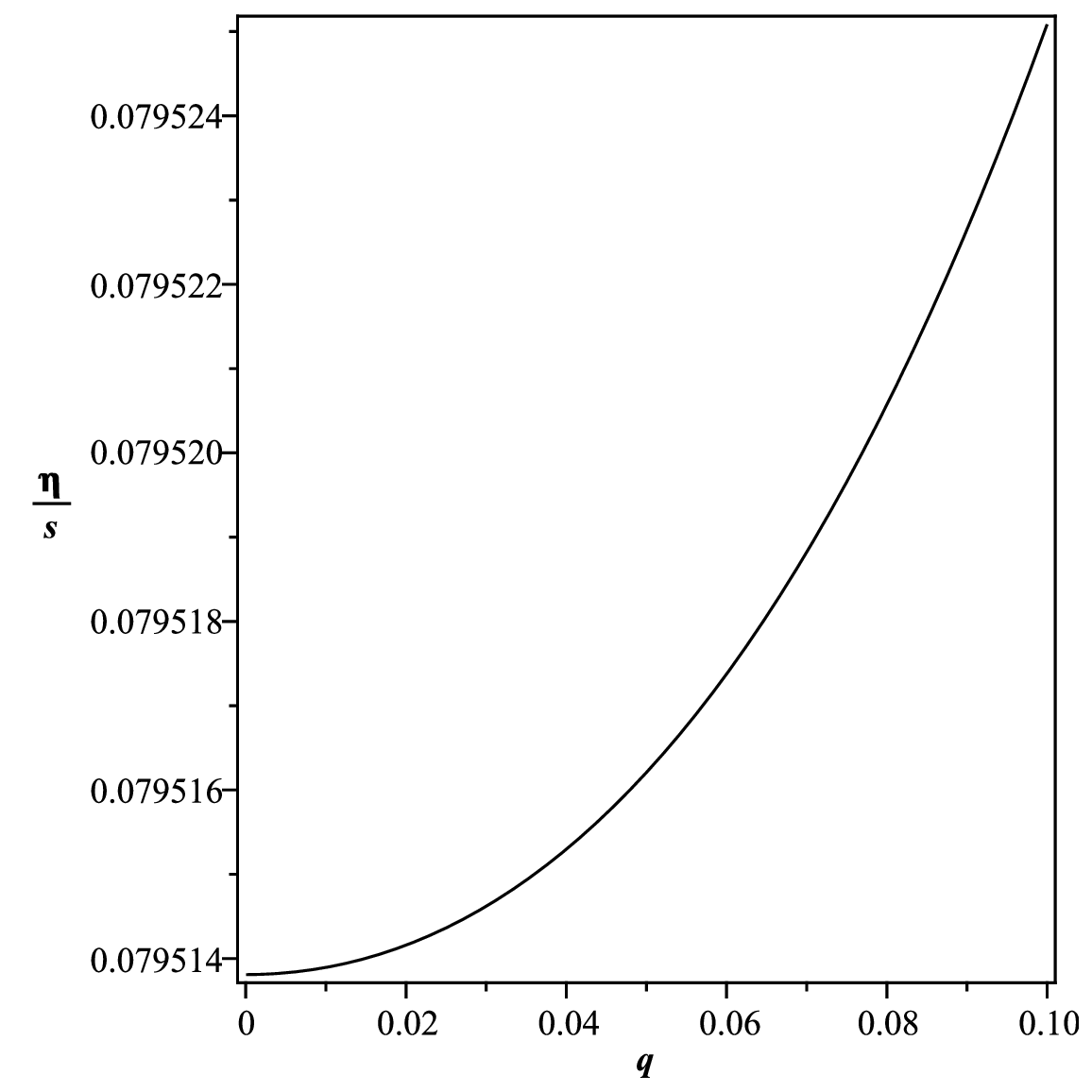}
\caption{The graphs of $\eta /s$ for the case of one-charged black hole by choosing $\mu=0.5$ and $R=0.5$, in terms of (left) higher derivative parameter
for $q=1$ and (right) black hole charge for $c_1=0.0001$.}
\end{center}
\end{figure}

\begin{figure}[th]
\begin{center}
\includegraphics[scale=.25]{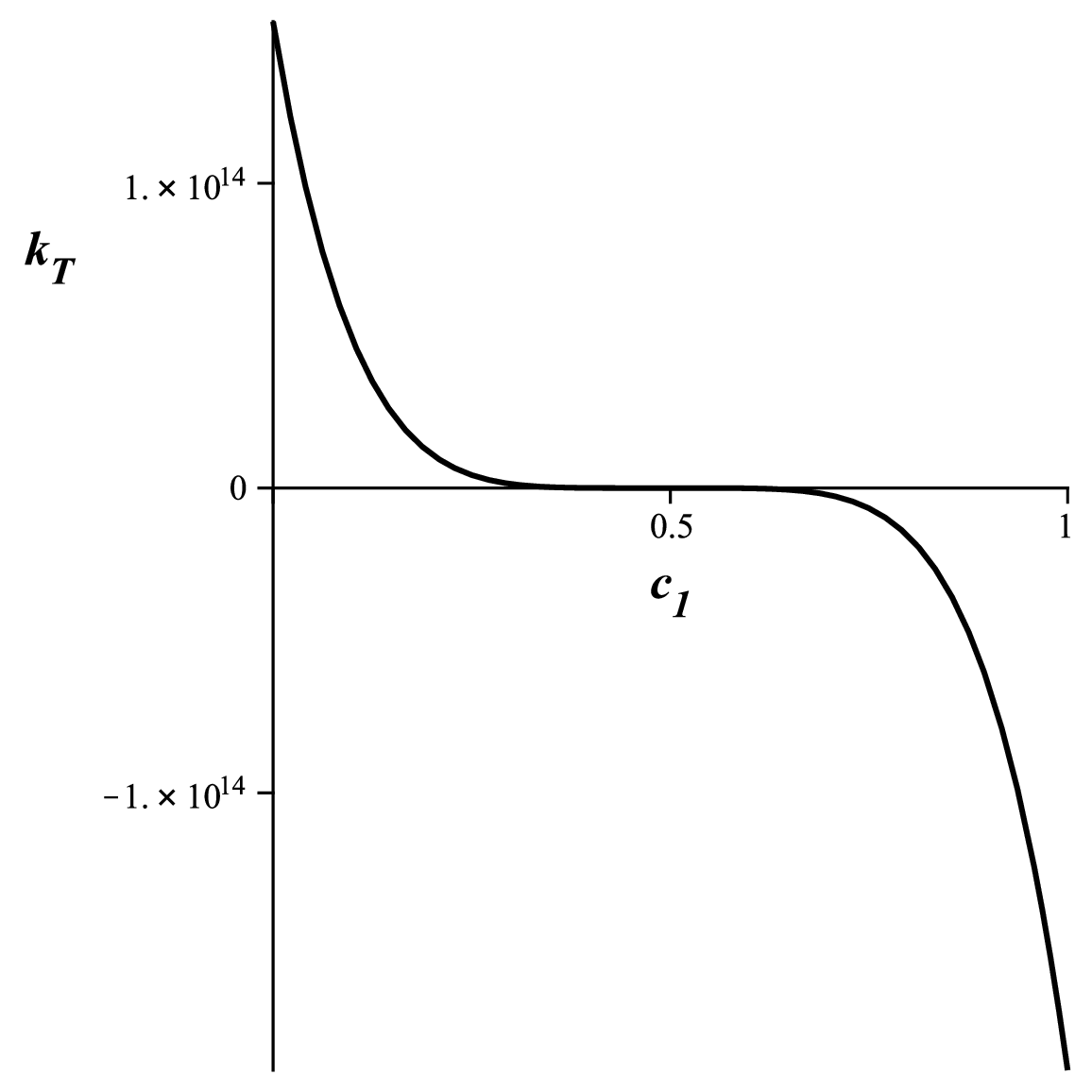}\includegraphics[scale=.25]{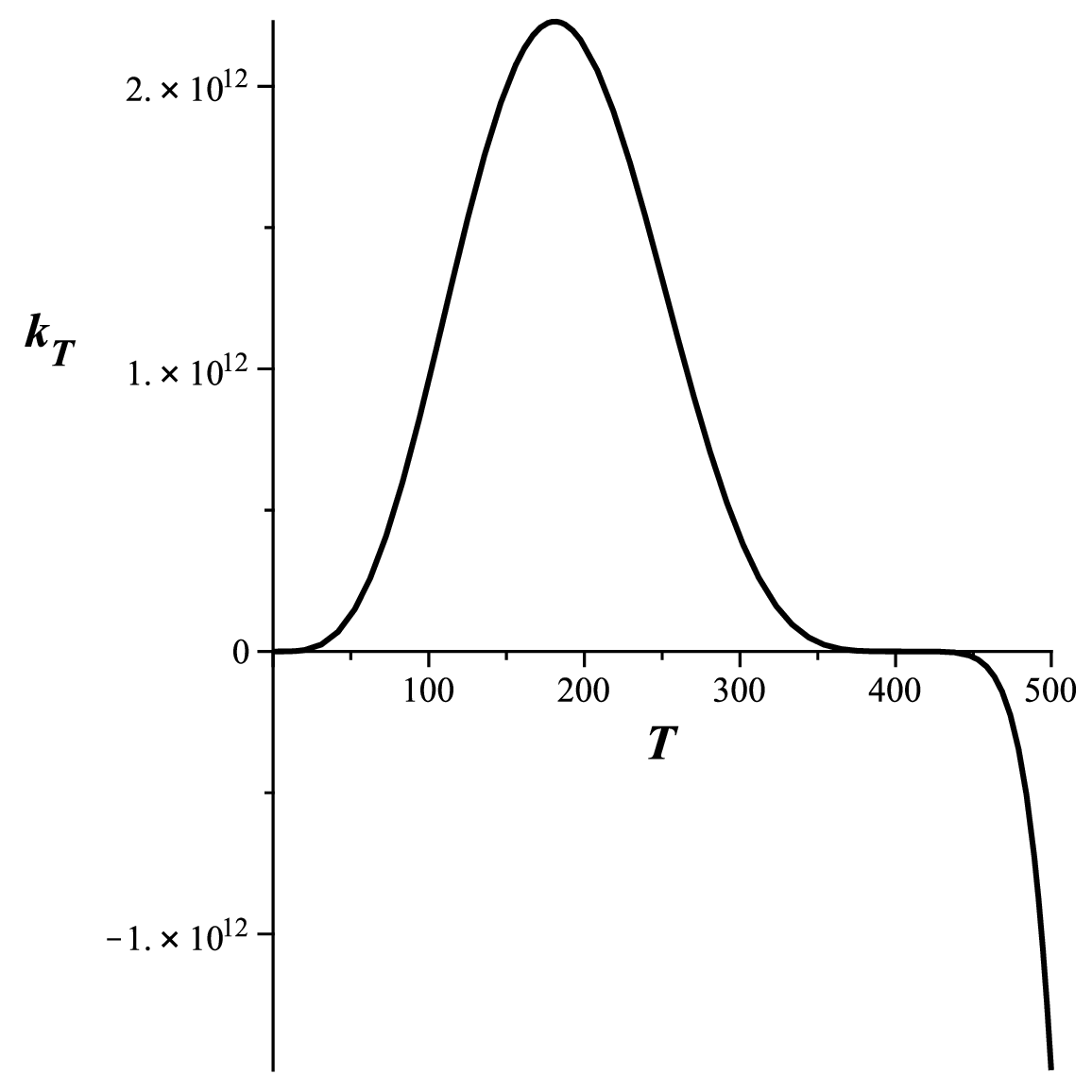}
\caption{The graphs of $k_T$ for the higher derivative correction by
choosing $\lambda=\sqrt{2}/2$, $\mu=0.5$. Left: Thermal conductivity
in terms of $c_{1}$ for $T=250$ MeV. Right: Thermal conductivity in
terms of $T$ for $c_{1}=0.3$}
\end{center}
\end{figure}

\section{Drag force}
Study of drag force on a moving heavy quark through a thermal plasma
is interesting point to understand physics of charm and bottom quark
at RHIC [85, 86, 87]. It is known that a moving quark in the
$\mathcal{N}=2$ thermal plasma corresponds to the stretched string
from $r=r_{m}$ on the D-brane to the black hole horizon. So,
calculating the energy loss of a heavy quark or drag force on the
moving quark reduces to find components of momentum density along
the string. The open string is described by the following Nambu-Goto
action,
\begin{equation}\label{s58}
S=-T_{0}\int{d\tau d\sigma \sqrt{-g}},
\end{equation}
where $T_{0}$ is the string tension. The coordinates $\tau$ and $\sigma$ are corresponding to the string world-sheet. Also, $g$ is determinant of the
world-sheet metric $g_{ab}$. We assume that the string moves along $x$ direction and use static gauge, where $\tau=t$ and $\sigma=r$. Therefore, the string
world-sheet is described by $x(r, t)$, so in order to write lagrangian density we use the metric (1) and find,
\begin{equation}\label{s59}
-g=\frac{1}{{\mathcal{H}}^{\frac{1}{3}}}
\left[1-\frac{{\mathcal{H}}r^{2}}{f_{k}R^{2}}\dot{x}^{2}+\frac{f_{k}r^{2}}{R^{2}}{x^{\prime}}^{2}\right],
\end{equation}
where dot and prime denote $t$ and $r$ derivatives respectively. By using Euler-Lagrange equation one can obtain the string equation of motion as the
following expression,
\begin{equation}\label{s60}
\frac{\partial}{\partial
r}(\frac{f_{k}r^{2}}{{\mathcal{H}}^{\frac{1}{3}}\sqrt{-g}}x^{\prime})=\frac{{\mathcal{H}}^{\frac{2}{3}}r^{2}}{f_{k}}\frac{\partial}{\partial
t}(\frac{\dot{x}}{\sqrt{-g}}),
\end{equation}
where $\sqrt{-g}$ is given by square of the relation (59). In order to obtain the total energy and momentum, drag force or energy loss of particle in the
thermal plasma, we have to calculate the canonical momentum densities. In that case one can obtain the following expressions,
\begin{eqnarray}\label{s61}
\left(\begin{array}{ccc}
\pi_{x}^{0} & \pi_{x}^{1}\\
\pi_{r}^{0}& \pi_{r}^{1}\\
\pi_{t}^{0} & \pi_{t}^{1}\\
\end{array}\right)=-\frac{T_{0}}{{\mathcal{H}}^{\frac{1}{3}}\sqrt{-g}} \left(\begin{array}{ccc}
-\frac{{\mathcal{H}}r^{2}}{f_{k}R^{2}}\dot{x} & \frac{f_{k}r^{2}}{R^{2}}x^{\prime}\\
\frac{{\mathcal{H}}r^{2}}{f_{k}R^{2}}\dot{x}x^{\prime} & 1-\frac{{\mathcal{H}}r^{2}}{f_{k}R^{2}}{\dot{x}}^{2}\\
1+f_{k}\frac{r^{2}}{R^{2}}{x^{\prime}}^{2} & -\frac{f_{k}r^{2}}{R^{2}}\dot{x}x^{\prime}\\
\end{array}\right).
\end{eqnarray}
Corresponding to the single quark, in CFT side, we have an open string in $AdS$ space which stretched from $r=r_{m}$ on D-brane to $r=r_{h}$ at the
horizon. In that case the total energy and momentum of string are obtained by the following integrals,
\begin{eqnarray}\label{s62}
E&=&-\int_{r_{h}}^{r_{m}}{\pi_{t}^{0}dr},\nonumber\\
P&=&\int_{r_{h}}^{r_{m}}{\pi_{x}^{0}dr}.
\end{eqnarray}
In this section we would like to obtain drag force for single quark and also quark-anti quark configurations. Also we discuss quasinormal modes of the
single quark solution. In that case we consider effects of adding B-field and higher derivative terms. Now we ready to obtain drag force for the single
quark solution.
\subsection{Single quark solution}
There is the simplest solution for the equation of motion (60), namely $x=x_{0}$, where $x_{0}$ is a constant and the string stretched straightforwardly
from D-brane at $r=r_{m}$ to the horizon at $r=r_{h}$. It means that in the dual picture there is a static quark in the thermal plasma. For such
configuration one can obtain $-g=(H_{1}H_{2}H_{3})^{-\frac{1}{3}}$ and $\pi_{x}^{0}=\pi_{r}^{0}=\pi_{t}^{1}=\pi_{x}^{1}=0$. It tells us that the drag force
is zero, as it expected for the static quark. Only non-zero components of momentum densities are
$\pi_{r}^{1}=\pi_{t}^{0}=-T_{0}[\prod_{i=1}^{3}(1+\frac{q_{i}}{r_{h}^{2}})]^{-\frac{1}{6}}$, so total energy of the string is obtained as,
\begin{equation}\label{s63}
E=T_{0}\left[r+\frac{1}{6r}\sum_{i}{q_{i}}+\frac{1}{36r^{3}}\sum_{i\neq
j}{q_{i}q_{j}}+\frac{1}{30r^{5}}\prod_{i}q_{i}\right]_{r_{h}}^{r_{m}},
\end{equation}
where we assume that the black hole charges $q_{i}$ are small. In zero temperature limit one can interpret $E$ as the rest mass of the quark, which is
obtained by the following expression,
\begin{equation}\label{s64}
M_{rest}=T_{0}\left[r_{m}-r_{h}+(\frac{1}{r_{m}}-\frac{1}{r_{h}})
\frac{\sum_{i}{q_{i}}}{6}+(\frac{1}{r_{m}^{3}}-\frac{1}{r_{h}^{3}})\frac{\sum_{i\neq
j}{q_{i}q_{j}}}{36}+(\frac{1}{r_{m}^{5}}-\frac{1}{r_{h}^{5}})\frac{\prod_{i}q_{i}}{30}\right].
\end{equation}
In STU model there is non-extremal black hole which is described by
non-extremality parameter $\mu$, but in the ${\mathcal{N}}=4$ SYM
theory there is near-extremal black hole. So, if we take
$\mu\rightarrow0$ ($q\rightarrow0$) limit, we have near-extremal
black hole, then the total energy of string obtained as
$E=T_{0}(r_{m}-r_{h})$. In the zero temperature limit ($r_{h}=0$)
the physical mass of quark (rest mass) becomes
$M_{rest}=T_{0}r_{m}$.\\
Now, we are going to consider most physical time-dependent solution of moving heavy quark through the thermal $\mathcal{N}=2$ plasma which is dual picture
of a curved string described by $x(r,t)=x(r)+vt$, where $v$ is the constant velocity of the single quark. In that case by using equation of motion (60) one
can find,
\begin{equation}\label{s65}
\frac{f_{k}r^{2}}{R^{2}v{\mathcal{H}}^{\frac{1}{3}}\sqrt{-g}}x^{\prime}=C,
\end{equation}
where $C$ is an integration constant and $\sqrt{-g}$ is obtained by
using the following equation,
\begin{equation}\label{s66}
-g=\frac{1}{{\mathcal{H}}^{\frac{1}{3}}}
\left[1-\frac{{\mathcal{H}}r^{2}}{f_{k}R^{2}}v^{2}+\frac{f_{k}r^{2}}{R^{2}}{x^{\prime}}^{2}\right].
\end{equation}
Solving the equation (65) for $x^{\prime}$ yields,
\begin{equation}\label{s67}
x^{\prime2}=\frac{C^{2}v^{2}R^{2}{\mathcal{H}}^{\frac{1}{3}}}{f_{k}^{2}r^{2}}
\frac{f_{k}R^{2}-{\mathcal{H}}r^{2}v^{2}}{f_{k}r^{2}-C^{2}v^{2}R^{2}{\mathcal{H}}^{\frac{1}{3}}}.
\end{equation}
By using these results in the canonical momentum densities (61) we find,
\begin{eqnarray}\label{s68}
\pi_{x}^{1}&=&-T_{0}Cv,\nonumber\\
\pi_{t}^{1}&=&T_{0}Cv^{2}.
\end{eqnarray}
These expressions exactly coincide with those obtained in the
$\mathcal{N}=4$ SYM theory [57]. These expressions construct the
rate of energy and momentum along the open string,
\begin{eqnarray}\label{s69}
\frac{d P}{dt}=\pi_{x}^{1}|_{r=r_{m}}=-T_{0}Cv,\nonumber\\
\frac{d E}{dt}=\pi_{t}^{1}|_{r=r_{m}}=T_{0}Cv^{2}.
\end{eqnarray}

\begin{figure}[th]
\begin{center}
\includegraphics[scale=.35]{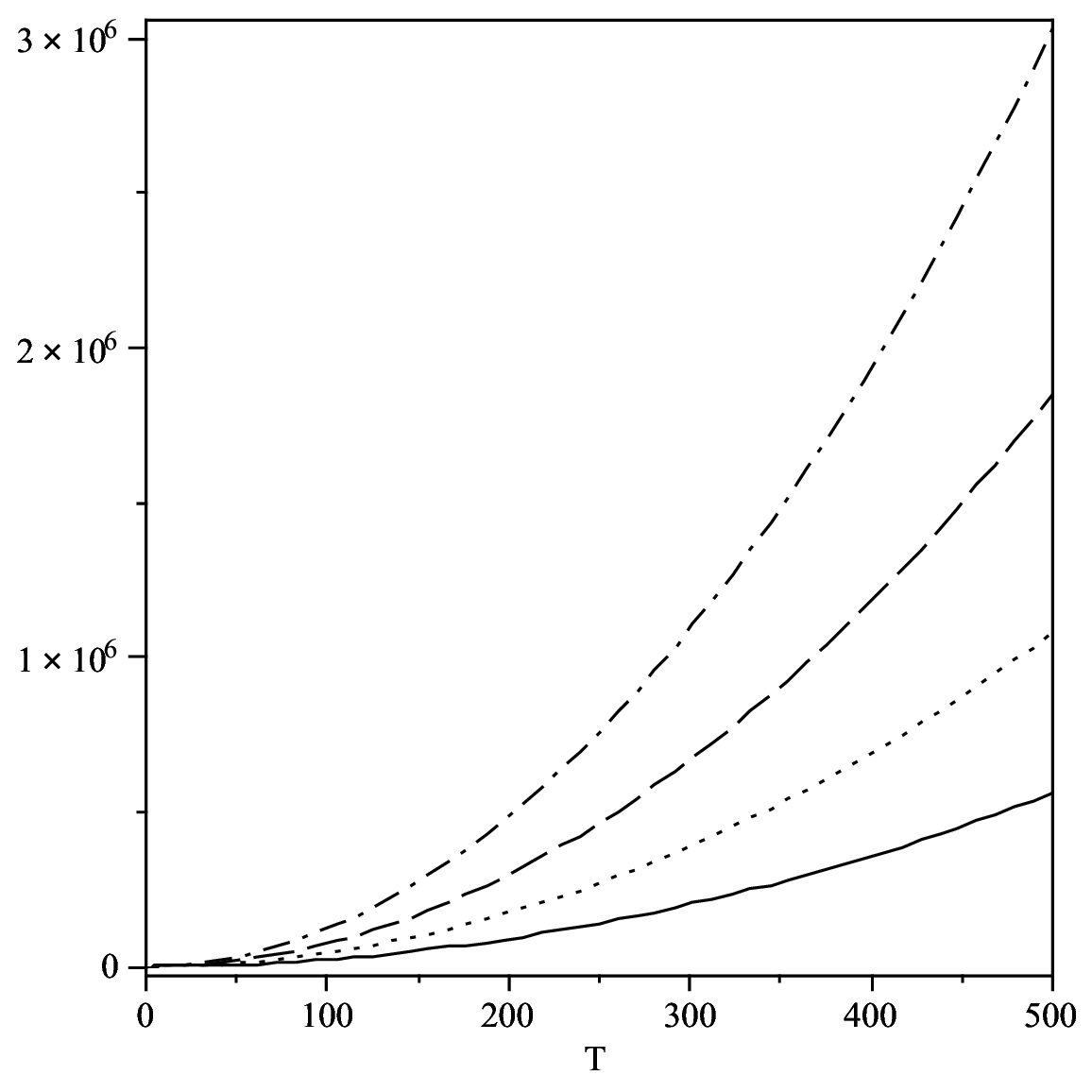}\includegraphics[scale=.35]{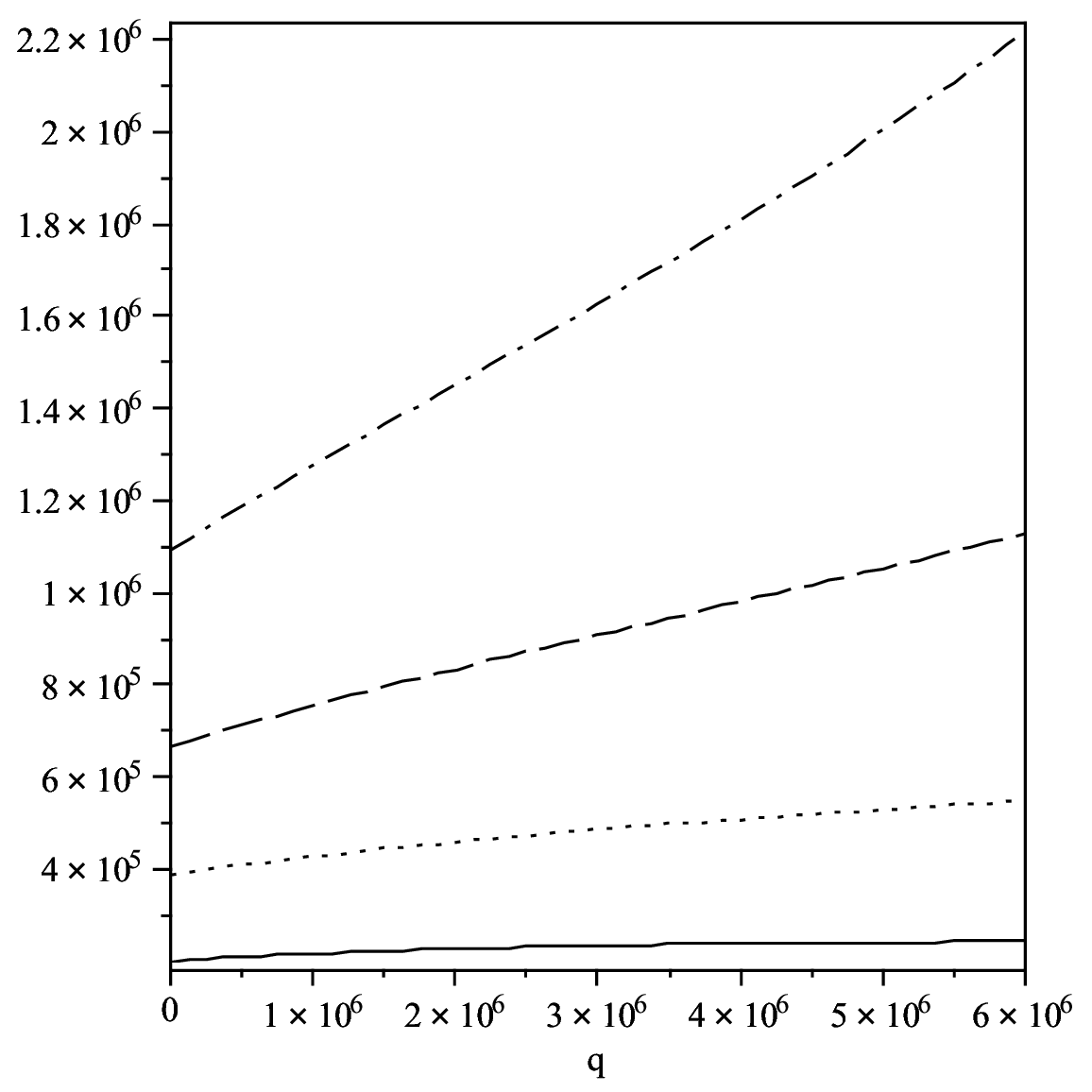}
\caption{The graphs of the drag force for $q_{1}=q, q_{2}=q_{3}=0$ and the small velocity limit. We set $\alpha^{\prime}=0.5$, $\lambda=6\pi$ and $\mu=1$.
The solid, dotted, dashed and dash dotted lines correspond to $v=0.3, 0.5, 0.7$ and $0.9$ respectively. These show that by increasing velocity, the drag
force increases. Left: drag force in terms of the temperature for $q=1$. Right: drag force in terms of the black hole charge for $T=300$ MeV. It tell us
that the black hole charge increases the value of the drag force.}
\end{center}
\end{figure}

Difference of our result with the $\mathcal{N}=4$ SYM theory is the
constant $C$. In order to find $C$ we use reality condition for
$x^{\prime2}$ and $\sqrt{-g}$. This condition tells that
$x^{\prime2}$ and $\sqrt{-g}$ have real value along the length of
the string. Therefore, we should find appropriate $r$, where
nominator and denominator of the relation (67) become positive. For
the small velocity we know that $fR^{2}-{\mathcal{H}}r^{2}v^{2}$ has
a zero at $r=r_{c}>r_{h}$. We set this root in the denominator of
the relation (67) and fix the constant $C$ as the following,
\begin{equation}\label{s70}
C=\left[\prod_{i=1}^{3}(1+\frac{q_{i}}{r_{c}^{2}})\right]^{\frac{1}{3}}\frac{r_{c}^{2}}{R^{2}},
\end{equation}
where,
\begin{equation}\label{s71}
r_{c}=r_{h}+\left(\frac{r^{2}v^{2}\mathcal{H}}{2R^{2}\left[\frac{\mu}{r^{3}}+\frac{r\mathcal{H}}{R^{2}}
-\frac{q_{1}H_{2}H_{3}+q_{2}H_{1}H_{3}+q_{3}H_{1}H_{2}}{rR^{2}}\right]}\right)_{r=r_{h}}+\mathcal{O}(v^{4}).
\end{equation}
It is important to note that this result is independent of curvature parameter $k$, however we should set $k = 1$ in the relation (2) to have $AdS_5 \times
S^5$ space. Combining the relations (69), (70) and (71) give us expression of the drag force which may be written as,
\begin{equation}\label{s72}
\frac{dP}{dt}=-T_{0}v\left[\prod_{i=1}^{3}(1+\frac{q_{i}}{r_{h}^{2}})\right]^{\frac{1}{3}}\frac{r_{h}^{2}}{R^{2}} (1+\mathcal{O}(v^{2})).
\end{equation}

\begin{figure}[th]
\begin{center}
\includegraphics[scale=.35]{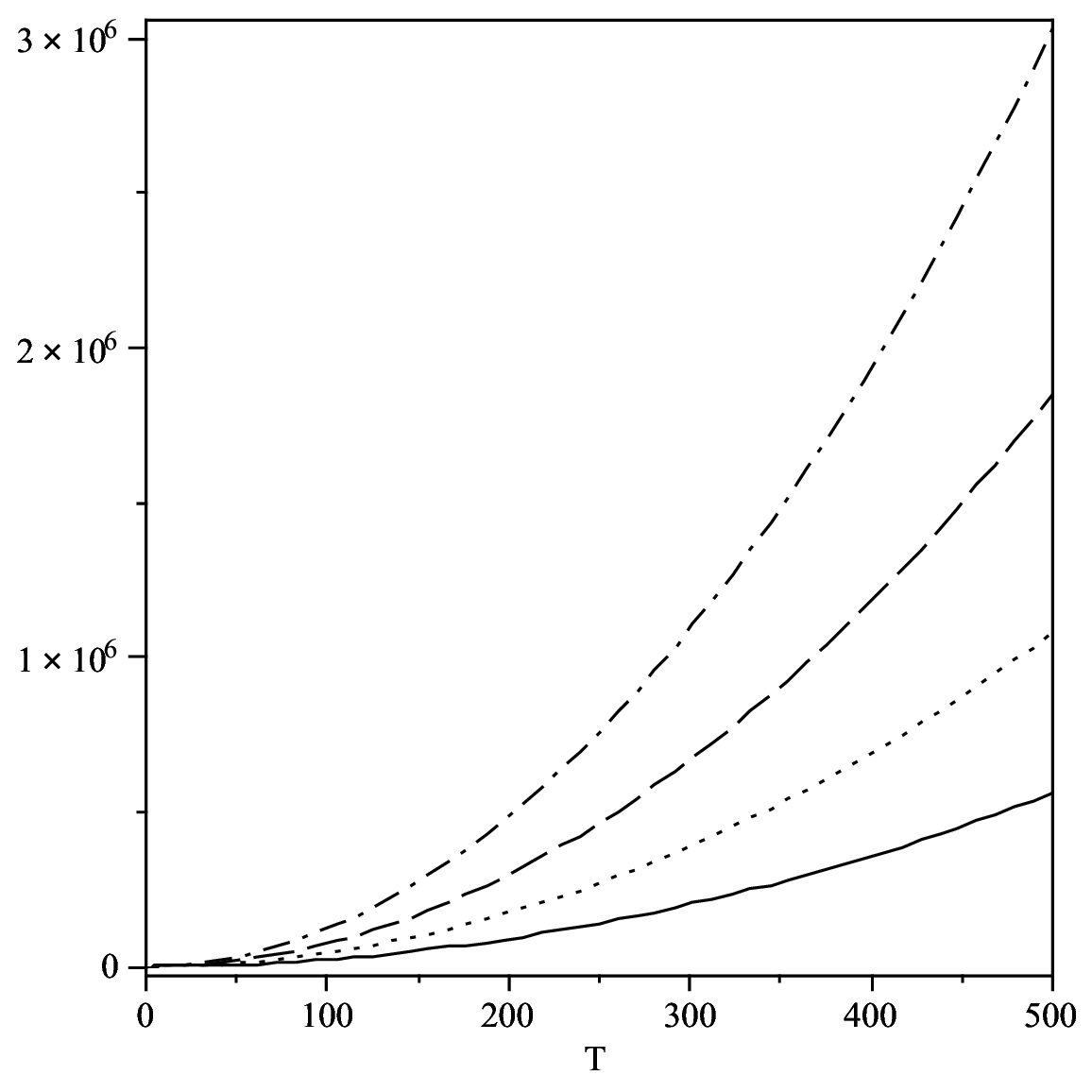}\includegraphics[scale=.35]{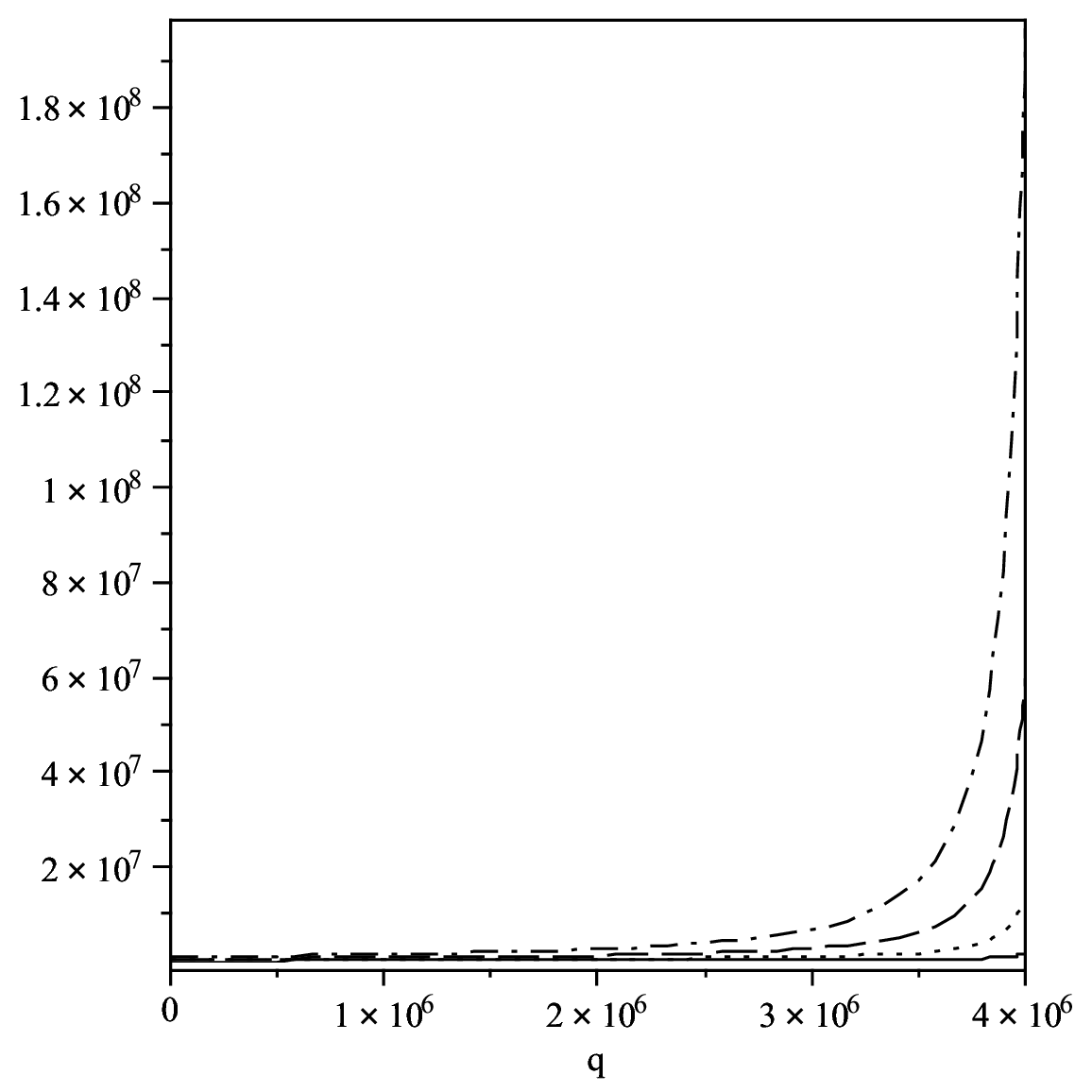}
\caption{The graphs of the drag force for $q_{1}=q_{2}=q, q_{3}=0$ and the small velocity limit. We set $\alpha^{\prime}=0.5$, $\lambda=6\pi$ and $\mu=1$.
The solid, dotted, dashed and dash dotted lines correspond to $v=0.3, 0.5, 0.7$ and $0.9$ respectively. These show that by increasing velocity, the drag
force increases. Left: drag force in terms of the temperature for $q=1$. Right: drag force in terms of the black hole charge for $T=300$ MeV. It tell us
that the black hole charge increases the value of the drag force.}
\end{center}
\end{figure}

Indeed the equation (72) is the momentum current into the horizon. Here, we have field theory interpretation of our system. One can image a single quark
moving in a constant external field with strength $\varepsilon=-\pi_{x}^{1}$. This external field keeps the curved string moving at the constant speed $v$.
We know that electromagnetic field lives on a D-brane on which this dragging string ends. The $\varepsilon$ changes the boundary conditions for the string.
Usually, the string should satisfy Dirichlet boundary conditions orthogonal to the D-brane and Neumann boundary conditions parallel with the D-brane. In
the presence of $\varepsilon$, the Neumann boundary
conditions can be altered.\\
Also by using the relations $\pi_{x}^{1}=-\zeta m v$ and,
\begin{equation}\label{s73}
D_{q}=\frac{T}{\zeta m},
\end{equation}
one can obtain diffusion coefficient ($D_{q}$) of the quark. We try
to discuss drag force and diffusion coefficient of the quark for
three cases of one, two and three charged black holes.\\
First, we assume $q_{1}=q, q_{2}=q_{3}=0$, so the horizon radius is given by the relation (26) and thermodynamical stability let us to choose $q\leq6\times
10^6$. In this case we draw plots of the drag force in terms of the temperature and the black hole charge in the Fig. 10. In that case diffusion
coefficient of the quark obtained as the following expression,
\begin{equation}\label{s74}
D_{q}=\frac{2r_{h}^{2}+q}{2\pi r_{h}^{3}(1+\frac{q}{r_{h}^{2}})^{\frac{5}{6}}}.
\end{equation}

\begin{figure}[th]
\begin{center}
\includegraphics[scale=.35]{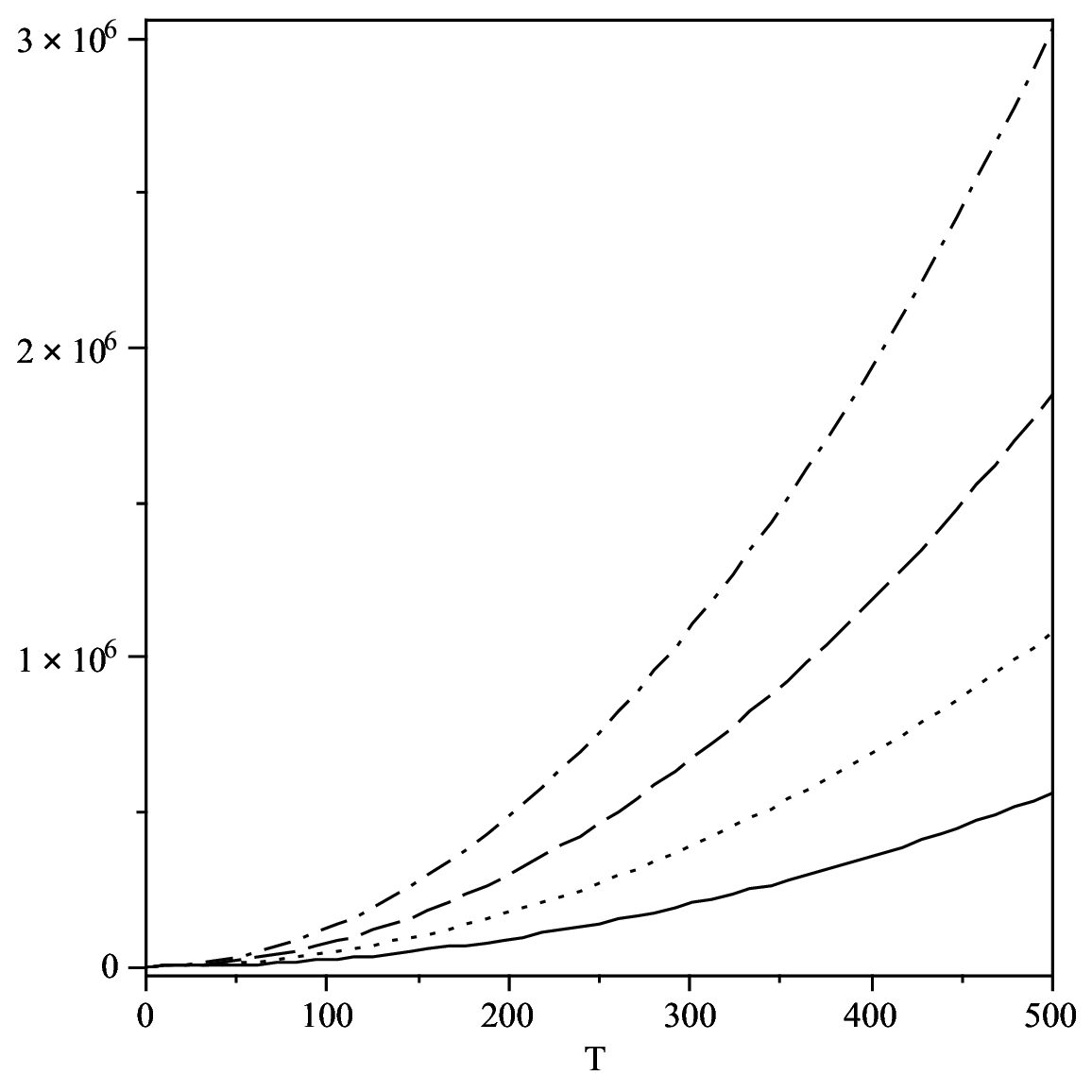}\includegraphics[scale=.35]{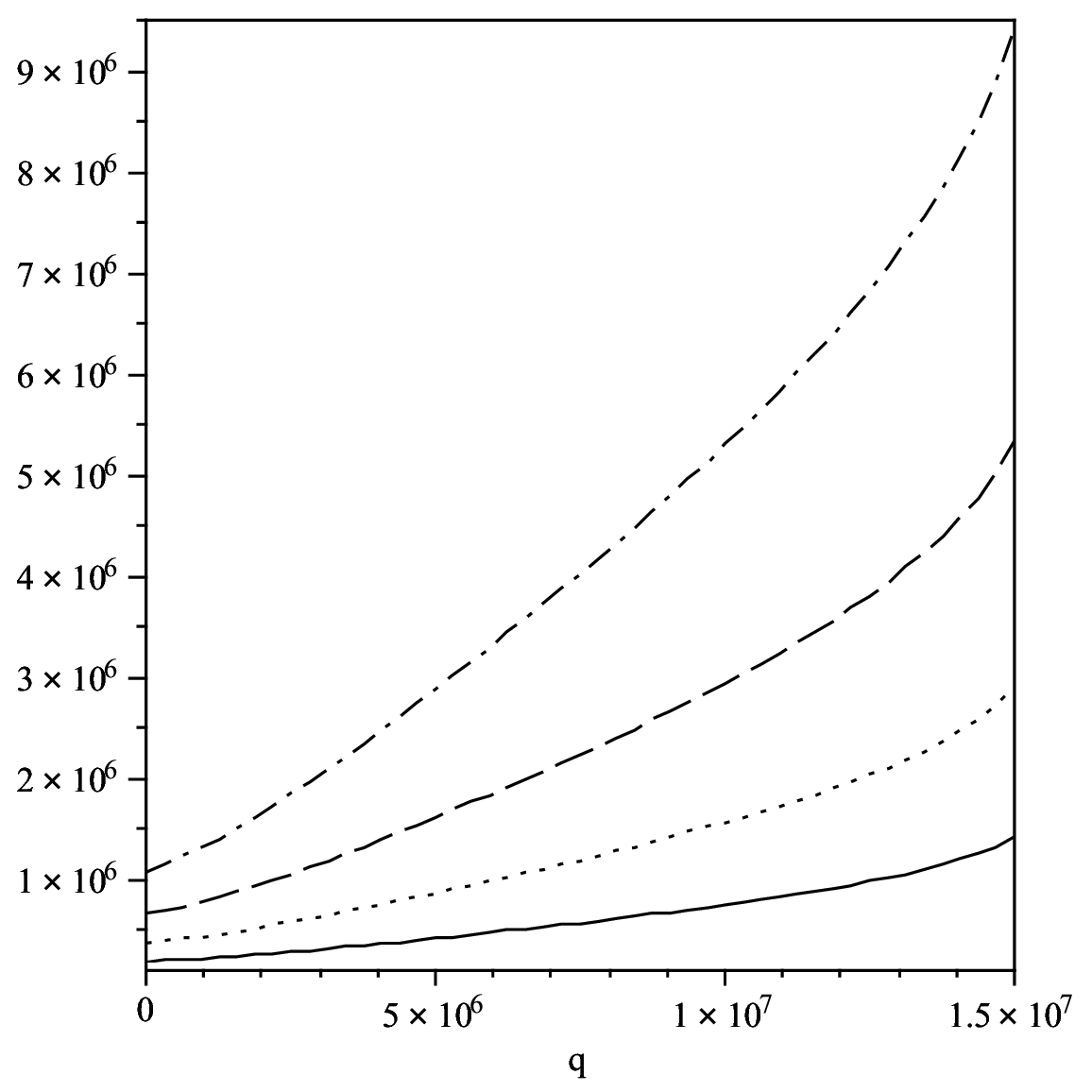}
\caption{The graphs of the drag force for $q_{1}=q_{2}=q_{3}=q$ and the small velocity limit. We set $\alpha^{\prime}=0.5$, $\lambda=6\pi$ and $\mu=1$. The
solid, dotted, dashed and dash dotted lines correspond to $v=0.3, 0.5, 0.7$ and $0.9$ respectively. These show that by increasing velocity, the drag force
increases. Left: drag force in terms of the temperature for $q=1$. Right: drag force in terms of the black hole charge for $T=300$ MeV. It tell us that the
black hole charge increases the value of the drag force.}
\end{center}
\end{figure}

So, for the $q\rightarrow0$ limit we have $D_{q}=\frac{1}{\pi r_{h}}$, which means that diffusion coefficient of the quark found proportional to inverse of
the temperature.\\
Second, we assume $q_{1}=q_{2}=q, q_{3}=0$, so the horizon radius is obtained from the relation (18) as $r_{h}=\pi R^{2} T$ and thermodynamical stability
let us to choose $q\leq4\times 10^6$.  In this case we draw plots of the drag force in terms of the temperature and the black hole charge in the Fig. 11.\\
In that case diffusion coefficient of the quark obtained as the following expression,
\begin{equation}\label{s75}
D_{q}=\frac{1}{\pi r_{h}(1+\frac{q}{r_{h}^{2}})^{\frac{2}{3}}}.
\end{equation}
So, for the $q\rightarrow0$ limit we have $D_{q}=\frac{1}{\pi r_{h}}$. It means that diffusion coefficient of the quark found proportional of inverse of
the temperature, which is expected.\\
Finally, we assume $q_{1}=q_{2}=q_{3}=q$, so the horizon radius is given by the relation (29) and thermodynamical stability let us to choose $q\leq15\times
10^6$. In this case we draw plots of the drag force in terms of the temperature and the black hole charge in the Fig. 12. We found that the black hole
charge increases the value of drag force. In that case diffusion coefficient of the quark obtained as the following expression,
\begin{equation}\label{s76}
D_{q}=\frac{2+\frac{3q}{r_{h}^{2}}-\frac{q^{3}}{r_{h}^{6}}}{2\pi
r_{h}(1+\frac{q}{r_{h}^{2}})^{\frac{5}{2}}}.
\end{equation}

\begin{figure}[th]
\begin{center}
\includegraphics[scale=.4]{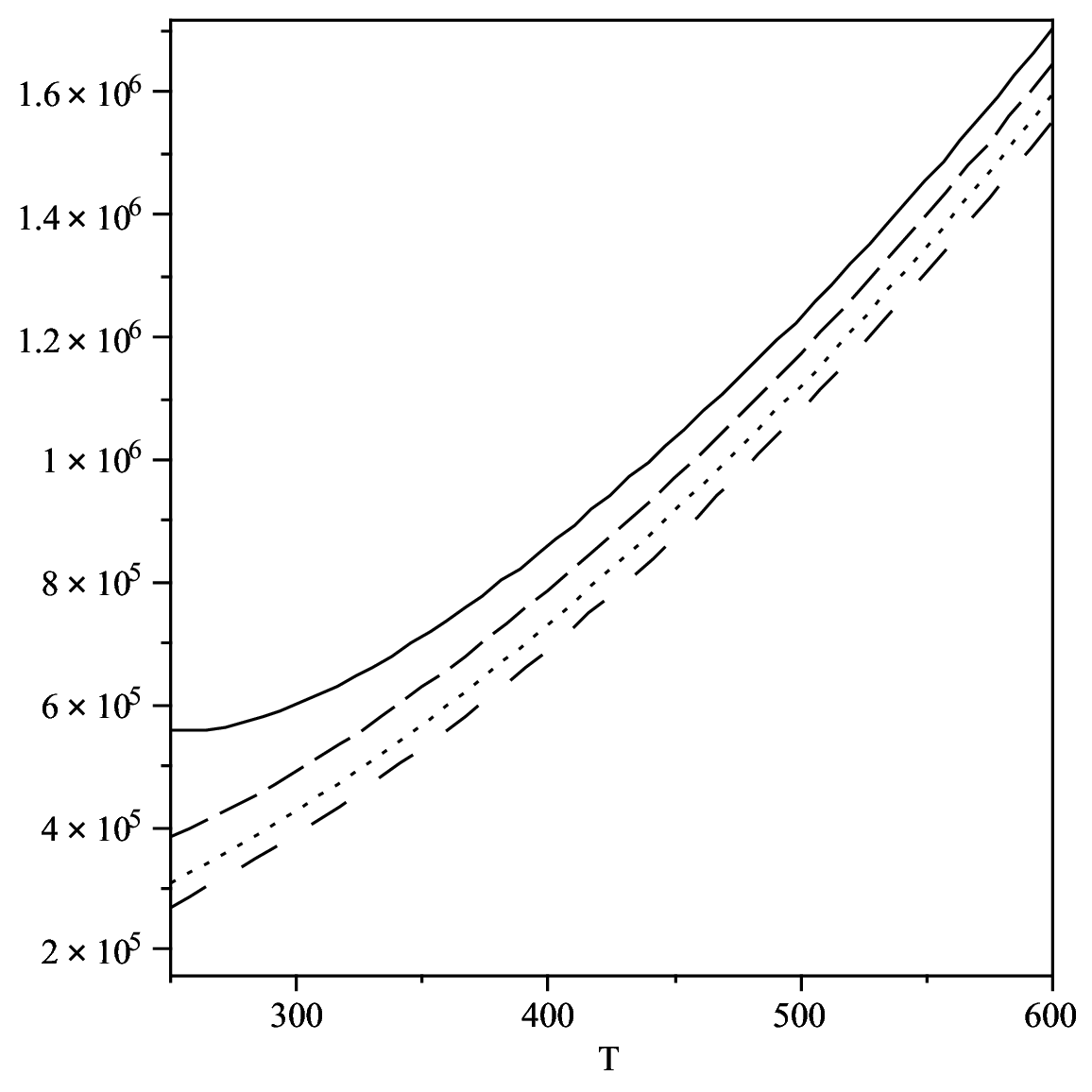}
\caption{The graphs of the drag force in terms of the temperature
for $v=0.5$, $\mu=1$, $\alpha^{\prime}=0.5$ and $\lambda=6\pi$.
Space dashed line drawn for the case of $q=0$. Dotted line drawn for
the case of one charged black hole with $q=10^6$. Dashed line drawn
for the case of Two charged black hole with $q=10^6$. Solid line
drawn for the case of three charged black hole with $q=10^6$.}
\end{center}
\end{figure}

Similar to the previous cases, for the $q\rightarrow0$ limit, we have $D_{q}=\frac{1}{\pi r_{h}}$, which means that diffusion coefficient
of the quark found proportional to inverse of the temperature.\\
As one can find from the Figs. 10-12 the behavior of the drag force
for small black hole charge approximately are the same. So, in order
to see difference of three cases we need to consider large black
hole charges. In that case it is interesting to compare above three
different configurations with each other and also with the case of
$q=0$ limit. It is easy to check that $q\rightarrow0$ limit of the
relation (72) reduced to the drag force of $\mathcal{N}=4$ SYM
theory [57]. Therefore, in the Fig. 13 we draw graph of the drag
force corresponding to four different situations. We found that the
black hole charge increases the value of drag force.
\subsection{Quasi-normal modes}
In this subsection we consider small perturbations of a straight string which stretched from $r=r_{m}$ to $r=r_{h}$ in STU background with three non-zero
charges. The quasi-normal modes give us information about the equilibrium state of the string after small perturbations. This allows us to obtain the
friction coefficient $\zeta$ in the non-relativistic regime of the quark. In that case we consider the static quark in the ${\mathcal{N}}=2$ supergravity
thermal plasma without any external fields. Indeed, we want to study the behavior of the string at the $t\rightarrow\infty$ and low velocity limits. The
small fluctuations around the straight string means that ${\dot{x}}^{2}$ and ${x^{\prime}}^{2}$ are small, so one can neglect them in the expression (66).
Then, under assumption of time-dependent solution of the form $x(r, t)=x(r) e^{-\zeta t}$, equation of motion reduces to the following relation,
\begin{equation}\label{s77}
\frac{f_{k}}{r^{2}{\mathcal{H}}^{\frac{5}{6}}}\partial_{r}\frac{f_{k}r^{2}}{{\mathcal{H}}^{\frac{1}{6}}}x^{\prime}=\zeta^{2}x.
\end{equation}
In order to obtain friction coefficient, we assume that $\zeta$ is
small, so one can use the following expansion,
\begin{equation}\label{s78}
x=x_{0}+\zeta x_{1}+\zeta^{2}x_{2}+\cdots.
\end{equation}
Also, by applying Neumann boundary condition we find,
\begin{equation}\label{s79}
x^{\prime}(r_{m})=\zeta
x_{1}^{\prime}(r_{m})+\zeta^{2}x_{2}^{\prime}(r_{m})=0.
\end{equation}
We should substitute the above relations to the equation (77) and compare appropriate coefficients, in that case the leading order yields to $x_{0}=A$,
where $A$ is a constant. Therefore, by using Neumann boundary condition and relation (79) one can obtain a quasinormal mode condition on $\zeta$ as the
following,
\begin{equation}\label{s80}
\zeta=\frac{r_{h}^{2}}{r_{m}R^{2}}
\left[(1+\frac{q_{1}}{r_{h}^{2}})(1+\frac{q_{2}}{r_{h}^{2}})(1+\frac{q_{3}}{r_{h}^{2}})\right]^{\frac{1}{3}}.
\end{equation}
Again, we may use this result to obtain drag force. In the large $r_{m}$ limit which corresponds to the heavy quark, from the relation (64), one can obtain
$M_{rest}=T_{0}r_{m}$. Also, we know that $\dot{P}=-\zeta M_{rest}v$. Therefor we find,
\begin{equation}\label{s81}
\frac{d P}{dt}\approx -T_{0}v\frac{r_{h}^{2}}{R^{2}}
\left[(1+\frac{q_{1}}{r_{h}^{2}})(1+\frac{q_{2}}{r_{h}^{2}})(1+\frac{q_{3}}{r_{h}^{2}})\right]^{\frac{1}{3}}.
\end{equation}
We see that the relation (81) exactly coincide with the relation
(72) which obtained for a slowly moving heavy quark.\\
Now, we can use these results to obtain the total energy and momentum of the string. By using the equation of motion (77), time dependent solution of the
form $\dot{x}=-\zeta x$, and momentum densities (61) one can obtain,
\begin{equation}\label{s82}
\pi_{x}^{0}= -\frac{T_{0}}{\zeta R^{2}}\partial_{r}\frac{f_{k}r^{2}}{\mathcal{H}^{\frac{1}{6}}}x^{\prime}.
\end{equation}
Using Neumann boundary condition together the total momentum integral (62) yields to the following result,
\begin{equation}\label{s83}
P= \frac{T_{0}}{\zeta R^{2}}\frac{f_{k}(r_{min})r_{min}^{2}}{\left[(1+\frac{q_{1}}{r_{min}^{2}})
(1+\frac{q_{2}}{r_{min}^{2}})(1+\frac{q_{3}}{r_{min}^{2}})\right]^{\frac{1}{6}}}x^{\prime}(r_{min}),
\end{equation}
where we insert $r_{min}>r_{h}$ as lower limit of integral to avoid
divergency. The reason is that the quasi-normal modes diverge close to the horizon. In addition they are rapidly oscillating.
So, quantities like $x(r_{h} )$ and $x^{\prime} (r_{h} )$ are not well defined right at $r=r_h$.
In order to regulate these divergences we cut-off the integrals at a finite $r_{min}$.\\
In order to obtain the total energy we keep second order of
velocities and expand $\sqrt{-g}$, then similar to the calculation
of momentum, one can find,
\begin{equation}\label{s84}
\pi_{t}^{0}= -\frac{T_{0}}{\mathcal{H}^{\frac{1}{6}}}\left[1+\frac{f_{k}r^{2}}{2R^{2}}x^{\prime2}
+\frac{r^{2}}{2R^{2}}\frac{\mathcal{H}}{f_{k}}\dot{x}^{2}\right].
\end{equation}
Then, by using the equation of motion and Neumann boundary condition we get,
\begin{equation}\label{s85}
E=T_{0}\int_{r_{min}}^{r_{m}}\frac{dr}{\mathcal{H}^{\frac{1}{6}}} -\frac{T_{0}}{2
R^{2}}\frac{f_{k}(r_{min})r_{min}^{2}}{\left[(1+\frac{q_{1}}{r_{min}^{2}})
(1+\frac{q_{2}}{r_{min}^{2}})(1+\frac{q_{3}}{r_{min}^{2}})\right]^{\frac{1}{6}}}x(r_{min})x^{\prime}(r_{min}).
\end{equation}
According to the relation (64) one can interpret the integral of the right hand side of the equation (85) as rest mass of quark. So, combining the
relations (83) and (85) yield to the relation between the energy and momentum as,
\begin{equation}\label{s86}
E=M_{rest}+\frac{P^2}{2M_{kin}},
\end{equation}
where the kinetic mass defined as the following expression,
\begin{equation}\label{s87}
M_{kin}\equiv \frac{T_{0}}{\zeta
R^{2}}r_{h}^{2}\left[(1+\frac{q_{1}}{r_{h}^{2}})
(1+\frac{q_{2}}{r_{h}^{2}})(1+\frac{q_{3}}{r_{h}^{2}})\right]^{\frac{1}{3}}.
\end{equation}
It is interesting to note that the equation (86) is valid for every theories such as $\mathcal{N}=4$ SYM theory and $\mathcal{N}=2$ gauged supergravity.

\subsection{Effect of the constant electromagnetic field}
In the previous sections the moving heavy quark through the $\mathcal{N}=2$ plasma considered without any external field. In this section we would like to
introduce a constant electromagnetic field on the brane which affects on the motion of heavy quark. In the description of the AdS/CFT correspondence the
endpoint of both fundamental and Dirichlet strings under influence of non-zero NS NS B-field background corresponds to the moving quark with a constant
electromagnetic field. We assume that the constant electromagnetic field is along $x^{1}$ and $x^{2}$ directions. Therefore, we add a constant $B$-field in
the form of $B=B_{01}dt\wedge dx_{1}+B_{12}dx_{1}\wedge dx_{2}$ to the line element (1), where $B_{01}$ is the constant electric field and $B_{12}$ is the
constant magnetic field. Also $B_{01}$ and $B_{12}$ are antisymmetric fields and other components of the $B$-field are
zero. We must note that the same work was done originally for $\mathcal{N}$=4 SYM theory [88].\\
Because of introducing $B_{01}$ and $B_{12}$, the curved string dual to the heavy quark may be described by the $x_{1}(r, t)=x_{1}(r)+v_{1}t$, $x_{2}(r,
t)=x_{2}(r)+v_{2}t$ and $x_{3}(r, t)=0$. Therefore, the square root of lagrangian density (59) takes the following form,
\begin{equation}\label{s88}
-g=\frac{1}{{\mathcal{H}}^{\frac{1}{3}}}\left[1-\frac{{\mathcal{H}}r^{2}}
{f_{k}R^{2}}\vec{v}^{2}+\frac{f_{k}r^{2}}{R^{2}}{x^{\prime}}^{2}-(B_{01}x_{1}^{\prime} +B_{12}(\vec{v}\times\vec{x}^{\prime}))^{2}\right],
\end{equation}
where $\vec{v}=(v_{1}, v_{2})$ is the vector of velocity and
$\vec{x}^{\prime}=(x_{1}^{\prime}, x_{2}^{\prime})$ is the projected
directions of string tail.\\
We consider three special cases as the following. First, we assume
that only electric field is exist and $B_{12}=0$. Second, we have
non-zero magnetic field and there is
$B_{01}=0$. Finally we discuss about the case where $\vec{v}\bot B_{01}$.\\
In order to study the effect of constant electric field we set $B_{12}=0$ and choose $x_{1}$ as the moving direction of the quark, so we have $v_{1}=v$,
$v_{2}=v_{3}=0$, $x_{1}(r, t)=x(r)+vt$ and $x_{2}(r, t)=x_{3}(r, t)=0$. Therefore, the equation (88) reduced to the following relation,
\begin{equation}\label{s89}
-g=\frac{1}{{\mathcal{H}}^{\frac{1}{3}}}\left[1-\frac{{\mathcal{H}}r^{2}} {f_{k}R^{2}}v^{2}+(\frac{f_{k}r^{2}}{R^{2}}-B_{01}^{2}){x^{\prime}}^{2}\right].
\end{equation}
Therefore, comparing the relations (65) and (89) yields to the following expression,
\begin{equation}\label{s90}
x^{\prime2}=\frac{C^{2}v^{2}R^{2}{\mathcal{H}}^{\frac{1}{3}}}{f_{k}^{2}r^{2}}
\frac{f_{k}R^{2}-{\mathcal{H}}r^{2}v^{2}}{f_{k}r^{2}(1-\frac{B_{01}^{2}R^{2}}{f_{k}r^{2}})^{2}-C^{2}v^{2}R^{2}{\mathcal{H}}^{\frac{1}{3}}
(1-\frac{B_{01}^{2}R^{2}}{f_{k}r^{2}})}.
\end{equation}
It means that $r_{c}$ is given by the relation (71), but the constant $C$ modified as the following,
\begin{equation}\label{s91}
C=\left[\prod_{i=1}^{3}(1+\frac{q_{i}}{r_{c}^{2}})\right]^{\frac{1}{3}}\frac{r_{c}^{2}}{R^{2}}
\sqrt{1-\frac{B_{01}^{2}R^{4}}{\prod_{i=1}^{3}(1+\frac{q_{i}}{r_{c}^{2}})r_{c}^{4}v^{2}}}.
\end{equation}
It yields us to the expression of drag force,
\begin{equation}\label{s92}
\frac{dP}{dt}=-T_{0}v\left[\prod_{i=1}^{3}(1+\frac{q_{i}}{r_{h}^{2}})\right]^{\frac{1}{3}}\frac{r_{h}^{2}}{R^{2}}
\sqrt{1-\frac{B_{01}^{2}R^{4}}{\prod_{i=1}^{3}(1+\frac{q_{i}}{r_{h}^{2}})r_{h}^{4}v^{2}}}+\mathcal{O}(v^{2}).
\end{equation}
It is clear that $B_{01}\rightarrow0$ limit of the equation (92)
reduces to the equation (72). Also, we find that the effect of the
constant electric field is decreasing the drag force, so this result
agree with the result of the Ref. [88]. It is also interesting to
write the linearized expression for small electric field of the
relation (92).
In that case the correction terms are of even powers of $B_{01}$, which is natural characteristic of the Nambu-Goto action.\\
In the second case we consider only constant magnetic field $B_{12}$. In that case one can choose $x_{1}(r, t)=x_{1}(r)+vt$, $x_{2}(r, t)=x_{2}(r)$ and
$x_{3}(r, t)=0$. Under these assumptions one can find,
\begin{eqnarray}\label{s93}
x_{1}^{\prime}(r)&=&\pi_{x_{1}}\left[\frac{\beta(\frac{1}{{\mathcal{H}}^{\frac{1}{3}}}
-\frac{{\mathcal{H}}^{\frac{2}{3}}r^{2}v^{2}}{f_{k}})}{\frac{f_{k}r^{2}}{{\mathcal{H}}^{\frac{1}{3}}}
\left((\pi_{x_{1}}^{2}-\frac{f_{k}r^{2}}{{\mathcal{H}}^{\frac{1}{3}}})(\pi_{x_{2}}^{2}-\beta)
-\pi_{x_{1}}^{2}\pi_{x_{2}}^{2}\right)}\right]^{\frac{1}{2}},\nonumber\\
x_{2}^{\prime}(r)&=&\pi_{x_{2}}\left[\frac{\frac{f_{k}r^{2}} {{\mathcal{H}}^{\frac{1}{3}}}(\frac{1}{{\mathcal{H}}^{\frac{1}{3}}}
-\frac{{\mathcal{H}}^{\frac{2}{3}}r^{2}v^{2}}{f_{k}})}{\beta \left((\pi_{x_{1}}^{2}-\frac{f_{k}r^{2}}{{\mathcal{H}}^{\frac{1}{3}}})(\pi_{x_{2}}^{2}-\beta)
-\pi_{x_{1}}^{2}\pi_{x_{2}}^{2}\right)}\right]^{\frac{1}{2}},
\end{eqnarray}
where we defined,
$\beta\equiv\frac{f_{k}r^{2}}{{\mathcal{H}}^{\frac{1}{3}}}-v^{2}B_{12}^{2}$
and set $\pi_{x_{i}}^{1}\equiv \pi_{x_{i}}$. Now, the reality
condition implies that $\pi_{x_{2}}=0$, and therefore we yield to
the expression (72). It tell us that there is no drag force in
$x^{2}$ direction and therefore the constant magnetic field have no
effect on the motion along $x^{1}$ direction. Actually vanishing of
$\pi_{x_{2}}$ is consequence of vanishing of $v_{2}$. According to
these two cases, (electric and magnetic fields), we found that the
constant magnetic field have no effect on the motion of string and
it is appropriate electric field which keeps the
string at constant speed $v$.\\
Finally, we consider the case of $\vec{v}\bot B_{01}$. It means that
one may choose the solutions of equation of motion as, $x_{1}(r,
t)=x_{1}(r)$, $x_{2}(r, t)=x_{2}(r)+vt$ and $x_{3}(r, t)=0$. A
possible drag force may be found as the following relation,
\begin{equation}\label{s94}
\frac{dP_{2}}{dt}= -T_{0}\sqrt{\frac{f_{k}(r_{h})r_{h}^{2}}{\left[\prod_{i}(1+\frac{q_{i}}{r_{h}^{2}})\right]^\frac{1}{3}}
-v^{2}B_{12}^{2}}+\mathcal{O}(v^{2}),
\end{equation}
and ${\dot{P}}_{1}=0$. In this case the constant electric field has
no effect on drag force. It should be mentioned that this situation
is not of our interesting case which considered in this paper. The
magnetic field on the brane has equivalent interpretation as the
following. One can consider a moving heavy quark in the
non-commutative plane. Both cases (ordinary theory with $B$-field
and non-commutative theory without $B$-field) yield to similar
result, which is decreasing the drag force or equivalently
decreasing the effective viscosity of QGP.\\
In the next step,
without any external fields, we try to obtain effect of higher
derivative terms on the drag force.

\subsection{Higher derivative correction}
Already we studied the effect of higher derivative correction on
shear viscosity and conductivities of QGP. Now, we are ready to
consider higher derivative effect on the drag force. By using the
solution (14) one can obtain,
\begin{equation}\label{s95}
\frac{d
P}{dt}=-T_{0}v\left[\prod_{i=1}^{3}(1+\frac{q_{i}}{r_{h}^{2}}-\frac{c_{1}q_{i}(q_{i}+\mu)}
{72r_{h}^{2}(r_{h}^{2}+q_{i})^{2}})\right]^{\frac{1}{3}}\frac{r_{h}^{2}}{R^{2}}
(1+\mathcal{O}(v^{2})),
\end{equation}
where $r_{h}$ is given by the relation (15), and we used,
\begin{equation}\label{s96}
r_{c}=r_{h}+\frac{r_{h}^{2}v^{2}\mathcal{H}(r_{h})}{R^{2}f_{k}^{\prime}(r_{h})}+\mathcal{O}(v^{4}),
\end{equation}
with $f_{k}$ and $\mathcal{H}$ are given by the relation (14), and
prime denotes derivative with respect to $r$. As expected, the
$c_{1}=0$ limit of the $\dot{P}$ coincides with the results of
subsection 5.1.

\begin{figure}[th]
\begin{center}
\includegraphics[scale=.33]{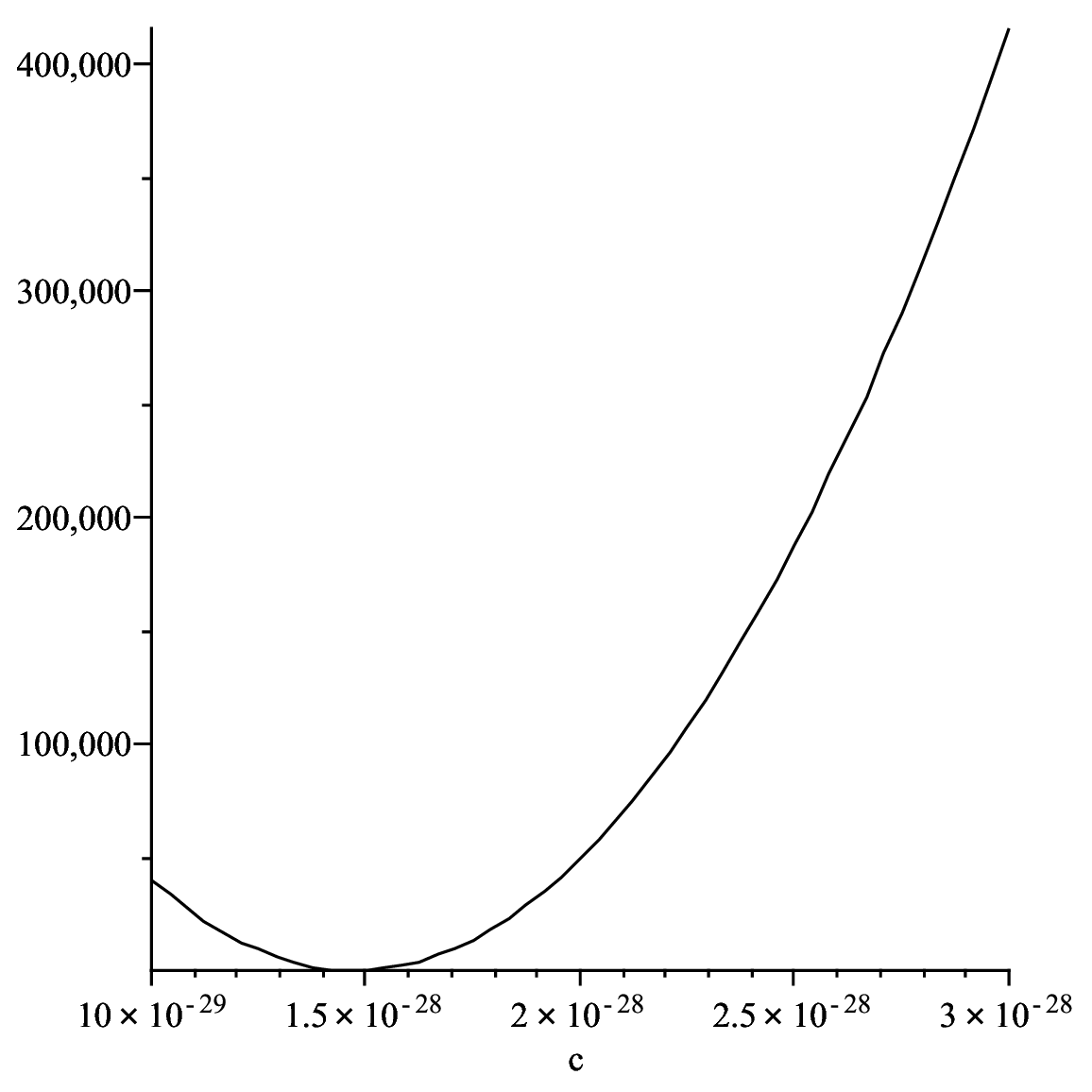}
\caption{The graph of the drag force in terms of the higher derivative parameter for $v=0.5$, $\mu=1$, $\alpha^{\prime}=0.5$, $\lambda=6\pi$, $T=300$ $MeV$
and $q=1$.}
\end{center}
\end{figure}

In that case we draw figure of the drag force in terms of the higher
derivative parameter (see Fig. 14) and find that, for
$c_{1}<1.5\times10^{-28}$ the effect of the higher derivative terms
is to decrease the drag force, then for $c_{1}>1.5\times10^{-28}$
the value of the drag force increases. So, if we set
$c_{1}\approx1.5\times10^{-28}$ drag force vanishes, also $c_{1}=0$
and $c_{1}\approx3\times10^{-28}$ yields to similar value of the
drag force. Therefore, in order to obtain enhanced drag force due to
higher derivative terms, we should set $c_{1}>3\times10^{-28}$. This
situation is similar for the three cases of one, two and three
charged black holes.

\subsection{Quark-anti quark solution}
Now, we consider a moving quark-antiquark pair which may be
interpreted as a meson. Indeed there is a moving meson with the
constant speed $v$ in the ${\mathcal{N}}=2$ supergravity thermal
plasma. Already the energy of a moving quark-antiquark pair in
${\mathcal{N}}=4$ SYM plasma calculated [89]. Now, we would like to
repeat same calculations in the STU background. The quark-antiquark
pair in the thermal QGP corresponds to an open string in $AdS_{5}$
space with two endpoints on the D-brane in the ($X, Y$) plan. Two
end points of string on the D-brane represent quark and antiquark
which separated from each other by a constant $l$. We assume that at
the $t=0$ string is straight and two endpoints of string move with
the constant velocity $v$ along the $X$ direction. The dynamics of
such configuration discussed in detail in the Ref. [89] for the
${\mathcal{N}}=4$ SYM plasma. Here, for the ${\mathcal{N}}=2$
supergravity thermal plasma one can obtain,
\begin{equation}\label{s97}
-g=\frac{1}{{\mathcal{H}}^{\frac{1}{3}}}\left[1+\frac{f_{k}r^{2}}{R^{2}}({x^{\prime}}^{2}+{y^{\prime}}^{2})
-\frac{{\mathcal{H}}r^{2}}{f_{k}R^{2}}(\dot{x}^{2}+\dot{y}^{2})
-\frac{r^{4}}{R^{4}}{\mathcal{H}}(\dot{x}^{2}{y^{\prime}}^{2}+\dot{y}^{2}{x^{\prime}}^{2}-2\dot{x}x^{\prime}\dot{y}y^{\prime})\right],
\end{equation}
where $f_{k}$ and $\mathcal{H}$ are given by the relation (2). The
equations of motion of $x$ and $y$  are given by the following
equations, respectively,
\begin{eqnarray}\label{s98}
\frac{\partial}{\partial r}\left[\frac{1}{\sqrt{-g}}
(\frac{r^{4}}{R^{2}}{\mathcal{H}}^{\frac{2}{3}}({\dot{y}}^{2}x^{\prime}-\dot{x}\dot{y}y^{\prime})+\frac{f_{k}r^{2}x^{\prime}}{{\mathcal{H}}^{\frac{1}{3}}})\right]
+r^{2}{\mathcal{H}}^{\frac{2}{3}}\frac{\partial}{\partial t}\left[\frac{1}{\sqrt{-g}}(\frac{\dot{x}}{f_{k}}
+\frac{r^{2}}{R^{2}}({y^{\prime}}^{2}\dot{x}-x^{\prime}\dot{y}y^{\prime}))\right]=0,\nonumber\\
\frac{\partial}{\partial r}\left[\frac{1}{\sqrt{-g}}
(\frac{r^{4}}{R^{2}}{\mathcal{H}}^{\frac{2}{3}}({\dot{x}}^{2}y^{\prime}-\dot{x}\dot{y}x^{\prime})-\frac{f_{k}r^{2}y^{\prime}}{{\mathcal{H}}^{\frac{1}{3}}})\right]
+r^{2}{\mathcal{H}}^{\frac{2}{3}}\frac{\partial}{\partial t}\left[\frac{1}{\sqrt{-g}}(\frac{\dot{y}R^{2}}{f_{k}}
+r^{2}({x^{\prime}}^{2}\dot{y}-x^{\prime}\dot{x}y^{\prime}))\right]=0,\nonumber\\
\end{eqnarray}
then momentum densities obtained by the following equation,
\begin{eqnarray}\label{s99}
&&\left(\begin{array}{ccc}
\pi_{x}^{0} & \pi_{x}^{1}\\
\pi_{y}^{0} & \pi_{y}^{1}\\
\pi_{r}^{0}& \pi_{r}^{1}\\
\pi_{t}^{0} & \pi_{t}^{1}\\
\end{array}\right)=-T_{0}\frac{r^{2}{\mathcal{H}}^{\frac{1}{3}}}{R^{2}\sqrt{-g}}\times\nonumber\\
&&\left(\begin{array}{ccc}
\frac{r^{2}}{R^{2}}{\mathcal{H}}^{\frac{1}{3}}x^{\prime}\dot{y}y^{\prime}
-(\frac{{\mathcal{H}}^{\frac{1}{3}}}{f_{k}}+
\frac{r^{2}}{R^{2}}{\mathcal{H}}^{\frac{1}{3}}{y^{\prime}}^{2})\dot{x}&
\frac{r^{2}}{R^{2}}{\mathcal{H}}^{\frac{1}{3}}y^{\prime}\dot{y}\dot{x}
-(\frac{{\mathcal{H}}^{\frac{1}{3}}r^{2}}{R^{2}}{\dot{y}}^{2}-\frac{f_{k}}{{\mathcal{H}}^{\frac{2}{3}}})x^{\prime}\\
\frac{r^{2}}{R^{2}}{\mathcal{H}}^{\frac{1}{3}}y^{\prime}\dot{x}x^{\prime}
-(\frac{{\mathcal{H}}^{\frac{1}{3}}}{f_{k}}+
\frac{r^{2}}{R^{2}}{\mathcal{H}}^{\frac{1}{3}}{x^{\prime}}^{2})\dot{y}&
\frac{r^{2}}{R^{2}}{\mathcal{H}}^{\frac{1}{3}}x^{\prime}\dot{y}\dot{x}
-(\frac{{\mathcal{H}}^{\frac{1}{3}}r^{2}}{R^{2}}{\dot{x}}^{2}-\frac{f_{k}}{{\mathcal{H}}^{\frac{2}{3}}})y^{\prime}\\
\frac{{\mathcal{H}}^{\frac{1}{3}}}{f_{k}}(\dot{x}x^{\prime}+\dot{y}y^{\prime})
&
\frac{R^{2}}{{\mathcal{H}}^{\frac{2}{3}}r^{2}}-\frac{{\mathcal{H}}^{\frac{1}{3}}}{f_{k}}({\dot{x}}^{2}+{\dot{y}}^{2})\\
\frac{R^{2}}{{\mathcal{H}}^{\frac{1}{3}}r^{2}}+\frac{f_{k}}{{\mathcal{H}}^{\frac{1}{3}}}({x^{\prime}}^{2}+{y^{\prime}}^{2})
&
-\frac{f_{k}}{{\mathcal{H}}^{\frac{2}{3}}}(\dot{x}x^{\prime}+\dot{y}y^{\prime})\\
\end{array}\right).
\end{eqnarray}
There are two interesting motions for the meson. The first one is
the moving quark-antiquark pair with constant speed $v$. The second
case
is the rotational motion of the quark-antiquark pair.\\
The first system may be described by the $x(r, t) =vt+x(r)$ and
$y(r, t)= y(r)$ profiles. These solutions satisfy boundary
conditions as $x(\infty, t)=vt$ and $y(\infty)=\pm l/2$. In this
case equation (99) reduces to the following expression,
\begin{eqnarray}\label{s100}
\left(\begin{array}{ccc}
\pi_{x}^{0} & \pi_{x}^{1}\\
\pi_{y}^{0} & \pi_{y}^{1}\\
\pi_{r}^{0}& \pi_{r}^{1}\\
\pi_{t}^{0} & \pi_{t}^{1}\\
\end{array}\right)=-T_{0}\frac{r^{2}}{R^{2}}\frac{{\mathcal{H}}^{\frac{2}{3}}}{\sqrt{-g}}
\left(\begin{array}{ccc}
-v(\frac{1}{f_{k}}+\frac{r^{2}}{R^{2}}{y^{\prime}}^{2})& \frac{f_{k}}{{\mathcal{H}}}x^{\prime}\\
\frac{r^{2}}{R^{2}}vy^{\prime}x^{\prime}&-(\frac{r^{2}}{R^{2}}v^{2}-\frac{f_{k}}{{\mathcal{H}}})y^{\prime}\\
\frac{v}{f_{k}}x^{\prime} &
\frac{R^{2}}{{\mathcal{H}}r^{2}}-\frac{v^{2}}{f_{k}}\\
(\frac{R^{2}}{f_{k}r^{2}}+{x^{\prime}}^{2}+{y^{\prime}}^{2})\frac{f_{k}}{{\mathcal{H}}}
&
-v\frac{f_{k}}{{\mathcal{H}}}x^{\prime}\\
\end{array}\right),
\end{eqnarray}
where,
\begin{equation}\label{s101}
-g=\frac{1}{{\mathcal{H}}^{\frac{1}{3}}}\left[1+\frac{f_{k}r^{2}}{R^{2}}({x^{\prime}}^{2}+{y^{\prime}}^{2})
-\frac{{\mathcal{H}}r^{2}v^{2}}{f_{k}R^{2}}
-\frac{r^{4}}{R^{4}}{\mathcal{H}}v^{2}{y^{\prime}}^{2}\right].
\end{equation}
In order to obtain drag force, we calculate $\pi_{x}^{1}$ and
$\pi_{y}^{1}$ components and solve them for $x^{\prime}$ and
$y^{\prime}$ respectively and obtain,
\begin{eqnarray}\label{s102}
x^{\prime}(r)&=&\pi_{x}^{1}\frac{R}{r}(1-\frac{{\mathcal{H}}r^{2}v^{2}}{f_{k}R^{2}})\left[(\frac{f_{k}}{{\mathcal{H}}}-\frac{r^{2}v^{2}}{R^{2}})
(T_{0}^{2}\frac{r^{2}}{R^{2}}f_{k}{\mathcal{H}}^{\frac{2}{3}}-{\mathcal{H}}(\pi_{x}^{1})^{2})-f_{k}(\pi_{y}^{1})^{2}
\right]^{-\frac{1}{2}},\nonumber\\
y^{\prime}(r)&=&\pi_{y}^{1}\frac{R}{r}\left[(\frac{f_{k}}{{\mathcal{H}}}-\frac{r^{2}v^{2}}{R^{2}})
(T_{0}^{2}\frac{r^{2}}{R^{2}}f_{k}{\mathcal{H}}^{\frac{2}{3}}-{\mathcal{H}}(\pi_{x}^{1})^{2})-f_{k}(\pi_{y}^{1})^{2}
\right]^{-\frac{1}{2}}.
\end{eqnarray}
As before, by using reality condition one can obtain,
\begin{equation}\label{s103}
(\pi_{y}^{1})^{2}=\left[(\frac{f_{k}}{{\mathcal{H}}}-\frac{r^{2}v^{2}}{R^{2}})
(T_{0}^{2}\frac{r^{2}}{R^{2}}{\mathcal{H}}^{\frac{2}{3}}-\frac{{\mathcal{H}}}{f_{k}}(\pi_{x}^{1})^{2})\right]_{r=r_{min}},
\end{equation}
where $r_{min}$ is turning point of string. One can check easily
that $r_{min}\geq r_{c}$ ($r_{c}$ is critical radius which
introduced in the subsection 5.1, but $r_{min}$ differs from UV cut
off which introduced in the relation (83) ). If $\pi_{y}^{1}=0$,
then $r_{min}= r_{c}$ and above solutions are similar to single
quark solution ($l=0$). Here, in order the string have a chance of
turning around smoothly, it requires that $\partial y/\partial
x=y^{\prime}/x^{\prime}=\infty$ at $r_{min}$ [89]. So, it is
necessary to have $\pi_{x}^{1}=0$. Therefore, one can find drag
force as,
\begin{equation}\label{s104}
\pi_{y}^{1}=\frac{T_{0}}{R}r_{min}{\mathcal{H}}^{\frac{1}{3}}(r_{min})\sqrt{\frac{f_{k}(r_{min})}{{\mathcal{H}}(r_{min})}
-\frac{r_{min}^{2}v^{2}}{R^{2}}}.
\end{equation}
In the second case we add a rotational motion with angular velocity
$\dot{\theta}$ to the motion of meson. Therefore, the string may be
described by the $x(r, t)=vt+x(r)\sin\theta$ and $y(r,
t)=y(r)\cos\theta$ profiles.\\
Fig. 15 shows the configuration of rotating string. The points $A$
and $B$ in the Fig. 15 represent quark and antiquark with separating
length $l$. The radial coordinate $r$ varies from $r_{h}$ (black
hole horizon radius) to $r=r_{m}$ on $D$-brane. $r_{c}$ is a
critical radius, obtained for single quark solution, which the
string can't penetrate beyond it and $r_{min}\geq r_{c}$.
$r_{min}=r_c$ is satisfied if points $A$ and $B$ located at origin
($l=0$), in that case there is the straight string which is dual
picture of the single static quark. $\theta$ is assumed to be the
angle with $Y$ axis and the string center of mass moves along $X$
axis with velocity $v$. Solutions of This configuration satisfy
boundary conditions $x(\infty, t)=vt\pm\frac{l}{2}\sin\theta$ and
$y(\infty, t)=\pm\frac{l}{2}\cos\theta$, where for $\theta=0$ reduce
to the boundary condition without rotational motion.\\
Also there is another condition due to our conjecture,
$y^{\prime}/x^{\prime}=\cot\theta$, which reduces to
$y^{\prime}/x^{\prime}\rightarrow\infty$ at the $\theta\rightarrow0$
limit, which is agree with the first case.\\
These boundary conditions can also satisfy with two separated string
which move at velocity $v$ along $X$ axis and simultaneously swing a
circle with radius $l/2$. Specifying these boundary conditions
doesn't lead to a unique solution for equation of motion, so we
should specify additional conditions for this motion.\\
Here, we assume that the string is initially upright, move at
velocity $v$
and rotates around its center of mass.\\
Now, by using above solutions in the equation (99) and solving
resulting equations with respect to $x^{\prime}$ and $y^{\prime}$
one can obtain following equations,
\begin{eqnarray}\label{s105}
A{x^{\prime}}^{2}+B{y^{\prime}}^{2}+Cx^{\prime}y^{\prime}+D&=&
0,\nonumber\\
A^{\prime}{x^{\prime}}^{2}+B^{\prime}{y^{\prime}}^{2}+C^{\prime}x^{\prime}y^{\prime}+D^{\prime}&=&
0,
\end{eqnarray}


where,
\begin{eqnarray}\label{s106}
A&=&\mathcal{R}^{2}\sin^{2}\theta
\left[{\pi_{x}^{1}}^{2}\left(\frac{f_{k}}{{\mathcal{H}}^{\frac{1}{3}}}
-{\mathcal{H}}^{\frac{2}{3}}y^{2}{\dot{\theta}}^{2}\mathcal{R}^{2}\sin^{2}\theta\right)
-T_{0}^{2}{\mathcal{H}}^{\frac{2}{3}}\mathcal{R}^{2}
\left({\mathcal{H}}^{\frac{1}{3}}y^{2}{\dot{\theta}}^{2}\mathcal{R}^{2}\sin^{2}\theta
-\frac{f_{k}}{{\mathcal{H}}^{\frac{2}{3}}}\right)^{2}\right],\nonumber\\
B&=&\mathcal{R}^{2}\cos^{2}\theta\left[{\pi_{x}^{1}}^{2}\left(\frac{f_{k}}{{\mathcal{H}}^{\frac{1}{3}}}
-{\mathcal{H}}^{\frac{2}{3}}(v+x\dot{\theta}\cos\theta)^{2}\mathcal{R}^{2}\right)
-T_{0}^{2}{\mathcal{H}}^{\frac{4}{3}}y^{2}{\dot{\theta}}^{2}
\mathcal{R}^{6}\sin^{2}\theta(v+x\dot{\theta}\cos\theta)^{2}\right],\nonumber\\
C&=&-2y\dot{\theta}\mathcal{R}^{4}\sin^{2}\theta\left[{\pi_{x}^{1}}^{2}{\mathcal{H}}^{\frac{2}{3}}\cos\theta
+T_{0}^{2}{\mathcal{H}}
\mathcal{R}^{2}\left({\mathcal{H}}^{\frac{1}{3}}y^{2}{\dot{\theta}}^{2}\mathcal{R}^{2}\sin^{2}\theta
-\frac{f_{k}}{{\mathcal{H}}^{\frac{2}{3}}}\right)\right](v+x\dot{\theta}\cos\theta),\nonumber\\
D&=&\mathcal{R}^{2}{\pi_{x}^{1}}^{2}\frac{{\mathcal{H}}^{\frac{2}{3}}}{f_{k}}
\left[\frac{f_{k}}{\mathcal{R}^{2}{\mathcal{H}}}
-y^{2}{\dot{\theta}}^{2}\sin^{2}\theta-(v+x\dot{\theta}\cos\theta)^{2}\right],\nonumber\\
A^{\prime}&=&\mathcal{R}^{2}\sin^{2}\theta
\left[{\pi_{y}^{1}}^{2}\left(\frac{f_{k}}{{\mathcal{H}}^{\frac{1}{3}}}
-{\mathcal{H}}^{\frac{2}{3}}y^{2}{\dot{\theta}}^{2}\mathcal{R}^{2}\sin^{2}\theta\right)
-T_{0}^{2}{\mathcal{H}}^{\frac{4}{3}}y^{2}{\dot{\theta}}^{2}\mathcal{R}^{6}
\sin^{2}\theta(v+x\dot{\theta}\cos\theta)^{2}\right],\nonumber\\
B^{\prime}&=&\mathcal{R}^{2}\cos^{2}\theta\left[{\pi_{y}^{1}}^{2}\left(\frac{f_{k}}{{\mathcal{H}}^{\frac{1}{3}}}
-{\mathcal{H}}^{\frac{2}{3}}(v+x\dot{\theta}\cos\theta)^{2}\mathcal{R}^{2}\right)
-T_{0}^{2}{\mathcal{H}}^{\frac{2}{3}}\mathcal{R}^{2}\left(\mathcal{R}^{2}{\mathcal{H}}^{\frac{1}{3}}(v+x\dot{\theta}\cos\theta)^{2}
-\frac{f_{k}}{{\mathcal{H}}^{\frac{2}{3}}}\right)^{2}\right],\nonumber\\
C^{\prime}&=&-2y\dot{\theta}\mathcal{R}^{4}\sin^{2}\theta\cos\theta\left[{\pi_{y}^{1}}^{2}{\mathcal{H}}^{\frac{2}{3}}
+T_{0}^{2}{\mathcal{H}}
\mathcal{R}^{2}\left(\mathcal{R}^{2}{\mathcal{H}}^{\frac{1}{3}}(v+x\dot{\theta}\cos\theta)^{2}
-\frac{f(r)}{{\mathcal{H}}^{\frac{2}{3}}}\right)\right](v+x\dot{\theta}\cos\theta),\nonumber\\
D^{\prime}&=&\mathcal{R}^{2}{\pi_{y}^{1}}^{2}\frac{{\mathcal{H}}^{\frac{2}{3}}}{f_{k}}
\left[\frac{f_{k}}{\mathcal{R}^{2}{\mathcal{H}}}
-y^{2}{\dot{\theta}}^{2}\sin^{2}\theta-(v+x\dot{\theta}\cos\theta)^{2}\right],
\end{eqnarray}
where we set $\frac{r}{R}\equiv\mathcal{R}$, so this is different
with Ricci scalar introduced in the relation (8). We must note that
the variable $C$ in equations (105) and (106) are different with
integration constant in equation (65), hence subsections 5.1 and
5.3. Therefore, from the equations (105) one can obtain,
\begin{eqnarray}\label{s107}
x^{\prime}(r)&=&2\left[\frac{D(B-\frac{{\pi_{y}^{1}}^{2}}{{\pi_{x}^{1}}^{2}}B^{\prime})}{C^{2}-{C^{\prime}}^{2}
-4(BA-B^{\prime}A^{\prime})}\right]^{\frac{1}{2}},\nonumber\\
y^{\prime}(r)&=&2\left[\frac{D(A-\frac{{\pi_{y}^{1}}^{2}}{{\pi_{x}^{1}}^{2}}A^{\prime})}{C^{2}-{C^{\prime}}^{2}
-4(BA-B^{\prime}A^{\prime})}\right]^{\frac{1}{2}}.
\end{eqnarray}
Here, if the rotational motion vanishes ($\dot{\theta}=0$), from
equations (104) one can see that coefficients of
$x^{\prime}y^{\prime}$ vanish ($C=C^{\prime}=0$) and our solutions
recover the motion of quark-antiquark pair without rotation. In
order to obtain drag force we use reality condition and find a
relation between variable (106) as
$\frac{A}{A^{\prime}}=\frac{B}{B^{\prime}}=\frac{C}{C^{\prime}}=\frac{D}{D^{\prime}}=(\frac{{\pi_{y}^{1}}}{{\pi_{x}^{1}}})^{2}$.
Then one can find two equations as, $C^{2}-4AB=0$ and
${C^{\prime}}^{2}-4A^{\prime}B^{\prime}=0$. These equations specify
$\pi_{x}^{1}$ and $\pi_{y}^{1}$ respectively. After some
calculations and simplifications we find,
\begin{eqnarray}\label{s108}
(\pi_{x}^{1})^{2}&=&\frac{1}{2a}
\left[\pm\sqrt{b^{2}-4ac}-b\right],\nonumber\\
(\pi_{y}^{1})^{2}&=&\frac{1}{2a^{\prime}}
\left[\pm\sqrt{{b^{\prime}}^{2}-4a^{\prime}c^{\prime}}-b^{\prime}\right],
\end{eqnarray}
where
\begin{eqnarray}\label{s109}
a&=&\prod_{i}(1+\frac{q_{i}}{r_{min}^{2}})^{\frac{1}{3}}\cos^{2}\theta(\mathcal{R}_{min}^{2}\varsigma+\xi\chi),\nonumber\\
b&=&T_{0}^{2}\prod_{i}(1+\frac{q_{i}}{r_{min}^{2}})^{\frac{2}{3}}\chi\left(2\mathcal{R}_{min}^{4}\varsigma+\cos^{2}\theta(\mathcal{R}_{min}^{2}\xi\chi-\varsigma)\right),\nonumber\\
c&=&T_{0}^{4}\prod_{i}(1+\frac{q_{i}}{r_{min}^{2}})\mathcal{R}_{min}^{6}\sin^{2}\theta\varsigma\chi^{2},\nonumber\\
a^{\prime}&=&\prod_{i}(1+\frac{q_{i}}{r_{min}^{2}})^{\frac{2}{3}}(\mathcal{R}_{min}^{2}\varsigma+\xi\chi),\nonumber\\
b^{\prime}&=&T_{0}^{2}\mathcal{R}_{min}^{2}\xi\left(\mathcal{R}_{min}^{4}\varsigma\prod_{i}
(1+\frac{q_{i}}{r_{min}^{2}})^{\frac{2}{3}}-2\mathcal{R}_{min}^{2}\varsigma-\prod_{i}(1+\frac{q_{i}}{r_{min}^{2}})^{\frac{2}{3}}\xi\chi\right),\nonumber\\
c^{\prime}&=&T_{0}^{2}\prod_{i}(1+\frac{q_{i}}{r_{min}^{2}})^{\frac{2}{3}}\mathcal{R}_{min}^{6}\varsigma\xi^{2}\left(\prod_{i}(1+\frac{q_{i}}{r_{min}^{2}})^{\frac{4}{3}}-T_{0}^{2}\right),
\end{eqnarray}
with,
\begin{eqnarray}\label{s110}
\varsigma&=&\prod_{i}(1+\frac{q_{i}}{r_{min}^{2}})y^{2}{\dot{\theta}}^{2}\mathcal{R}_{min}^{2}\sin^{2}\theta(v+x\dot{\theta}\cos\theta)^{2},\nonumber\\
\xi&=&\frac{f_{k}(r_{min})}{\prod_{i}(1+\frac{q_{i}}{r_{min}^{2}})^{\frac{1}{3}}}
-\prod_{i}(1+\frac{q_{i}}{r_{min}^{2}})^{\frac{2}{3}}(v+x\dot{\theta}\cos\theta)^{2}\mathcal{R}_{min}^{2},\nonumber\\
\chi&=&\prod_{i}(1+\frac{q_{i}}{r_{min}^{2}})^{\frac{1}{3}}y^{2}{\dot{\theta}}^{2}R_{min}^{2}\sin^{2}\theta
-\frac{f_{k}(r_{min})}{\prod_{i}(1+\frac{q_{i}}{r_{min}^{2}})^{\frac{2}{3}}},
\end{eqnarray}
where $\mathcal{R}_{min}=\frac{r_{min}}{R}$ and $r_{min}$ is the turning point. The direct consequence of rotational motion is that drag force is no longer
constant. From equation (108) one can see that the momentum densities
of string vary with respect to $x(r)$ and $y(r)$.\\
But, this result is not appropriate description of a meson.
According to previous works [88, 89] the $q\bar{q}$ pair should be
close enough together and not moving too quickly. The presence of
functions $x(r)$ and $y(r)$ in relations (110) is consequence of
relativistic motion, which is not acceptable. On the other hand,
because of non-vanishing drag forces, it is expected that the
velocity of a $q\bar{q}$ pair decreases. So, we consider a moving
heavy $q\bar{q}$ pair with non-relativistic speed, which rotates by
angel $\theta=\omega t$ around the center of mass. Indeed this
situation is corresponding to the motion of the heavy meson with
large spin. Actually, in the very large angular momentum limit, a
classical approximation is reliable. In this case, the angular
velocity of the string is very small. Therefore, we are going to
discuss the case of non-relativistic motion
($\dot{\theta}^{2}\rightarrow0$ and $\dot{\theta}v\rightarrow0$). In
that case $\varsigma=c=c^{\prime}=0$ and we have,
\begin{eqnarray}\label{s111}
(\pi_{x}^{1})^{2}&=&\frac{r_{min}^{2}}{R^{2}}T_{0}^{2}
f_{k}(r_{min})
\mathcal{H}^{-\frac{1}{3}}(r_{min}),\nonumber\\
(\pi_{y}^{1})^{2}&=&\frac{r_{min}^{2}}{R^{2}}T_{0}^{2}\left(f_{k}(r_{min})-\mathcal{H}(r_{min})\frac{r_{min}^{2}v^{2}}{R^{2}}\right)
\mathcal{H}^{-\frac{1}{3}}(r_{min}).
\end{eqnarray}
Now, we assume that $v^{2}\rightarrow0$ and angular velocity is infinitesimal constant ($\dot{\theta}=\omega\ll1$), and the quark-antiquark pair rotates
around origin. In that case we neglect $\omega^{4}$ terms and obtain values of momentum densities as the following,
\begin{equation}\label{s112}
\pi_{x}^{1}=\pi_{y}^{1}=T_{0}\frac{r_{min}}{R}\frac{\left[k-\frac{\mu}{r_{min}^{2}}
+\frac{r_{min}^{2}}{R^{2}}{\prod_{i}(1+\frac{q_{i}}{r_{min}^{2}})}\right]^{\frac{1}{2}}}
{\prod_{i}(1+\frac{q_{i}}{r_{min}^{2}})^{\frac{1}{6}}}.
\end{equation}
In order to obtain the non-zero components of momentum densities
(111) and (112) we should use negative sign in the relations (108).
Therefore, correct sign in the equations (108) is minus sign, and we
yield to constant drag forces as expected for the non-relativistic
motion. In order to extend this work one may consider more quarks,
such as four quarks in the baryon [90, 91] through the thermal
plasma.
\section{Jet-quenching parameter}
One of the interesting properties of the strongly-coupled plasma at RHIC is the jet quenching of partons produced with high transverse momentum. This
parameter controls the description of relativistic partons and it is possible to employ the gauge/gravity duality and determine this quantity in the finite
temperature gauge theories. In order to obtain the jet-quenching parameter one needs to rewrite the metric (1) in the light-cone coordinates. Therefore,
one can introduce light-cone coordinates $x^{\pm}=\frac{t\pm x^{1}}{\sqrt{2}}$, and rewrite the metric (1) in the following form,
\begin{eqnarray}\label{s113}
ds^{2}&=&\frac{1}{2}(\frac{\mathcal{H}^{\frac{1}{3}}r^{2}}{R^{2}}-\frac{f_{k}}{\mathcal{H}^{\frac{2}{3}}})
\left((dx^{+})^{2}+(dx^{-})^{2}\right)
-(\frac{\mathcal{H}^{\frac{1}{3}}r^{2}}{R^{2}}+\frac{f_{k}}{\mathcal{H}^{\frac{2}{3}}})dx^{+}dx^{-}\nonumber\\
&+&\mathcal{H}^{\frac{1}{3}}\left(\frac{r^{2}}{R^{2}}(dx_{2}^{2}+dx_{3}^{2})+\frac{dr^{2}}{f_{k}}\right).
\end{eqnarray}
We begin with the general relation for the jet-quenching parameter
[65],
\begin{equation}\label{s114}
\hat{q}\equiv8\sqrt{2}\frac{S_{I}}{L^{-}L^{2}},
\end{equation}
where $S_{I}=S-S_{0}$ ($S$ denotes $q\bar{q}$ pair action and
$S_{0}$ denotes the action of isolated $q$ and $\bar{q}$). It means
that the jet-quenching parameter is proportional to energy of the
string, so we expect that this quantity will be opposite of the drag
force which is indeed energy loss of the string. Therefore,
calculation of the jet-quenching parameter reduces to obtain actions
$S$ and $S_{0}$.\\
One can image the situation with an open string
whose endpoints lie on the brane. In the light-cone coordinates, the
string may be described by $r(\tau, \sigma)$. We use the static
gauge where $\tau=x^{-}$ and $\sigma=x^{2}\equiv y$, and all other
coordinates considered as constants. In that case $-\frac{L}{2}\leq
y \leq\frac{L}{2}$, and $L^{-}\leq x^{-} \leq0$, and because of
$L^{-}\gg L$ one can assume that the world-sheet is invariant along
the $x^{-}$ direction. Therefore, the string may described by the
function $r(y)$, so the boundary condition is
$r(\pm\frac{L}{2})=\infty$. In this configuration, the induced
metric on the string world-sheet obtained as the following,
\begin{equation}\label{s115}
2g=(\frac{\mathcal{H}^{\frac{2}{3}}r^{2}}{R^{2}}-\frac{f_{k}}{\mathcal{H}^{\frac{1}{3}}})
(\frac{r^{2}}{R^{2}}+\frac{{r^{\prime}}^{2}}{f_{k}}).
\end{equation}
Since equation (115) is dependent of coordinate $x^{-}$, one can integrate over $x^{-}$ and then the Nambu-Goto action is given by,
\begin{equation}\label{s116}
S=\frac{\sqrt{2}L^{-}}{2\pi\alpha^{\prime}}\int_{0}^{\frac{L}{2}}{dy
\sqrt{(\frac{\mathcal{H}^{\frac{2}{3}}r^{2}}{R^{2}}-\frac{f_{k}}{\mathcal{H}^{\frac{1}{3}}})
(\frac{r^{2}}{R^{2}}+\frac{1}{f_{k}}{r^{\prime}}^{2})}}.
\end{equation}
One can remove the $r^{\prime}$ by using the equation of motion. In that case, since the lagrangian density is time-dependent, one can write,
\begin{equation}\label{s117}
{\mathcal{H}}=\frac{\partial{\mathcal{L}}}{\partial
r^{\prime}}r^{\prime}-{\mathcal{L}}=Const.\equiv E.
\end{equation}
Therefore, the following relation is obtained,
\begin{equation}\label{s118}
r^{\prime2}=\frac{f_{k}r^{2}}{R^{2}E^{2}}\left[\frac{\mathcal{H}^{\frac{1}{3}}}{2R^{2}}
(\frac{\mathcal{H}^{\frac{1}{3}}r^{2}}{R^{2}}-\frac{f_{k}}{\mathcal{H}^{\frac{2}{3}}})r^{2}-E^{2}\right].
\end{equation}
Equation (118) has two important poles where $r^{\prime}=0$. The main pole exist at the horizon. So, it is clear that the equation (118) has a zero at the
horizon where $f_{k}=0$. In this case the string comes from infinity ($r(L/2)=\infty$) and touches the horizon and returns to infinity ($r(-L/2)=\infty$).
The second pole of equation (118) obtained by the following relation,
\begin{equation}\label{s119}
\frac{f_{k}r^{2}}{\mathcal{H}^{\frac{1}{3}}R^{2}}-\frac{\mathcal{H}^{\frac{2}{3}}r^{4}}{R^{4}}+2E^{2}=0.
\end{equation}
In the Ref. [92] found that the string world sheet has one end at a
Wilson line at the boundary with $Im[t]=0$, and the other end at a
Wilson line the boundary with $Im[t]=-i\epsilon$. The only way that
the string world-sheet linking these two Wilson lines can meet is if
the string world-sheet hangs down to the horizon. Therefore, the
only physical situation is the first case where the string touches
the horizon. Also in our case, drawing the $r^{\prime2}$ in terms of
$r$ tells that the turning point of
string should be $r_{h}$.\\
By using equation (118) in (116), and also the new definition of $B\equiv 1/E^{2}$, one can rewrite the Nambu-Goto action in the following form,
\begin{equation}\label{s120}
S=\frac{L^{-}\sqrt{B}}{2\pi\alpha^{\prime}}\int_{r_{h}}^{\infty}{dr
\frac{r(\frac{\mathcal{H}^{\frac{2}{3}}r^{2}}{R^{2}}-\frac{f_{k}}{\mathcal{H}^{\frac{1}{3}}})}
{\sqrt{\frac{\mathcal{H}^{\frac{1}{3}}}{2}(\frac{\mathcal{H}^{\frac{1}{3}}
r^{2}}{R^{2}}-\frac{f_{k}}{\mathcal{H}^{\frac{2}{3}}})Bf_{k}r^{2}-f_{k}R^{2}}}}.
\end{equation}
For the low energy limit ($E\rightarrow0$) we expand equation (120) to leading order in $1/B$. This is reasonable since the determination of $\hat{q}$
demands the study of the small separation limit of $L$. Then at the first order of $1/B$ one can obtain,
\begin{equation}\label{s121}
S=\frac{L^{-}}{2\pi\alpha^{\prime}}\int_{r_{h}}^{\infty}{dr
\sqrt{\frac{2\mathcal{H}^{\frac{1}{3}}}{f_{k}}(\frac{\mathcal{H}^{\frac{1}{3}}r^{2}}{R^{2}}-\frac{f_{k}}
{\mathcal{H}^{\frac{2}{3}}} )}
\left[1+\frac{R^{2}}{(\frac{\mathcal{H}^{\frac{2}{3}}r^{2}}{R^{2}}-\frac{f_{k}}{\mathcal{H}^{\frac{1}{3}}})Br^{2}}\right]}.
\end{equation}
Now, one can extract action $S_{0}$ which can be interpreted as the self-energy of the isolated quark and the isolated antiquark. In that case by using the
result of the Ref. [56] one can obtain,
\begin{equation}\label{s122}
S_{0}=\frac{L^{-}}{2\pi\alpha^{\prime}}\int_{r_{h}}^{\infty}{dr
\sqrt{\frac{2\mathcal{H}^{\frac{1}{3}}}{f_{k}}(\frac{\mathcal{H}^{\frac{1}{3}}r^{2}}{R^{2}}-\frac{f_{k}}
{\mathcal{H}^{\frac{2}{3}}})}}.
\end{equation}
Therefore, we can extract $S_{I}$ as the following,
\begin{equation}\label{s123}
S_{I}=\frac{1}{\sqrt{B}}\frac{L^{-}}{2\pi\alpha^{\prime}}\int_{r_{h}}^{\infty}{dr
\sqrt{\frac{2R^{4}}{(\frac{\mathcal{H}^{\frac{2}{3}}r^{2}}{R^{2}}-\frac{f_{k}}{\mathcal{H}^{\frac{1}{3}}})Bf_{k}r^{4}}}}.
\end{equation}
On the other hand, one can integrate equation (118) and obtain the following relation for infinitesimal $1/B$,
\begin{equation}\label{s124}
\frac{L}{2}=R^{2}\int_{r_{h}}^{\infty}{dr
\frac{1}{\sqrt{\frac{B}{2}(\frac{\mathcal{H}^{\frac{2}{3}}r^{2}}{R^{2}}
-\frac{f_{k}}{\mathcal{H}^{\frac{1}{3}}})f_{k}r^{4}}}}.
\end{equation}
Therefore, by using relations (115), (123) and (124) we can specify
the jet-quenching parameter as the following,
\begin{equation}\label{s125}
\hat{q}=\frac{(I(q))^{-1}}{\pi\alpha^{\prime}}.
\end{equation}
where,
\begin{equation}\label{s126}
I(q)=R^{2}\int_{r_{h}}^{\infty}{\frac{dr}
{\sqrt{(\frac{\mathcal{H}^{\frac{2}{3}}r^{2}}{R^{2}}-\frac{f_{k}}{\mathcal{H}^{\frac{1}{3}}})f_{k}r^{4}}}}.
\end{equation}
In order to obtain the explicit expression of the jet-quenching
parameter we set $k=1$ and consider three special cases of one, two and three charged black hole.\\
\subsection{One-charged black hole}
In the case of one-charged black hole we set $q_{1}=q,
q_{2}=q_{3}=0$ in the integral (126) and yield to the following
expression,
\begin{equation}\label{s127}
I(q_{1})=R^{4}\int_{r_{h}}^{\infty}{\sqrt{\frac{(1+\frac{q}{r^{2}})^{\frac{1}{3}}}
{(r^{2}-\mu)(r^{4}+(q+R^{2})r^{2}-\mu R^{2})}}dr},
\end{equation}
where $r_{h}$ is given by the equation (26). In order to compare our
result with the results of $\mathcal{N}=4$ SYM plasma we should use
re-scaling (31), in that case it is easy to check that our results
are agree with the case of $\mathcal{N}=4$ SYM plasma. We show this
point later for the special
case of three-charged black hole.\\
By using the numerical study, we draw the curves of the
jet-quenching parameter in terms of the black hole charge and the
temperature in Fig. 16 and Fig. 17 respectively. These figures show
that the jet-quenching parameter of the $\mathcal{N}=2$ theory is
larger than the jet-quenching parameter of the $\mathcal{N}=4$
theory.\\
For example by choosing $R^{2}=\alpha^{\prime}\sqrt{\lambda}$,
$\alpha^{\prime}=0.5$, $\lambda=6\pi$, $q=10^6$ and $T=300$ $MeV$
one can obtain, $\hat{q}=42$ $GeV^2/fm$ in STU model, while
$\mathcal{N}=4$ SYM plasma gives $\hat{q}\approx4.5$ $GeV^2/fm$. In
that case the thermodynamical stability lets us choose
$q\sim\times10^6$ for $T=300 MeV$. On the other hand, for the small
black hole charge, by taking $\alpha^{\prime}=0.5$ and
$\lambda=6\pi$ one can obtain $\hat{q}=37.5$ $GeV^2/fm$. It means
that the black hole charge
increases the jet-quenching parameter.\\
In order to obtain $\hat{q}=5$ $GeV^2/fm$ the corresponding
temperature of the QGP is $155 MeV$, which is smaller than expected
[93].
\begin{figure}[th]
\begin{center}
\includegraphics[scale=.4]{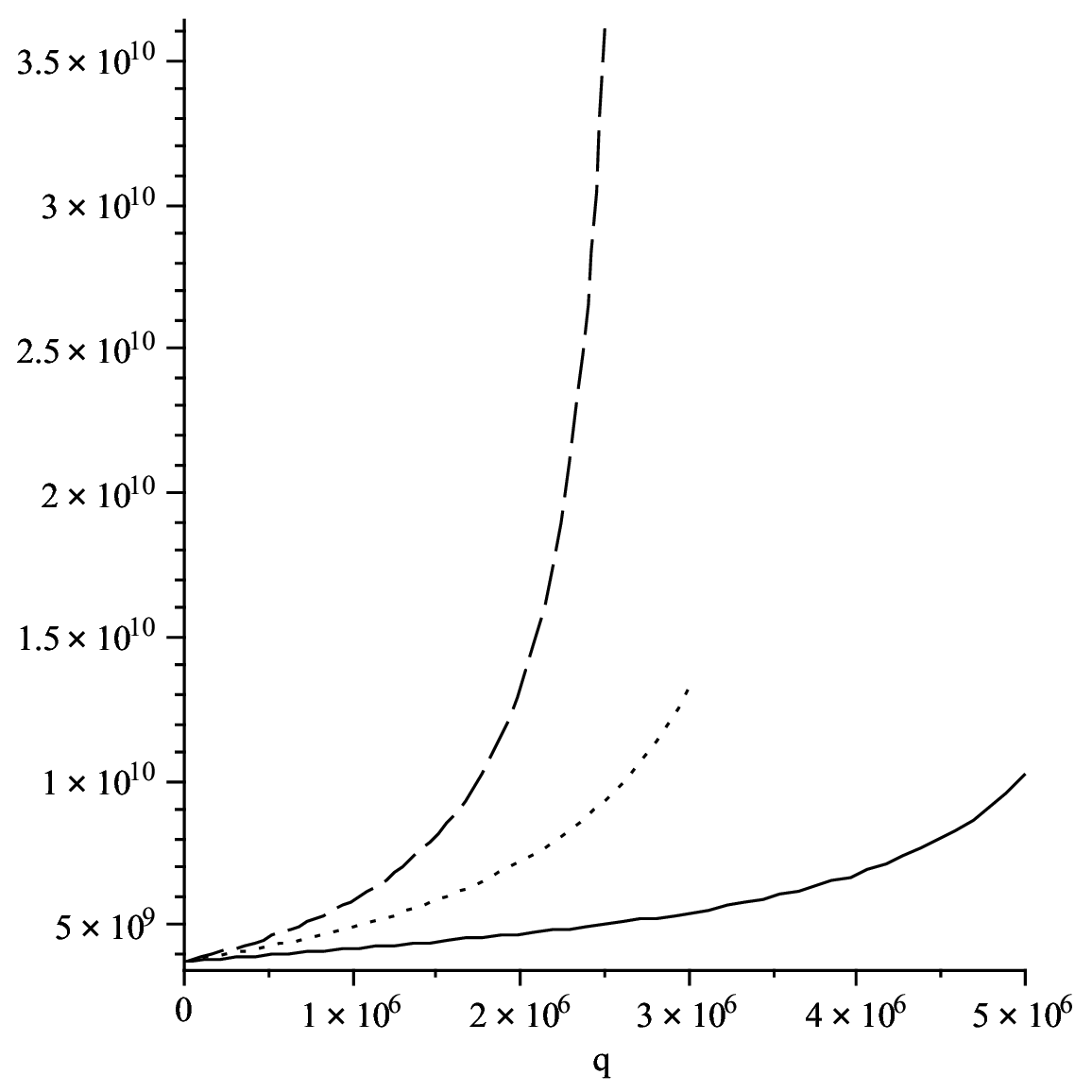}
\caption{Plot of the jet-quenching parameter in terms of the black hole charge. We fixed our parameters as $\alpha^{\prime}=0.5$, $\lambda=6\pi$, and
$T=300$ $MeV$. The solid line represents the case of $q_{1}=q, q_{2}=q_{3}=0$. The dotted line represents the case of $q_{1}=q_{2}=q, q_{3}=0$. The dashed
line represents the case of $q_{1}=q_{2}=q_{3}=q$. It show that increasing the number of black hole charges increases the value of the jet-quenching
parameter.}
\end{center}
\end{figure}

\begin{figure}[th]
\begin{center}
\includegraphics[scale=.4]{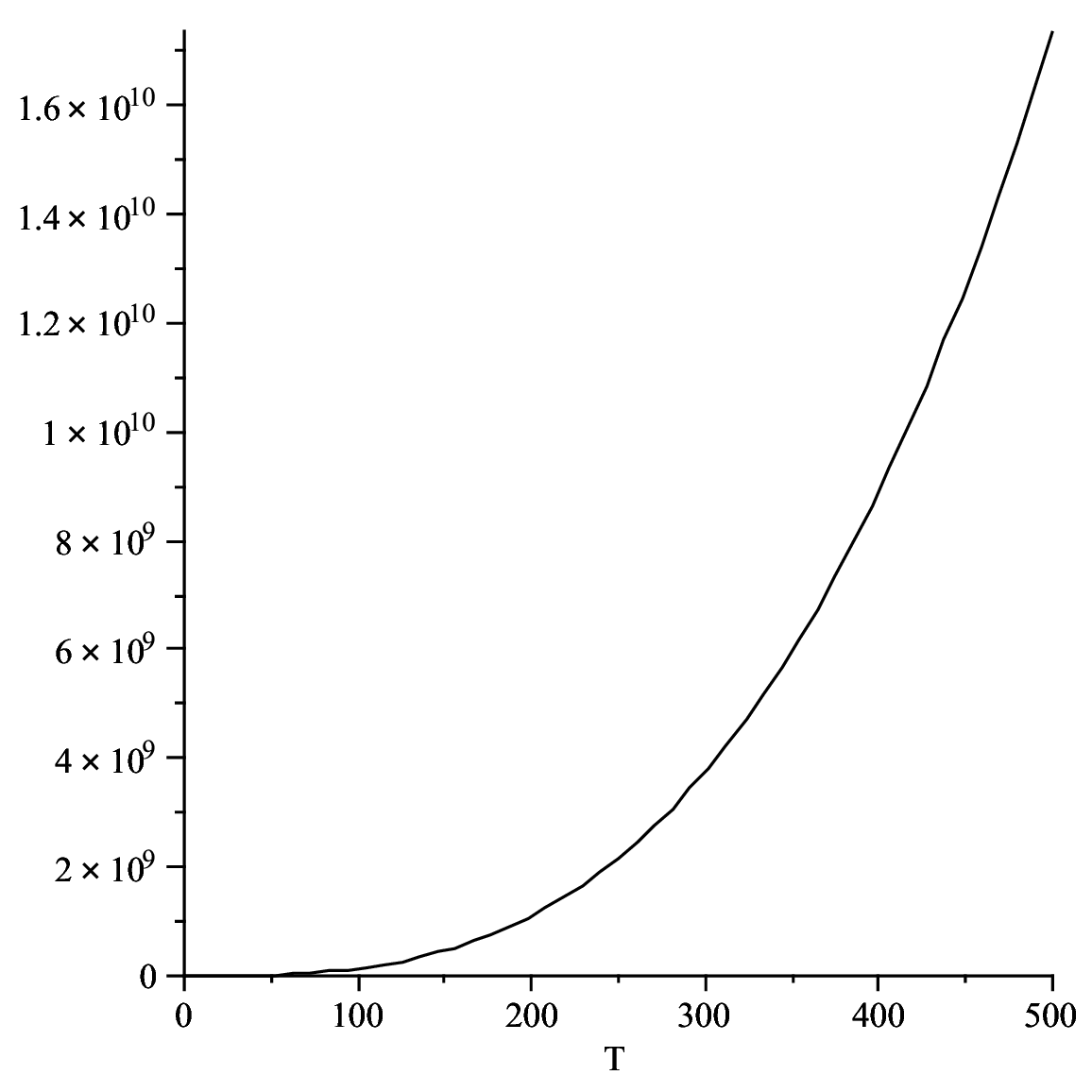}
\caption{Plot of the jet-quenching parameter in terms of the temperature for small black hole charge. We fixed the parameters as $\alpha^{\prime}=0.5$,
$\lambda=6\pi$. In that case three different cases of one, two, and three-charged black hole have similar behavior.}
\end{center}
\end{figure}

\subsection{Two-charged black hole}
In the case of two-charged black hole we set $q_{1}=q_{2}=q,
q_{3}=0$. So, the integral (126) reduces to the following
expression,
\begin{equation}\label{s128}
I(q_{1,2})=R^{4}\int_{r_{h}}^{\infty}{\sqrt{\frac{(1+\frac{q}{r^{2}})^{\frac{2}{3}}}
{\rho(r^{4}+(2q+R^{2})r^{2}-\mu R^{2}+q^{2})}}dr},
\end{equation}
where $r_{h}=\pi R^{2}T$, and we defined,
\begin{equation}\label{s129}
\rho\equiv((R^{2}-1)r^{4}+(2qR^{2}-R^{2}-q)r^{2}+R^{2}q^{2}+\mu
R^{2}-q^{2}).
\end{equation}
By using the numerical study, we find the jet-quenching parameter in
terms of the black hole charge and the temperature in the Fig. 16
and Fig. 17 respectively. These plots show that the jet-quenching
parameter of the $\mathcal{N}=2$ theory is larger than the
jet-quenching parameter of the $\mathcal{N}=4$ theory. Also, we find
that the jet-quenching parameter of the two-charged black hole is
larger than the jet-quenching parameter of the one-charge black
hole. For example by choosing $R^{2}=\alpha^{\prime}\sqrt{\lambda}$,
$\alpha^{\prime}=0.5$, $\lambda=6\pi$, $q=10^6$ and $T=300$ $MeV$
one can obtain $\hat{q}=49$ $GeV^2/fm$. In that case the
thermodynamical stability lets us choose $q\sim\times10^6$ for
$T=300 MeV$. If we consider small value of the black hole charge
then find the same value of the jet-quenching parameter as the
previous case, and this point is illustrated in Fig. 17. Therefore,
in order to obtain $\hat{q}=5$ $GeV^2/fm$, the corresponding
temperature of the QGP is $155 MeV$ for small black hole charge.
\subsection{Three-charged black hole}
In the last case we set three charges equal ($q_{1}=q_{2}=q_{3}=q$).
As we know, this configuration of STU model is identical to the
Reissner-Nordstrom-$AdS_{5}$ black hole [94]. In that case the
integral (126) reduces to the following expression,
\begin{equation}\label{s130}
I(q_{1,2,3})=R^{4}\int_{r_{h}}^{\infty}{\sqrt{\frac{r^{2}(r^{2}+q)}
{\varrho(r^{6}+(R^{3}+3q)r^{4}+(3q^{2}-\mu R^{2})r^{2}+q^{3})}}dr},
\end{equation}
where we defined,
\begin{equation}\label{s131}
\varrho\equiv((R^{2}-1)r^{6}+(3qR^{2}-R^{2}-3q)r^{4}+(3R^{2}q^{2}+\mu
R^{2}-3q^{2})r^{2}+(R^{2}-1)q^{3}),
\end{equation}
and $r_{h}$ is given by the relation (29). Numerically, we give plots of the jet-quenching parameter in terms of the black hole charge and the temperature
in Fig. 16 and Fig. 17 respectively. These plots show that the jet-quenching parameter of the $\mathcal{N}=2$ theory is larger than the jet-quenching
parameter of the $\mathcal{N}=4$ theory. Also we find that the jet-quenching parameter of the three-charged black hole is larger than the jet-quenching
parameter of the one-charge and two-charged black holes. For example by choosing $R^{2}=\alpha^{\prime}\sqrt{\lambda}$, $\alpha^{\prime}=0.5$,
$\lambda=6\pi$, $q=10^6$ and $T=300$ $MeV$ one can obtain $\hat{q}=58$ $GeV^2/fm$. In that case the thermodynamical stability lets us choose
$q\sim\times10^6$ for $T=300$ $MeV$. If we consider small value of the black hole charge then find the same value of the jet-quenching parameter as the
previous cases, and this point is illustrated in Fig. 17. Therefore, in order to obtain $\hat{q}=5$ $GeV^2/fm$ the corresponding temperature of the QGP is
$155 MeV$ for a small black hole
charge.\\
As we promised already in order to compare our results with the case of $\mathcal{N}=4$ SYM we also perform the re-scaling (31) which yields us to obtain
the following result,
\begin{equation}\label{s132}
\hat{q}=\frac{r_{0}^{2}}{\pi\alpha^{'}R^{4}}\left[\int_{r_{h}}^{\infty}{\frac{dr}
{r^{2}\sqrt{\frac{f}{H}}}}\right]^{-1},
\end{equation}
where
\begin{eqnarray}\label{s133}
f&=&H^{3}-\frac{r_{0}^{4}}{r^{4}} ,\nonumber\\
H&=&1+\frac{q}{r^{2}},
\end{eqnarray}
which agree with the results of the Refs. [67, 68], where the
jet-quenching parameter calculated with the chemical potential. The
horizon radius $r_{0}$ obtained for the case of zero-charge black
hole. For the black hole with non-vanishing charges, it is clear
that the horizon radius decreases ($r_{h}<r_{0}$). From the relation
(19) we know that the $q=0$ limit is equal to $\phi=0$ limit and one
can say that the jet-quenching parameter from the $\mathcal{N}=2$
supergravity theory with zero chemical potential is equal to the
jet-quenching parameter from the $\mathcal{N}=4$ SYM theory.

\subsection{Effect of the constant electric field}
In this subsection, similar to the subsection 5.3, we would like to
find effect of the constant electric field by adding a two form
$F=B_{01}dt\wedge dx_{1}$ as a constant electric field to the line
element (1). Antisymmetric field $B_{01}\equiv e$ is the constant
electric field. Now, we are going to obtain the effect of the
constant electric field on the jet-quenching parameter. In that case
the Nambu-Goto action is given by using the following equation,
\begin{equation}\label{s134}
2g=(\frac{\mathcal{H}^{\frac{2}{3}}r^{2}}{R^{2}}-\frac{f_{k}}{\mathcal{H}^{\frac{1}{3}}}+e)
(\frac{r^{2}}{R^{2}}+\frac{{r^{\prime}}^{2}}{f_{k}}).
\end{equation}
Therefore, one can obtain the jet-quenching parameter as the
following,
\begin{equation}\label{s135}
\hat{q}=\frac{(I(q, e))^{-1}}{\pi\alpha^{\prime}},
\end{equation}
where,
\begin{equation}\label{s136}
I(q, e)=R^{2}\int_{r_{h}}^{\infty}{\frac{dr} {\sqrt{(\frac{\mathcal{H}^{\frac{1}{3}}r^{2}}{R^{2}}-\frac{f_{k}}{\mathcal{H}^{\frac{2}{3}}}+e)
\mathcal{H}^{\frac{1}{3}}f_{k}r^{4}}}},
\end{equation}
and $f$ and $\mathcal{H}$ are given by the relation (2). In order to
find the effect of the constant electric field on the jet-quenching
parameter we examine above integral for three different cases of
one, two and three-charged black hole.\\
Numerically, and under near boundary approximation, we draw graph of
the jet-quenching parameter in terms of the constant electric field
and find that the constant electric field increases the value of the
jet-quenching parameter.\\
In the Fig. 18 we draw the jet-quenching parameter in terms of the
constant electric field for the large black hole charge. It shows
that the effect of the constant electric field is increasing the
jet-quenching parameter.
\begin{figure}[th]
\begin{center}
\includegraphics[scale=.4]{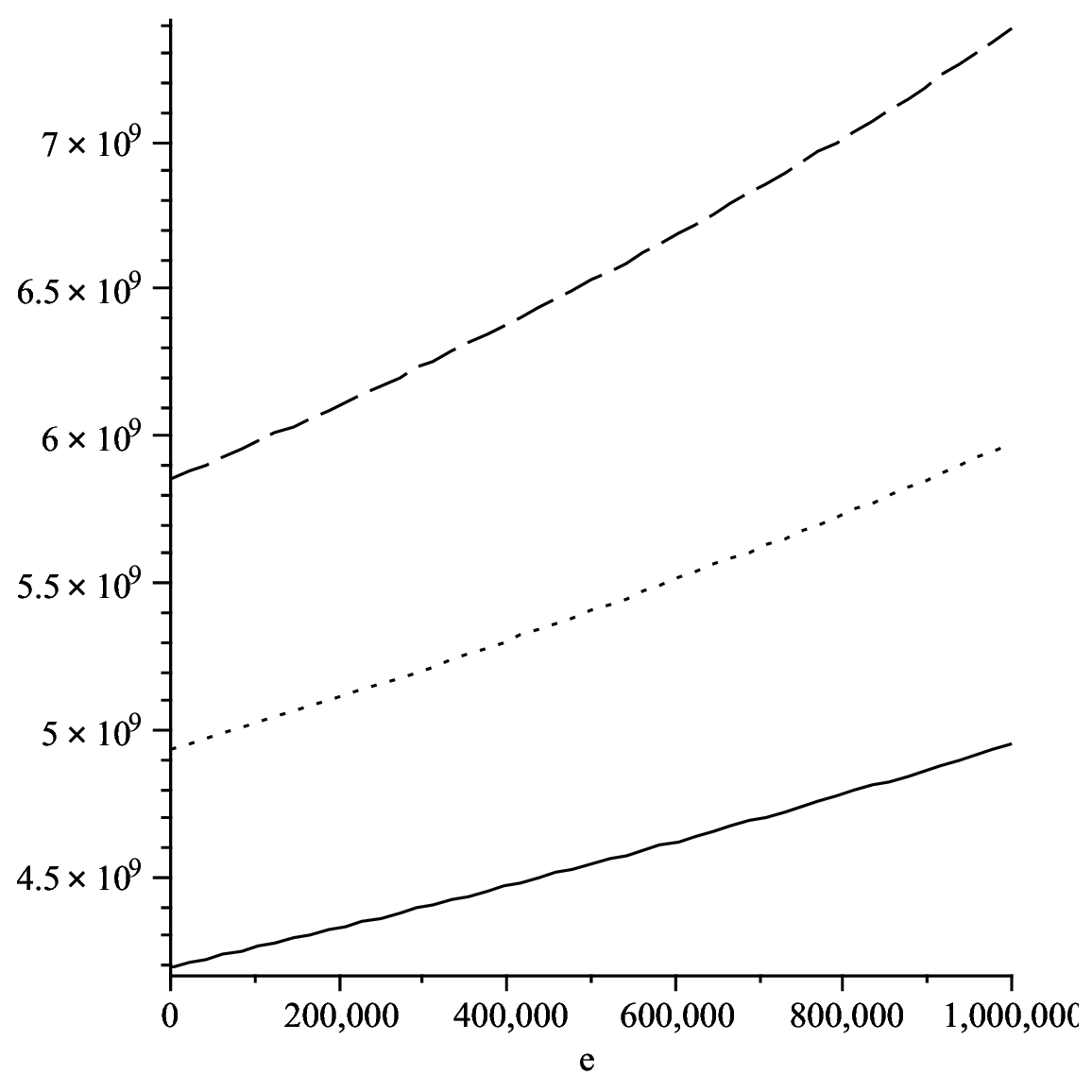}
\caption{Plot of the jet-quenching parameter in terms of the
constant electric field. We fixed our parameters as
$\alpha^{\prime}=0.5$, $\lambda=6\pi$, $q=10^{6}$ and $T=300$ $MeV$.
The solid line represents the case of $q_{1}=q, q_{2}=q_{3}=0$. The
dotted line represents the case of $q_{1}=q_{2}=q, q_{3}=0$. The
dashed line represents the case of $q_{1}=q_{2}=q_{3}=q$. It shows
that the jet-quenching parameter increased by the constant electric
field.}
\end{center}
\end{figure}
\subsection{Higher derivative correction}
Finally, in absence of any external field, we would like to
calculate the effect of higher derivative terms on the jet-quenching
parameter. In that case the jet-quenching parameter obtained as the
following expression,
\begin{equation}\label{s137}
\hat{q}=\frac{(I(q, c_{1}))^{-1}}{\pi\alpha^{\prime}},
\end{equation}
where,
\begin{equation}\label{s138}
I(q, c_{1})=\int_{r_{h}}^{\infty}{\frac{dr}
{\sqrt{(\frac{H^{2}r^{2}}{R^{2}}-\frac{f_{k}}{H})f_{k}r^{4}}}},
\end{equation}
also, we used relations (14) and (15) for the case of
$q_{1}=q_{2}=q_{3}=q$. One can study near boundary behavior of the
jet-quenching parameter and find that the higher derivative terms
include at $\mathcal{O}(\frac{c_{1}}{T^9})$. In that case we find
that the higher derivative terms decrease the value of the
jet-quenching parameter.\\
So, for the fixed parameters such as
$\alpha^{\prime}=0.5$, $\lambda=6\pi$, $T=300$ $MeV$ and small black
hole charge, we obtain $c_{1}<0.00021$ to have positive
jet-quenching parameter, which is agree with the result of the
subsection 5.4. For example with the above fixed parameters and
$c_{1}=0.0001$ one can obtain $\hat{q}=4.6$ $GeV^2/fm$ which is
approximately value of the jet-quenching parameter of the
$\mathcal{N}=4$ SYM theory. In order to obtain $\hat{q}=5$
$GeV^2/fm$ the corresponding higher derivative
parameter should be $c_{1}\approx97\times10^{-4}$ at $T=300 MeV$.\\
Again, we can use re-scaling (31) and obtain,
\begin{equation}\label{s139}
\hat{q}=\frac{r_{0}^{2}}{\pi\alpha^{'}R^{4}}\left[\int_{r_{h}}^{\infty}{\sqrt{\frac{H}{f}}\frac{dr}{r^{2}}}\right]^{-1},
\end{equation}
where,
\begin{eqnarray}\label{s140}
f&=&(1+\frac{q}{r^{2}})^3-\frac{r_{0}^{4}}{r^{4}}+\frac{c_{1}
r_{0}^{4}}{24R^{2}r^4}
\left[\frac{r_{0}^{4}}{4r^{2}(r^{2}+q)}-\frac{8q}{3}\right] ,\nonumber\\
H&=&1+\frac{q}{r^{2}}-\frac{c_{1} q
r_{0}^{4}}{24R^{2}r^{4}(r^{2}+q)},
\end{eqnarray}
and radius $r_{h}$ is the root of the $f=0$ from the equation (140).
The equation (139) may be solved numerically, and explicit
expression of the jet-quenching parameter can be obtained. But it is
clear that the effect of higher derivative correction is to decrease
the jet-quenching parameter. One can check this statement by taking
$q=0$ limit. In this limit the jet-quenching parameter derived as,
\begin{equation}\label{s141}
{\hat{q}}_{0}=\frac{r_{0}^{2}}{\pi\alpha^{\prime}R^{4}}
\left[\int_{r_{h}}^{\infty}{4\sqrt{\frac{6R^{2}r^{4}}{96R^{2}r^{4}(r^{4}-r_{0}^{4})+c_{1}r_{0}^{8}}}
\frac{dr}{r}}\right]^{-1},
\end{equation}
where,
\begin{equation}\label{s142}
r_{h}^{4}=\frac{r_{0}^{4}}{2}\left(1+\sqrt{1-\frac{c_{1}}{24R^{2}}}\right).
\end{equation}
In that case it is necessary that $c_{1}<24\alpha^{\prime}\sqrt{\lambda}$. Comparing equation (141) with the jet-quenching parameter of the $\mathcal{N}=4$
SYM theory tell us that the effect of $c_{1}$ is decreasing the jet-quenching parameter.
\section{Conclusion}
In this paper we studied some important quantities to understand the
nature of QGP more exactly. Indeed, we considered thermal QGP
include a chemical potential. This chemical potential comes from
$\mathcal{N}=2$ supergravity in 5 dimensions. This theory contains a
non-extremal black hole with three electrical charges and well known
as STU model. First of all we reviewed properties of STU model and
extracted their equations. We studied thermodynamics of STU
background and extracted the Hawking temperature, entropy density,
specific heat and free energy of QGP. We found that the black hole
charge increase the value of specific heat. In order to compare our
results with the $\mathcal{N}=4$ SYM plasma we used special
re-scaling which actually was a transformation to the flat space.\\
We investigated the ratio of shear viscosity to entropy density and found that the universality of $\eta/s$ is valid also in STU model. Also, we found that
the shear viscosity is decreasing for the cases of one-charged and three-charged black holes and is increasing for the case of two-charged black hole. We
discussed briefly about
thermal and electrical conductivities of QGP.\\
Then, we considered problem of the drag force and found energy loss
of single quark and quark-antiquark pair. We showed that the value
of the drag force enhanced due to the black hole charges. Also we
calculated diffusion coefficient of the quark for the three special
cases of one, two and three-charged black holes. We found that the
effect of constant electric field is decreasing of the drag force,
but higher derivative terms may be increases or decreases the value
of drag force. It depend to the value of higher derivative
parameter.\\
Here, we found interesting relation between drag force
of the single quark (72) and entropy density (20) which is
$s^{2}\propto\dot{P}^{3}$. This relationship is valid also in the
case of $\mathcal{N}=4$ SYM plasma. We discuss about this relation
and also other interesting
relations end of this section.\\
Finally we studied the jet-quenching parameter and found that the jet-quenching parameter like the drag force enhanced due to the black hole charges. It
means that the energy of the string in $\mathcal{N}=2$ thermal plasma is larger than the string in $\mathcal{N}=4$ thermal plasma, hence the string in
$\mathcal{N}=2$ thermal plasma lose more energy than the string in $\mathcal{N}=4$ thermal plasma. In this case we found that the constant electric field
enhances the jet-quenching parameter, but higher derivative terms decreases the value of jet-quenching parameter. We examine our solution for three special
cases of one, two and three-charged black holes. All cases yield to the same value of the jet-quenching parameter for the small black hole charge. However,
thermodynamical stability allow to choose the black hole charge of order $10^6$. In that case we found $\hat{q} = 42, 49$ and $58 GeV^{2}/fm$ for one, two
and three-charged black hole respectively. These values of the jet-quenching parameter are far from experiments of RHIC (experimental data tell us that ($5
< \hat{q} < 25$). There is no worry for this statement because the temperature of the $\mathcal{N}=2$ supergravity theory should given smaller than the
$\mathcal{N}=4$ SYM theory. In that case with the temperature about $155 MeV$ we obtained the jet-quenching parameter in the experimental
range.\\
Let us now compare some interesting ratios of QGP quantities. First,
we summarize results of the $\mathcal{N}=4$ SYM theory.
The entropy density, drag force of moving heavy quark and jet-quenching
parameter of $\mathcal{N}=4$ SYM QGP are given by,
\begin{eqnarray}\label{s143}
s&=&\frac{\pi^{2}}{2}N^{2}T^{3},\nonumber\\
\dot{P}&=&\frac{\pi}{2}v\sqrt{\lambda}T^{2},\nonumber\\
\hat{q}&=&\frac{\pi^{2}}{a}\sqrt{\lambda}T^{3},
\end{eqnarray}
where $a=1.311$ is a constant and $\lambda$ is 't Hooft coupling.
Now, it is clear that,
\begin{eqnarray}\label{s144}
\frac{s}{\dot{P}}&\propto& T,\nonumber\\
\frac{s}{\hat{q}}&\propto& Const.\nonumber\\
\frac{\hat{q}}{\dot{P}}&\propto& T.
\end{eqnarray}

\begin{figure}[th]
\begin{center}
\includegraphics[scale=.25]{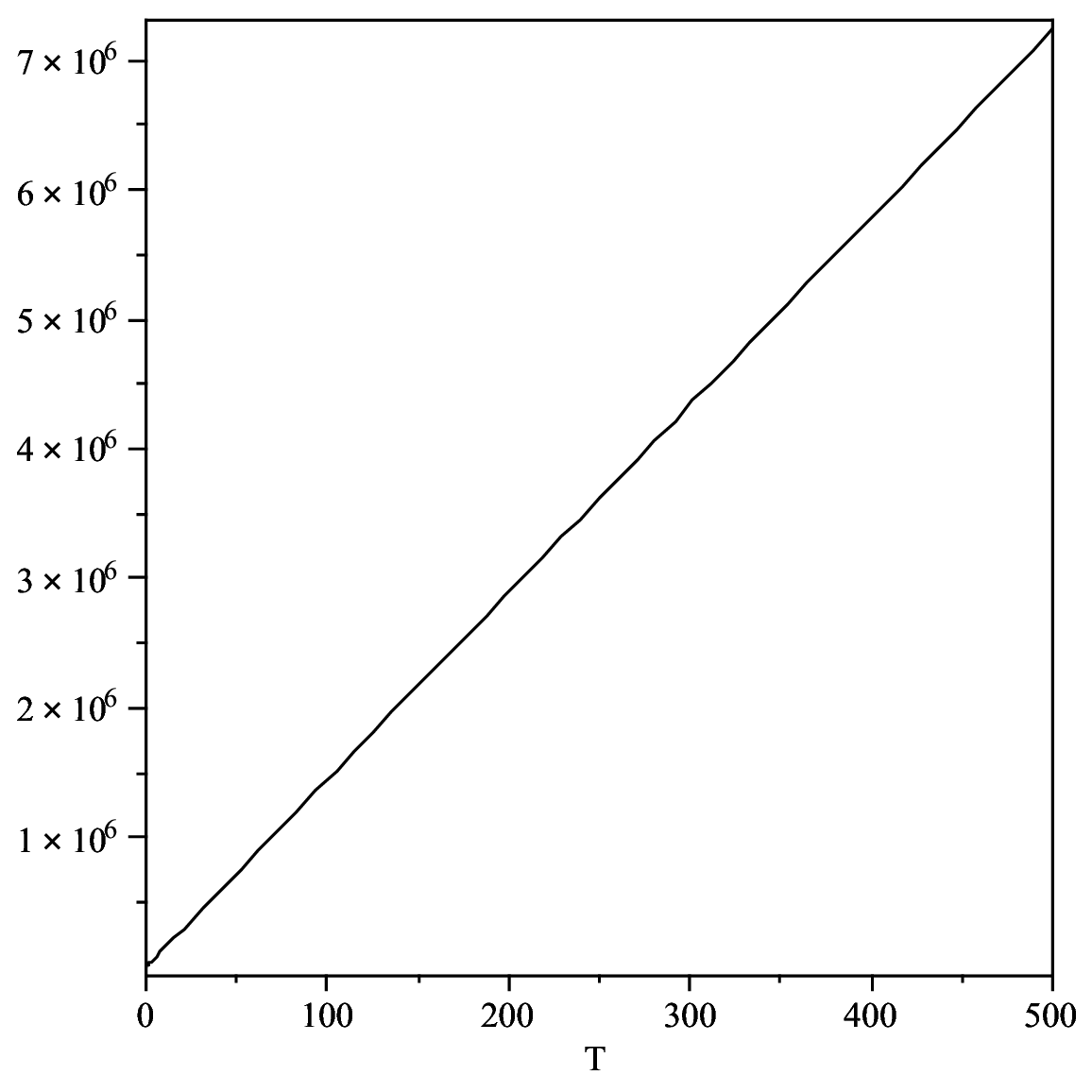}
\caption{Plot of the $s/\dot{P}$ in terms of the temperature $T$. We fixed our parameters as $\alpha^{\prime}=0.5$, $\lambda=6\pi$ and $q=1$. Three cases
of one, two and three-charged black holes have similar manner for small black hole charge. It show that the $s/\dot{P}$ is linear in $T$.}
\end{center}
\end{figure}

It is interesting to study such relations in the $\mathcal{N}=2$
supergravity thermal plasma. We obtained entropy density (20), drag
force of moving heavy quark (72) and jet-quenching parameter (125)
of $\mathcal{N}=2$ QGP. We can draw graph of $s/\dot{P}$,
$s/\hat{q}$ and $\hat{q}/\dot{P}$ to investigate behavior of these
ratios. In the Fig. 19 we give $s/\dot{P}$ in terms of the
temperature and find linear behavior of $T$. So, it is in agreement
of
$\mathcal{N}=4$ case, therefore we can claim $s/\dot{P}\propto T$ is valid at the both $\mathcal{N}=4$ and $\mathcal{N}=2$ cases.\\
Also we draw $s/\hat{q}$ and $\hat{q}/\dot{P}$ in terms of the
temperature in the Fig. 20, and find that $s/\hat{q}$ yields to a
constant, and
$\hat{q}/\dot{P}$ has linear behavior of $T$ which are in agreement of $\mathcal{N}=4$ case.\\
In the recent works a general non-extremal rotating charged AdS
black holes in five-dimensional $U(1)^{3}$ gauged supergravity [95]
and also higher dimensional one studied [96, 97]. Now, it is
interesting to generalized results of this paper to these cases.\\
Also, it is interesting to check validity of the relations (144) for
some different models such as thermal non-relativistic
non-commutative Yang-Mills plasma [98, 99, 100].\\

\begin{figure}[th]
\begin{center}
\includegraphics[scale=.25]{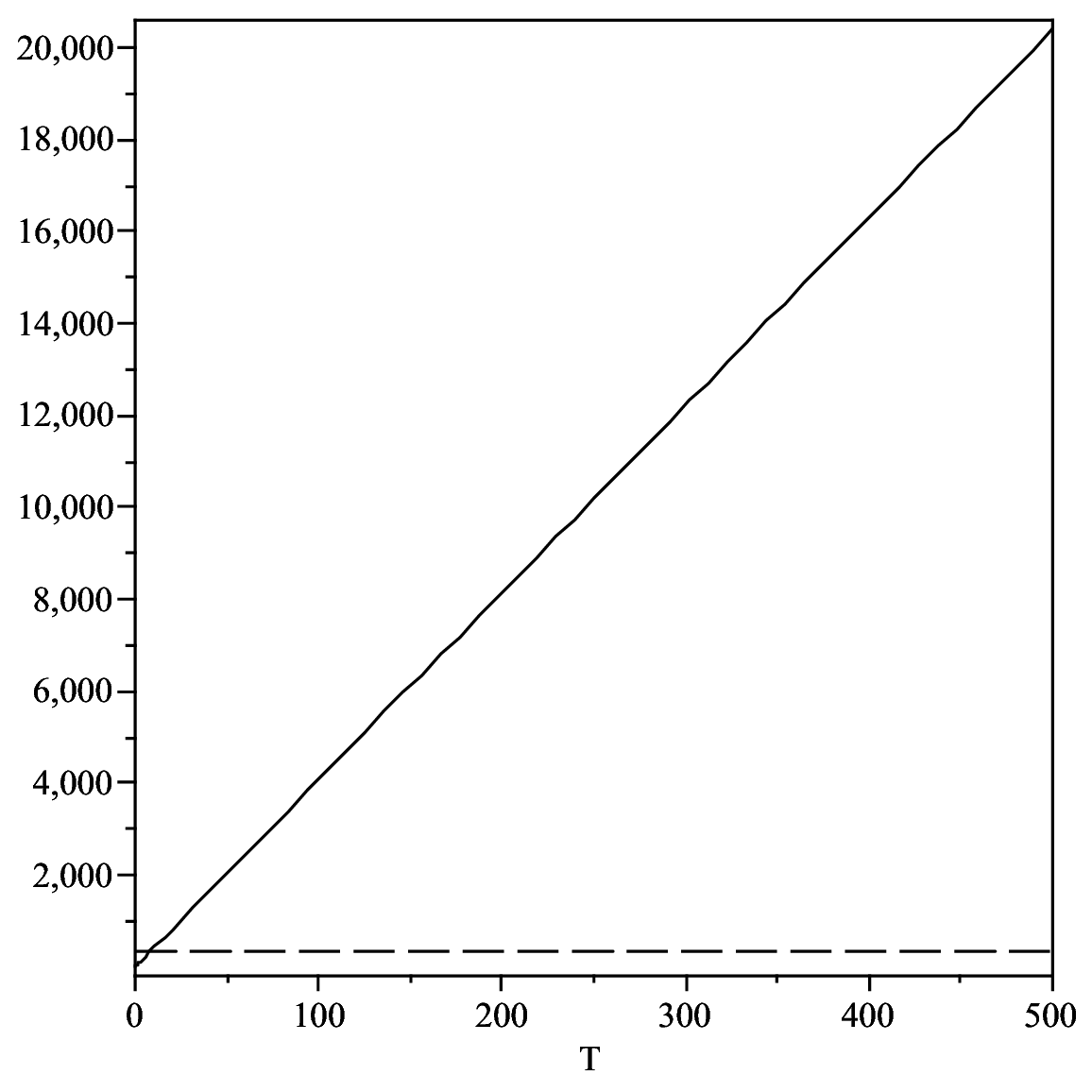}
\caption{Plot of the $s/\hat{q}$ (dashed line) and $\hat{q}/\dot{P}$ (solid line) in terms of the temperature $T$. We fixed our parameters as
$\alpha^{\prime}=0.5$, $\lambda=6\pi$ and $q=1$. Three cases of one, two and three-charged black holes have similar manner for small black hole charge. It
show that the $s/\hat{q}$ is a constant and $\hat{q}/\dot{P}$ is linear in $T$.}
\end{center}
\end{figure}

{\bf Acknowledgments} It is pleasure to thanks C. P. Herzog for reading manuscript and giving good suggestions. Also we would like to thanks Jose Edelstein
and A. R. Amani for discussion about shear viscosity, and K. B. Fadafan for his collaboration about the jet-quenching parameter.

\end{document}